%% file: coda1.0.mn.tex
\documentclass[usenatbib,fleqn]{mnras}
\usepackage{amsmath,amssymb}
\usepackage{graphics}
\usepackage{delarray}
\usepackage{graphicx}
\usepackage[table]{xcolor}
\usepackage{url}
\usepackage{breakurl}
\usepackage{rotating}
\usepackage{lscape}
\usepackage[T1]{fontenc}
\usepackage{ae,aecompl}

\newcommand{\Zsun}{{{\rm Z_{\odot}}}}

\newcommand{\zcoda}{$z_{\rm sim}$}

\newcommand{\fesc}{\rm f_{esc,\star}}

\newcommand{\Lbox}{\rm L_{box}}

\newcommand{\hmpc}{$\rm{h^{-1}Mpc}$ }
\newcommand{\hmkpc}{$\rm{h^{-1}kpc}$ }

\newcommand{\pier}[1]{\noindent{}}
\newcommand{\rev}[1]{\noindent{}}

\newcommand{\Msun}{\ensuremath{\mathrm{M}_{\odot}} }
\newcommand{\Msunnospace}{\ensuremath{\mathrm{M}_{\odot}}}

\newcommand{\hmo}{{\rm h^{-1}}}

\newcommand{\nicefrac}[2]{\leavevmode\kern.1em
            \raise.5ex\hbox{\the\scriptfont0 #1}\kern-.1em
      /\kern-.15em\lower.25ex\hbox{\the\scriptfont0 #2}}

\title[CoDa: Cosmic Dawn]{Cosmic Dawn (CoDa): the First Radiation-Hydrodynamics Simulation of Reionization and Galaxy Formation in the Local Universe}

\author[P. Ocvirk et al.]{
Pierre Ocvirk,$^{1}$ 
Nicolas Gillet,$^{1}$ 
Paul R. Shapiro,$^{2}$ 
Dominique Aubert,$^{1}$
\newauthor
Ilian T. Iliev,$^{3}$ 
Romain Teyssier,$^{4}$ 
Gustavo Yepes,$^{5,6}$
Jun-Hwan Choi,$^{2}$
\newauthor
David Sullivan,$^{3}$ 
Alexander Knebe,$^{5,6}$
Stefan Gottl\"{o}ber,$^{7}$ 
Anson D'Aloisio,$^{2,8}$ 
\newauthor
Hyunbae Park,$^{2,9}$
Yehuda Hoffman,$^{10}$ 
and Timothy Stranex$^{4}$
\\
$^1$Observatoire Astronomique de Strasbourg, Universit\'e de Strasbourg, CNRS UMR 7550, 11 rue de l'Universit\'e, 67000 Strasbourg, France\\
$^2$Department of Astronomy, University Texas, Austin, TX 78712-1083, USA\\
$^3$Astronomy Center, Department of Physics \& Astronomy, Pevensey II Building, University of Sussex, Falmer, Brighton BN1 9QH, United Kingdom\\
$^4$Institute for Theoretical Physics, University of Zurich, Winterthurerstrasse 190, CH-8057 Z\"urich, Switzerland\\
$^5$Grupo de Astrof\'{i}sica, Departamento de Fisica Teorica, Modulo C-8, Universidad Aut\'{o}noma de Madrid, Cantoblanco E-280049, Spain\\
$^6$Astro-UAM, UAM, Unidad Asociada CSIC\\
$^7$Leibniz-Institute f\"{u}r Astrophysik Potsdam (AIP), An der Sternwarte 16, D-14482 Potsdam, Germany\\
$^8$University of Washington, Dept. of Astronomy, 3910 15th Ave NE, Seattle, WA 98195-0002, USA\\
$^9$Korea Astronomy and Space Science Institute, Daejeon, 305-348, Korea\\
$^{10}$Racah Institute of Physics, Hebrew University, Jerusalem 91904, Israel
}
\date{Accepted XXX. Received YYY; in original form ZZZ}

\pubyear{2015}

\begin{document}
\label{firstpage}
\pagerange{\pageref{firstpage}--\pageref{lastpage}}
\maketitle

\begin{abstract}
\include{abstract_paul.mn}
\end{abstract}



%




\section{Introduction}
{The epoch of reionization (hereafter, "EoR") resulted from 
the escape of ionizing radiation from the first generation of 
star-forming galaxies into the otherwise cold, neutral gas of the
primordial intergalactic medium ("IGM").  This radiation created 
intergalactic H II regions of ever-increasing size, surrounding the 
galaxies that created them, leading eventually to the complete overlap
of neighboring H II regions and the end of the EoR within the first
billion years after the big bang.  The growth and geometry of the 
H II regions reflected that of the underlying galaxies as the radiation
sources and their spatial clustering, and that of the
density fluctuations in the intergalactic gas as the absorbing medium.
The story of cosmic reionization is inseparable, therefore, from 
that of the emergence of galaxies and large-scale structure in the 
universe.  Recently, the observational frontier has begun to push
the look-back-time horizon accessible to direct observation back into
this period.  High-redshift galaxies and quasars have been observed,
for example, which constrain the evolution of the IGM opacity to
H Lyman alpha resonance scattering, the mean ionizing flux density,
the UV luminosity function of galaxies, and the cosmic star formation
history, while observations of the cosmic microwave background ("CMB")
constrain the mean elecron scattering optical depth integrated
thru the entire IGM over time, back to the epoch of recombination, and 
give some information about the evolution of the mean ionized fraction.
Observational probes of the large-scale structure of reionization and 
its evolving "patchiness" are just beginning.  The cosmic 21cm background
from the evolving patchwork of intergalactic H I regions, not yet reionized,
is a prime example, which is an important science-driver for the development
of a new generation of low-frequency radio telescope arrays, such as LOFAR, MWA, PAPER, and SKA.}



The EoR is also thought to be important for its impact on galaxy formation. It has been suggested that the rising
intergalactic UV radiation field is responsible for suppressing the star formation of low-mass galaxies,
by removing gas from the smallest galaxies --- minihalos --- by photo\-evaporation \citep{shapiro2004,iliev2005} and through the suppression of gas infall onto low-mass galaxies above the minihalo mass range, by photoheating the intergalactic gas and raising its pressure \citep{shapiro1994,gnedin2000,hoeft2006}, affecting their star formation efficiency.

This process could provide a credible solution to the ``missing satellites problem'' \citep{klypin1999,moore1999}, by inhibiting star formation in low mass galaxies at early times \citep{bullock2000,benson2002a,benson2002b,benson2003}. In this framework, a number of semi-analytical models (hereafter SAMs) have been shown to reproduce well the satellite population of the Milky Way (hereafter MW), such as \cite{koposov2009,munoz2009,busha2010,maccio2010,li2010,font2011}. 
They suggest that the  ultra-faint dwarf galaxies (hereafter UFDs) discovered by the SDSS \citep{martin2004,willman2005short,zucker2006,belokurov2007short, irwin2007short, walsh2007} are effectively reionisation fossils, living in dark matter sub-haloes of about $10^{6-9} \Msun$. Deep HST observations have recently made it possible to start testing this idea. While this is currently a matter of debate, it seems that at least the low-mass satellites of the MW and M31 may have star formation histories compatible with a very early suppression \citep{brown2014,weisz2015}, e.g. by reionization.
In this context, we are led to the exciting possibility that low-mass satellites in the Local Group can be used as probes of the local reionization process.

While the reionization of the Universe is often considered to be complete at $z=6$ \citep{fan2006}, further quasar Lyman $\alpha$ observations have shown that extended opaque hydrogen troughs may still exist below this redshift, highlighting the patchiness of the process and suggesting that the reionization may actually still be ongoing at $z=6$ and end as low as $z\sim5$ \cite{becker2015}.
Given this observational evidence, one can only expect that the reionization of the Local Group must have been very different from a uniform or instantaneous event. 
Indeed, \cite{iliev2011} described two distinct reionization scenarios for the Local Group, internal (and slow, driven by sources within the Local Group) or external and fast, driven by a massive, rare object such as Virgo. Many more scenarios can be conceived, depending on the physics considered (e.g. quasars, X-ray binaries), and it would be desirable, as a long term goal, to establish which properties of the local satellites population could allow us to discriminate between these scenarios.

Already, \cite{ocvirk2011} showed that the structure of the UV background during reionization has a strong impact on the properties of the satellite population of galaxies. In particular, they showed that an internally-driven reionization led to significant changes in the radial distribution of the satellites of the MW. This prediction was then further supported by using more refined models \citep{ocvirk2013,ocvirk2014}, built upon high resolution zoom simulations of the formation of the Local Group performed by the CLUES project\footnote{Constrained Local UniversE Simulations, \url{http://www.clues-project.org/index.html}}. However, these studies suffered from several limitations. 
First of all, these studies were performed in the so-called ``post-processing'' framework, and therefore only crudely account for stellar feedback (radiative and SNe). Furthermore, since the gas density field is frozen in the post-processing paradigm, the effect of radiation on the distribution of gas in the IGM (smoothing of gas overdensities due to photo-heating) can not be accounted for either, while it has been shown to have a significant impact on photon consumption \citep{pawlik2015}. Second, the gas distribution is only known in the zoom region, which is only a few Mpc wide, and mainly contains the progenitors of the MW, M31 and M33, and the IGM in-between. Beyond this region, the box is populated with low resolution pure dark matter particles. This dual description makes it very challenging to describe external (i.e. pure dark matter, low resolution) and internal (i.e. zoom region, including gas particles) reionization sources in a homogeneous, consistent way.




{Our ultimate challenge for modelling reionization, then, is to be able
to simulate the coupled multi-scale problem of global reionization and individual
galaxy formation, with gas and gravitational dynamics and radiative transfer, 
simultaneously.  And to predict the relic impact of this global reionization on the
universe today, to make comparisons possible with the observationally-accessible
nearby universe, including the Local Group and its satellites, it is further necessary
to start from initial conditions preselected to produce the observed galaxies and
large-scale structure of the local universe.  For this simulation to characterize the
evolving "patchiness" of reionization in a statistically meaningful way requires a
a comoving simulation volume as large as $\sim (100$ Mpc$)^3$ [e.g. \citep{iliev2006,iliev2014}].
Accounting for the millions of galaxies in this volume over the full mass range of 
galactic haloes which are thought to contribute most significantly to reionization,
the so-called "atomic-cooling haloes" (henceforth, "ACHs"), those with virial temperatures
above $\sim 10^4$ K (corresponding to halo masses above $\sim 10^8$ \Msun) while modelling
the impact of reionization on these individual galaxies and the IGM, requires a physical resolution of a few kpc's over the entire volume.

To meet this challenge and
satisfy all these requirements, we
have developed a new, hybrid CPU-GPU, fully-coupled, cosmological radiation-hydrodynamics-    
gravity code, RAMSES-CUDATON, capable of simulating the EOR in a comoving volume 91 Mpc
on a side, with $4096^3$ dark matter particles and a cubic-lattice of $4096^3$ cells for the
gas and radiation field.  Our simulation self-consistently models BOTH global reionization
AND the formation and reionization of the Local Group, by starting from a "constrained
realization" of initial conditions for the local universe which reproduces all the familiar
features of the local universe in a volume centered on the Local Group, such as the Milky
Way, M31, and the Fornax and Virgo clusters --- i.e. CLUES initial conditions. We call this
simulation "CoDa", for Cosmic Dawn.
}



{The main goal of this first paper is to introduce the simulation and compare the results with current observational constraints on the EOR, such as the evolution of the cosmic neutral fraction, the cosmic ionizing flux density, the cosmic star formation rate, the Thomson optical depth measured by cosmic microwave background experiments and high redshift UV luminosity functions (Sec. \ref{s:results}). }
It is laid out as follows: first we describe the code and the simulation setup (Sec. \ref{s:methodology}). We then proceed to our results and compare to available observational constraints of the EoR (Sec. \ref{s:results}), and finish with a short summary.

\section{Methodology}
\label{s:methodology}

The Cosmic Dawn simulation uses the fully coupled radiation hydrodynamics code RAMSES-CUDATON. This section describes the principles of the code and its deployment. For quick reference, the parameters of the simulation are summarized in Tab. \ref{t:sum}.

\begin{figure*}
  {\includegraphics[width=1.\linewidth,clip]{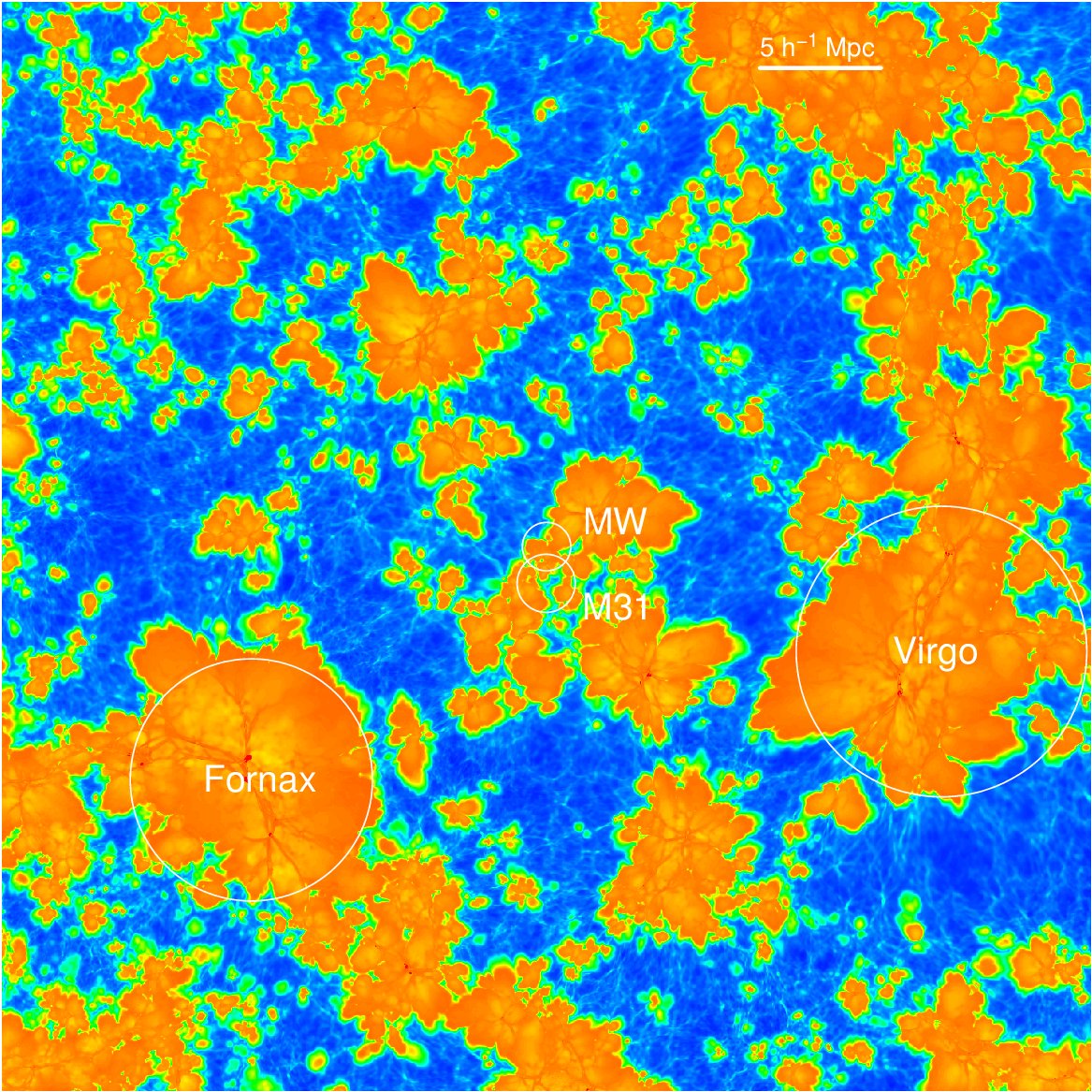}}
  \caption{Temperature distribution in a 45 \hmpc x 45 \hmpc x 0.03 \hmpc slice of the simulation at redshift \zcoda$=6.15$. Orange regions show photo-heated, ionized material, while the cold, still neutral medium appears in blue. The slice is situated in the supergalactic YZ plane with its origin shifted to the center on the Local Group progenitor at \zcoda$=7$. The circles of decreasing radius indicate the positions and approximate sizes of the progenitors of, respectively, Virgo, Fornax, the Milky Way and M31.}
\label{f:bigmap}
\end{figure*}

\begin{table}
\begin{tabular}{lr}
\hline
\multicolumn{2}{c}{Cosmology (WMAP5+BAO+SN)} \\
\hline
Dark energy density $\Omega_{\Lambda}$  & 0.721 \\
Matter density $\Omega_{\rm{m}}$  & 0.279 \\
Baryonic matter density $\Omega_{\rm{b}}$  & 0.046 \\
Hubble constant $h={\rm H}_0/(100 \, {\rm km/s})$  & 0.70 \\
Power spectrum & \\
$\,$ Normalization $\sigma_8$       &       0.817 \\
$\,$ Index $n$  & 0.96 \\
\\
\hline
\multicolumn{2}{c}{Setup} \\
\hline
Number of nodes (GPUs)  & 8192 (8192) \\
Grid size   &	$4096^{3}$ \\
Box size $\Lbox$  &    	91 Mpc   \\
Grid cells per node  & 128x256x256 \\
DM particle number $N_{DM}$   &	$4096^3$ \\
DM particle mass $M_{DM}$	 & 3.49 x $10^5$ \Msun \\
Initial redshift $z_{start}$  & 150 \\
End redshift $z_{end}$   & 4.23 \\
\\
\hline
\multicolumn{2}{c}{Star formation}\\
\hline
Density threshold $\delta_{\star}$  & $50 \, \langle \rho \rangle$ \\
Temperature threshold $T_{\star}$ &	$2 \times 10^4$K \\
Efficiency $\epsilon_{\star}$  & $10^{-2}$  \\
Stellar particle mass (post-SN) $M_{\star}$  & 3194 \Msun \\
Massive star lifetime $t_{\star}$  & 10 Myr \\
\\
\hline
\multicolumn{2}{c}{Supernova feedback}\\
\hline
Mass fraction $\eta_{SN}$  & 10\% \\
Energy $E_{SN}$  &  $10^{51}$ erg \\
\\
\hline
\multicolumn{2}{c}{Radiation} \\
\hline
Stellar ionizing emissivity  & 1.824$\times 10^{47}$ ph/s/\Msun \\
Stellar particle escape fraction $\fesc$  & 0.5 \\
Effective photon energy   & 29.61 eV \\
Effective H cross-section $\sigma_E$ & 1.097 x $10^{-22}$m$^2$ \\ 
Speed of light $c$  & 299 792 458 m/s \\
\end{tabular}
\caption{Cosmic Dawn simulation parameters summary}
\label{t:sum}
\end{table}


\subsection{RAMSES-CUDATON}
\label{s:ramsescudaton}

\subsubsection{RAMSES}
\label{s:ramses}
RAMSES is a code for simulating the formation of large scale structures, galaxy formation and self-gravitating hydrodynamics in general \citep{teyssier02}. Collisionless N-body dynamics are solved via a particle-mesh integrator. The adaptive mesh refinement was turned off in this run (see Sec. \ref{s:rhd} for more explanations on this).
Gas dynamics is modeled using a second-order unsplit
Godunov scheme \citep{teyssier2006,fromang2006} based on the HLLC Riemann solver \citep{toro1994}. 
We assume a perfect gas Equation of State (hereafter EoS) with $\gamma = 5/3$. 



We consider star
formation using a phenomenological approach. In each
cell with gas density larger than a gas overdensity $\delta_{\star}=50$, we spawn new star particles
at a rate given by
\begin{equation}
\dot{\rho_{\star}} = \epsilon_{\star} \frac{\rho_{gas}}{t_{ff}} \,\, {\rm with} \, \, t_{ff}=\sqrt{\frac{3 \pi}{32 G \rho}}
\end{equation}

where $t_{ff}$ is the free-fall time of the gaseous component and
$\epsilon {\star}$= 0.01 is the star formation efficiency. 

We also require the cell temperature to be lower than $T_{\star}=2 \times 10^4$ K in order to form stars: above this temperature, the gas is fully ionized, suffers inefficient hydrogen cooling and can not form stars. This temperature criterion is widely used in hydrodynamical simulations \citep{stinson2006,agertz2013}. A common variation in simulations of the EoR, producing a similar effect, is to use a threshold in ionized fraction, as in \cite{iliev2007,iliev2011,ocvirk2014,gillet2015}.

The star particle mass at birth depends on the cell gas density, but is always a multiple of a fixed elementary mass $M_{\star}^{birth}$, chosen to be a small fraction, $\sim 5$\% of the baryonic mass resolution. In this framework, with the box size and resolution of CoDa (see Sec. \ref{s:setup}), we have $M_{\star}^{birth}=3549$ \Msunnospace. This mass is small enough to sample adequately the star formation history of even low mass galaxies, and still large enough to mitigate stochastic variations in the number of massive stars per star particle.

For each star particle, we assume that $\eta_{SN}$=10\% of its mass is in the form of massive stars which will go supernova after a lifetime $t_{\star}=10$ Myr, leaving no remnant. We consider a supernova energy $E_{SN}=10^{51}$ erg per 10 \Msun of progenitor, and this feedback was implemented in the RAMSES code using the kinetic feedback of \cite{Dubois08}. After the massive stars have exploded, the remaining low mass stellar population is represented by a long-lived stellar particle of mass
\begin{equation}
M_{\star}=(1-\eta_{SN})M_{\star}^{birth}=3194 M_{\odot} \, .
\end{equation} 
No chemical enrichment was implemented: indeed, extrapolating the results of \cite{maiolino2008} to $z=6-10$ yields very low metallicities for the gas ($Z/\Zsun = 1/100 - 1/1000$), in a similar range to the simulations of \cite{pallottini2014}. At these metallicities, the cooling rates are still rather close to those for metal-free gas in most of the temperature range \citep{SD93} and therefore we do not expect chemical enrichment to make a dramatic impact on our results.

In cosmological simulations, it is customary to rely on subgrid models,
providing an effective EoS that captures the basic turbulent and
thermal properties of the gas in turbulent, multiphase, centrifugally supported disks. Models with various degrees of
complexity have been proposed in the literature, for instance using a polytropic EoS \citep{Yepes97,springel2003,Schaye07}.
We refrain from using this approach, because these subgrid models account for star formation and radiation from massive stars in star forming complexes, which are already explicitly accounted for in our feedback models. Therefore we set the RAMSES parameters so that the polytropic EoS does never kick in.

To summarize, we used for this simulation rather standard galaxy formation recipes, which have proven quite successful in reproducing various properties of galaxy evolution \citep{ocvirk2008,governato2009,governato2010}. The main novelty of this study is the inclusion of the coupling between the hydrodynamics and the radiation produced by the stars, treated by our radiative transfer module ATON, described in Sec. \ref{s:ATON}.

At the galactic scale, since galaxies can be smaller than one CoDa cell, their ISM is not resolved. However, the goal of our star formation model is to produce sources with suitable number density and ionizing photon output to reionize the box in a reasonable time, and we must calibrate it to do so.


{The overall star formation efficiency and feedback 
effect that result from our modelling are subject to adjustment of the
adopted efficiency parameters, $\delta_\star$ and $\epsilon_\star$.  While the previous
experience of simulating galaxy formation with such prescriptions is a useful
guide, there is some additional tuning required to simulate the EoR. For a 
self-consistent simulation of the EoR, these local parameters for the 
 internal efficiency of individual galaxies should be chosen so as to achieve
whatever global efficiency over time is required to make the simulated EoR
satisfy the known observational constraints on the EoR.  We discuss some of
these global EoR constraints in more detail in Sec. \ref{s:glob}.  Unfortunately, it  
is not possible to perform multiple runs of the size and resolution of the CoDa
simulation, each with different values for these efficiency parameters, since the 
computational cost is prohibitive.  Each such simulation would have to run all
the way to the end of reionization, and one cannot substitute for this by doing
coarse-grained, lower-resolution versions of this large volume to reduce the
computational cost enough per simulation, since lower resolution simulations do not 
yield the same outcome as the CoDa simulation with its higher resolution.  
In practice, we solve this problem, instead, by performing a large suite of smaller-box
simulations (e.g. as small as 4 \hmpc on a side) with the same particle mass and
grid resolution as CoDa but much smaller volumes, since each of these can be run on
the same computer as CoDa but require many fewer CPU-GPU hours.  In this way, we were able
to adjust the efficiency parameters for CoDa in advance, to best guess the outcome of the
much larger-volume simulation.  

     There are limitations that are inevitable in this procedure, however.  
For example, small boxes tend to reionize much more quickly
(i.e. over a more narrow redshift range) than do large-boxes and to have fewer
rare haloes, thereby delaying the first appearance of sources.  Periodic boundary
conditions for a small box also tend to make its reionization history differ from that
in a large box, since, e.g., radiation from sources inside the box that escapes from
it must re-enter that box, but there is no realistic accounting of more distant
sources.  In addition, cosmic variance makes it difficult to guarantee that the outcome
of a random realization of Gaussian-random-noise initial conditions in such a small
volume predicts the outcome of a different realization in a much larger volume.
Finally, 
current observational constraints on the global EoR are still highly uncertain and 
continue to change with time as more data becomes available, and these do not, 
themselves, exclude the possibility that the reionization of the local universe 
differed in some way from that of the universe-at-large.
}

\subsubsection{ATON}
\label{s:ATON}
ATON is a UV continuum radiative transfer code, which relies on a moment-based description of
the radiative transfer equations and uses the M1 closure relation \citep{gonzalez2007}. It tracks the out-of-equilibrium ionisations
and cooling processes involving atomic hydrogen \citep{aubert2008}. Radiative quantities (energy
density, flux and pressure) are described on a fixed grid and evolved
according to an explicit scheme under the constraint of a Courant-Friedrich-Lewy condition (hereafter CFL), i.e. timesteps must be no larger than some fraction of a cell-crossing time for a signal travelling at the characteristic speed for radiative transport. In the case of interest here, involving ionization fronts in the low-density IGM, this speed can approach the speed of light.
Each stellar particle is considered to radiate for one massive star lifetime $t_{\star}=10$ Myr, after which the massive stars die (triggering a supernova explosion) and the particle becomes UV-dark. We used a mono-frequency treatment of the
radiation with an effective frequency of 29.61 eV for a $10^5$ K black body
spectrum as in \cite{aubert2008}. This corresponds to a $\sim 100 $\Msun pop III star, although the mass scale of such primordial stars is currently very much debated (cf the introduction of \cite{jeon2014} for a summary).
Finally, we assume each star particle to have an intrinsic emissivity of 4800 ionizing photons/Myr per stellar baryon (i.e. $1.824\times 10^{47}$ photons/s/\Msunnospace), as in \cite{baek2009}, although the latter used a 50 000 K spectrum. Here the exact temperature of the stellar sources is actually of little importance: this temperature affects the effective ionizing photon frequency, which sets the effective cross-section of the photo-ionization reaction $\sigma_E$, and therefore the penetration of UV photons. In such EoR simulations, all UV, H-ionizing photons are consumed within a few cells of the ionization front (hereafter I-front), and variations of the UV photons' energy mostly affects the thickness of the I-front, but not its average position. Therefore, the emissivity of the stellar sources turns out to be more important than the effective frequency. Using a $50 000 $K black body spectrum would yield results rather similar to the $10^5$K we used here, provided the emissivity in ionizing photon number is kept constant.

Because of the relatively high spatial resolution of the CoDa simulation for an EoR simulation, we do not make any correction in terms of a clumping factor, to account for small-scale inhomogeneity in the gas density that was unresolved by the grid spacing, as was done for the largest boxes of \cite{aubert2010}. 
{A price which must be paid for such a high spatial resolution, however, is that the CFL
 timestep limiter of the light-crossing time per cell is exceedingly small and would 
 normally make the number of timesteps required to span the evolution of the system
 over a given physical time prohibitively large if all of the gas dynamical quantities
 had to be advanced with this same small timestep.  The sound-crossing and gravity 
 time-scales are orders of magnitude larger than this radiation transport time,
 however, so there is a great mismatch between the number of timesteps required to account
 accurately for mass motions and that required to evolve the radiation field and ionization
 state of the gas.  We have solved this computational problem without sacrificing accuracy
 by separately advancing the radiative transfer and ionization balance rate equations
 on GPUs while simultaneously advancing those for the gas and gravitational dynamics 
 on the CPUs.  By programming those GPUs so that each performs hundreds of these 
 smaller radiative transfer timesteps in the same wall-clock time as it takes the CPU 
 to perform one dynamical step, we are able to overcome this computational barrier.}
ATON has been ported on multi-GPU architecture, where each GPU handles a Cartesian sub-domain and communications are dealt with using the MPI protocol \citep{aubert2010}. By achieving an x80 acceleration factor compared to CPUs, the CFL
condition is satisfied at high resolution within short wall-clock computing
times. As a consequence, no reduced speed of light approximation is required.

We then have to determine an escape fraction for our stellar particles, i.e. the fraction of UV radiation able to escape the stellar birth cloud represented by the particle, into the interstellar medium (hereafter ISM) of the cell it belongs to, which is different from the escape fraction from the entire galaxy. Here we discuss two potential sources of absorption in the ISM: Hydrogen, and dust. We will therefore consider the two corresponding escape fractions, $\fesc^{H}$ and $\fesc^{dust}$ (we use the ${}_{,\star}$ subscript to remind the reader at all times that this escape fraction applies to a stellar particle and its cell and not to a galaxy and the IGM). Their product gives the stellar particle escape fraction $\fesc=\fesc^{H} \times \fesc^{dust}$. While there is some guidance (although somehow controversial) in the literature for the choice of the escape fraction, we can consider two limiting cases:
\begin{itemize}
\item{the large scale, low spatial resolution regime, where the radiative transfer grid cells size is larger than the galaxies, e.g. dx$=0.44$ \hmpc as in \cite{iliev2014}. This regime prescribes the use of a ``galactic escape fraction'', which accounts for the ionizing photons lost to the unresolved interstellar medium of each galaxy. In this case, $\fesc$ can span a wide range of values, from almost 0 to about $40\%$, as shown by high resolution simulations \citep{wise2009,razoumov2010,yajima2011,wise2014}.}
\item{the very high resolution regime, which is closer to resolving the interstellar medium and star forming molecular clouds, such as \cite{kim2013a,kimm2014,rosdahl2015}, where the cell can be as small as dx$\sim$ 1 pc in the most refined regions. In this case the totality of the stellar photons produced reaches the interstellar medium. Therefore the stellar particle escape fraction $\fesc=1$ although the {\em galactic} escape fraction can be much smaller.}
\end{itemize}

{{In CoDa, as galaxies form, their progenitor dark matter haloes span volumes
  from several to hundreds of grid cells across.  Even after they are fully collapsed,
  these haloes are still typically a few to a few tens of cells across.  As a result,
  the effects of internal radiative transfer of UV photons through the interstellar
  hydrogen are accounted for, although in a rather coarse way.  Hence, we set $\fesc^{H}=1$, so as not to over-count the attenuation caused by the interstellar hydrogen  gas inside the galaxy.}}
Dust may also play an important role: according to \cite{gnedin2014a}, dust optical depths as high as $0.8$ in the UV could be common within galaxies by the end of the EoR, which amounts to a throughput of about $\fesc^{dust}=0.5$ of the UV flux. With these assumptions, we obtain a stellar particle escape fraction of $\fesc=0.5$.



Finally, we neglect any possible Active Galactic Nuclei (hereafter AGN) phase of our galaxies. Such sources could already be in place in rare massive proto-clusters during reionization \citep{dubois2012a,dubois2012}. They are very rare and thought to be minor contributors to the cosmic budget of hydrogen-ionizing photons \citep{haardt2012,haardt2015}, although they could be important for explaining the line of sight variations of the properties of the Ly $\alpha$ forest just after reionization \citep{chardin2015a}.

\subsubsection{Radiation-hydrodynamics coupling}
\label{s:rhd}

We developed an interface that enables data exchange between RAMSES and ATON on the fly, leading effectively to a coupling between dynamics (handled by RAMSES) and radiation (handled by ATON). 

At the end of a dynamical time-step, RAMSES sends to ATON the gas density, temperature and ionized fraction. Radiative transfer is then performed by ATON using these inputs, via sub-cycling of 100s to 1000s of radiative sub-steps. Once this sub-cycling is completed, ATON sends the temperature and ionized fraction back to RAMSES.
However, CUDATON (the GPU port of ATON) can only handle regular Eulerian grids. For this reason, AMR is not used for this simulation, and RAMSES is effectively used as a hydro-PM code. Furthermore, CUDATON assumes a Cartesian domain decomposition for the MPI-parallelism when similar domains are treated by each GPU. Meanwhile, RAMSES relies on a space-filling curve decomposition. As a consequence, data must also be reorganized at each transfer between the 2 applications.

This coupling method is similar to RAMSES-RT \citep{rosdahl2013} but with one photon group only and no AMR. RAMSES-RT has been extensively tested and passes the test suite of \cite{iliev2009}. RAMSES-CUDATON was similarly tested, as described in \cite{stranex2010}\footnote{\url{http://www.physikstudium.uzh.ch/fileadmin/physikstudium/Masterarbeiten/Stranex_2010.pdf}}.
As expected due to the similarity of the codes, RAMSES-CUDATON performs similarly well with these tests:
various instances of Str\"omgren spheres and the photo-evaporation of a gas cloud, the hallmark of radiation-hydrodynamics coupling in galaxy formation.

One of the great advantages of RAMSES-CUDATON is its ability to run with the true speed of light, which stems from the combination of a moment-based method with GPU acceleration. This is quite unique among the handful of existing radiation-hydrodynamics codes. Some of them, mostly those involving ray-based photon propagation schemes, use infinite speed of light. Others, in particular those using moment-based methods such as ATON for the RT (e.g. RAMSES-RT), or \cite{gnedin2014}, use the reduced speed of light approximation (as low as 1/100 of the real speed of light) in order to reduce the computational cost of the simulation. This is because in such a framework, the time-step of the code is set by the fastest physical process (gravitational, hydrodynamical and radiative). Since the speed of light is typically 100 to 1000 times faster than the speed of sound or bulk matter motions in cosmological simulations, the time-step must be about 100's of times shorter and therefore the simulation will be 100's of times slower than its pure hydrodynamical counterpart. While this is probably a good enough approximation in the dense interstellar medium of galaxies, it is not valid in the low density intergalactic medium, i.e. most of the volume of a cosmological simulation as we perform in the present paper\citep{rosdahl2013}. Not only would the speed of ionization fronts in the IGM be misrepresented by the reduced speed of light approximation, but long-range feedback effects, like the impact of one massive object or a distant cluster of objects on the properties of low-mass galaxies may be impossible to capture in such a "slow light" framework.
In RAMSES-CUDATON this problem is alleviated thanks to the GPU optimization. The x80 boost almost cancels out the added cost of the RT, and allows us to work efficiently with the real speed of light.



\subsection{Simulation setup}
\label{s:setup}

\subsubsection{Initial conditions}
\label{s:ics}
The initial conditions (hereafter ICs) were produced by the CLUES project (Constrained simulations of the local universe), assuming a WMAP5 cosmology \citep{hinshaw2009}, i.e. $\Omega_{\rm{m}}=0.279$, $\Omega_{\rm{b}}=0.046$, $\Omega_{\Lambda}=0.721$ and $h=0.7$. A power spectrum with a normalization of $\sigma_8=0.817$ and $n=0.96$ slope was used. 
The comoving box is 91 (i.e.64$h^{-1}$) Mpc on a side, with 4096$^3$ dark matter particles on 4096$^3$ cells and with the same cubic-lattice of 4096$^3$ cells for the gas and radiation properties.  The mass of each dark matter particle is then $3.49$ x $10^5$ \Msun. { The CoDa initial conditions have the average universal density for the chosen cosmology.}

These ICs are tailored so as to reproduce the structure of the local universe at z=0, using constrained realizations of a Gaussian random field, in the spirit of \cite{hoffman1991}.  
In essence, the method takes as its input, galaxy observations of the local universe today: galaxy catalogs with distances and peculiar velocities, galaxy group catalogs, such as the MARKIII \citep{willick1997}, SBF \citep{tonry2001} and the Karachentsev catalog \citep{karachentsev2004}. These are used to infer the z $=$ 0 dark matter distribution. Using the Hoffman-Ribak algorithm \citep{hoffman1991}, a set of ICs (particle positions and velocities) at high redshift are produced, which will evolve into this present matter distribution over a Hubble time. This technique reproduces the large scale structure around the LG, although with some scatter, whereas small scales below 1 \hmpc are essentially random. In order to make sure that a realistic LG is obtained, a second step was performed.
A large number of similarly constrained realizations of the initial conditions, of order 200, was produced and a selection made for the one which best fit the LG and its relation to other known features of the local universe at z  = 0, with masses and locations of objects like the LG and its largest galaxies, the MW and M31, and a cluster such as Virgo. The best realization is then enriched with higher order random k-modes to an arbitrary resolution, limited in practice by computer memory. The resulting ICs therefore contain:
\begin{itemize}
\item{constrained low k-modes, driving the emergence of the large scale structures and galaxy clusters,}
\item{unconstrained but {\em ``chosen''} intermediate k-modes, picked from a sample of many realizations to produce a LG,}
\item{fully random high k-modes, added by the resolution enhancement procedure.}
\end{itemize}
This procedure is detailed in \cite{gottloeber2010}, and it produced the original CLUES ICs up to a resolution of 2048$^3$ in the full box. Here we have used Ginnungagap\footnote{\url{https://github.com/ginnungagapgroup/ginnungagap}}  to increase further the resolution of the ICs up to that required by CoDa, i.e. 4096$^3$.

A CoDa temperature map at z=6.15 in the supergalactic YZ plane is shown in  Fig. \ref{f:bigmap}. The white circles of decreasing radius denote the approximate locations and sizes of the progenitors of Virgo and Fornax galaxy clusters, M31 and the MW. Massive objects such as galaxy cluster progenitors appear as large bubbles of photo-heated gas in the temperature maps. Two galaxy group progenitors also appear on the map between the Local Group and Virgo, reminiscent of the CenA and M81 groups, as described for instance in \cite{courtois2013}. The exact occurence, position and growth of these particular groups are not strongly constrained in our ICs generation procedure, but such groups are statistically expected to emerge in the filament that harbors the LG and that points to the Virgo cluster \cite{libeskind2015}. The impact of such groups on the formation and evolution of the LG could be quite significant and therefore their existence in the simulation is one of the virtues of using such tailored initial conditions.


Because of its location with respect to Fornax and Virgo, the LG shown in Fig. \ref{f:bigmap} can be considered as ``the'' LG of the box. While the fairly large size of the box is a requirement in order to reproduce well enough the local Universe without artifacts due to the periodic boundary conditions, and properly model reionization sources external to the Local Group, it has a third benefit: it will allow us find many galaxy pairs similar to the Local Group in mass ratios and separation through the box. This should therefore allow us to study ``what if'' scenarios, where analogs of the MW-M31 system would have evolved in a different environment.

The baryonic ICs were generated assuming a uniform temperature and the gas and dark matter are assumed to have identical velocity fields. The startup ionized fraction was set to the freeze-out value at $z=150$. They are computed following standard recipes such as in RECFAST \citep{seager1999}.

\subsubsection{Code deployment}
The CoDa simulation was performed on the Titan supercomputer at Oak Ridge National Laboratory.
The computational domain spans a comoving cube 91 Mpc on a side, sampled on a fixed grid of 4096$^3$ cells. The code was deployed on 8192 cores, with each core coupled to 1 GPU, therefore requiring 8192 Titan nodes (1 GPU per node). Each node hosted one MPI process which managed a 128x256x256 volume.


\subsection{Run and data management}
The run was performed from August to December 2013, and took about 10 days wall clock, using a total of about 2 million node hours, i.e. 60 million hours according to the Titan charging policy (1 node hour = 30 core hours). A total of 138 snapshots were written, every 10 Myr, from z=150 down to z=4.23, each one of size $\sim$16 TBytes, summing up to more than 2 PBytes of data in total. 
Since it is not possible to save such a large amount of data in long-term storage, we decided to keep the following reduced data products:
\begin{itemize}
\item{Full-box, low resolution (2048$^3$ grid) versions of the gas field (mass and volume-weighted temperature, volume-weighted pressure, density, velocities, ionized fraction), ionizing flux density, and of the dark matter density field, plus a list of all the star particles in the box, with their mass, coordinates and ages.}
\item{Cutouts: for a set of 495 cubic subvolumes of 4 h$^{-1}$ Mpc on a side, centered on regions of interest, the original, full-resolution data of the simulation, each with 256$^3$ cells, was saved.  The coordinates of these cutout regions were pre-computed from a pure N-body simulation (no gas dynamics) by GADGET, with 2048$^3$ particles, using the same CLUEs initial conditions, but coarsened for this lower resolution (i.e. instead of CoDa's $4096^3$ particles and cells), which we ran down to z $=$ 0, to identify interesting regions in the present-day local universe. We will refer to this dark matter only simulation as DM2048. Each cutout contains all the gas quantities (same as full-box), the ionizing flux density, and the dark matter particles, at maximum resolution. The regions of interest we picked include the LG as determined by the CLUES constrained initial conditions. They also include 62 LG analogs, i.e. pairs of galaxies with masses, separations and relative velocities compatible with the present-day MW-M31 system. They will be used in the future as examples of LGs forming in different environments. Each of these systems require 2 cutouts, one for the MW and one for M31. Finally, we added 399 regions following halo progenitors of various masses, from large galaxies to dwarfs, embedded in different environments, from cluster and galaxy groups to voids.}
\item{Full box halo catalogs (see Sec. \ref{s:fof}) containing the number of particles, position and velocity of each halo, as well as the list of dark matter particles it contains (ids and positions).}
\end{itemize}
This strategy allowed us to bring down the data volume to a more manageable $\sim 200$ TBytes, while still allowing us to achieve our scientific goals.

\subsection{Processing: friends-of-friends halo catalogs}
\label{s:fof}
We used the massively parallel friends-of-friends (hereafter FoF) halo finder of \cite{roy2014} with a standard linking length of b=0.2 to detect dark matter haloes in the CoDa simulation. They are reliably detected down to $\sim 10^8$ \Msun. More details on the resulting CoDa mass functions can be found in Appendix \ref{s:aMFs}.

\subsection{Online data publication}
We plan to make a subset of the data and higher level products publicly available through the cosmosim database hosted by Leibnitz Institut f\"ur Astrophysik Potsdam\footnote{\url{https://www.cosmosim.org/cms/simulations/cosmic-dawn/}} and the VizieR database at CDS Strasbourg\footnote{\url{http://vizier.u-strasbg.fr/viz-bin/VizieR?-source=VI/146}}.

\section{Results}
\label{s:results}

In this section we show that the simulation effectively captures the main physical processes we intended to describe, and show that the global behaviour of the simulation is correct. We emphasize the radiative-hydrodynamic processes, since it is the main novelty of this simulation, and finish by discussing the role of various halo mass ranges in the overall reionization process. Our intention is to introduce these results and highlight a few of them, while leaving further analysis and interpretation for future papers. {Note: In what follows, in order to make comparisons possible between simulation results at a given redshift and some observable properties of the universe at a different redshift, we will, henceforth, refer to simulation redshift as \zcoda, while z will refer to the observed, or "real world" redshift.}


\subsection{Maps: Zoom-In on Galaxies and the Cosmic Web during the EoR}
\subsubsection{Example: The M31 Progenitor Cut-out}


\begin{figure*}
\begin{tabular}{ccc}
{\includegraphics[height=5.7cm,clip=true]{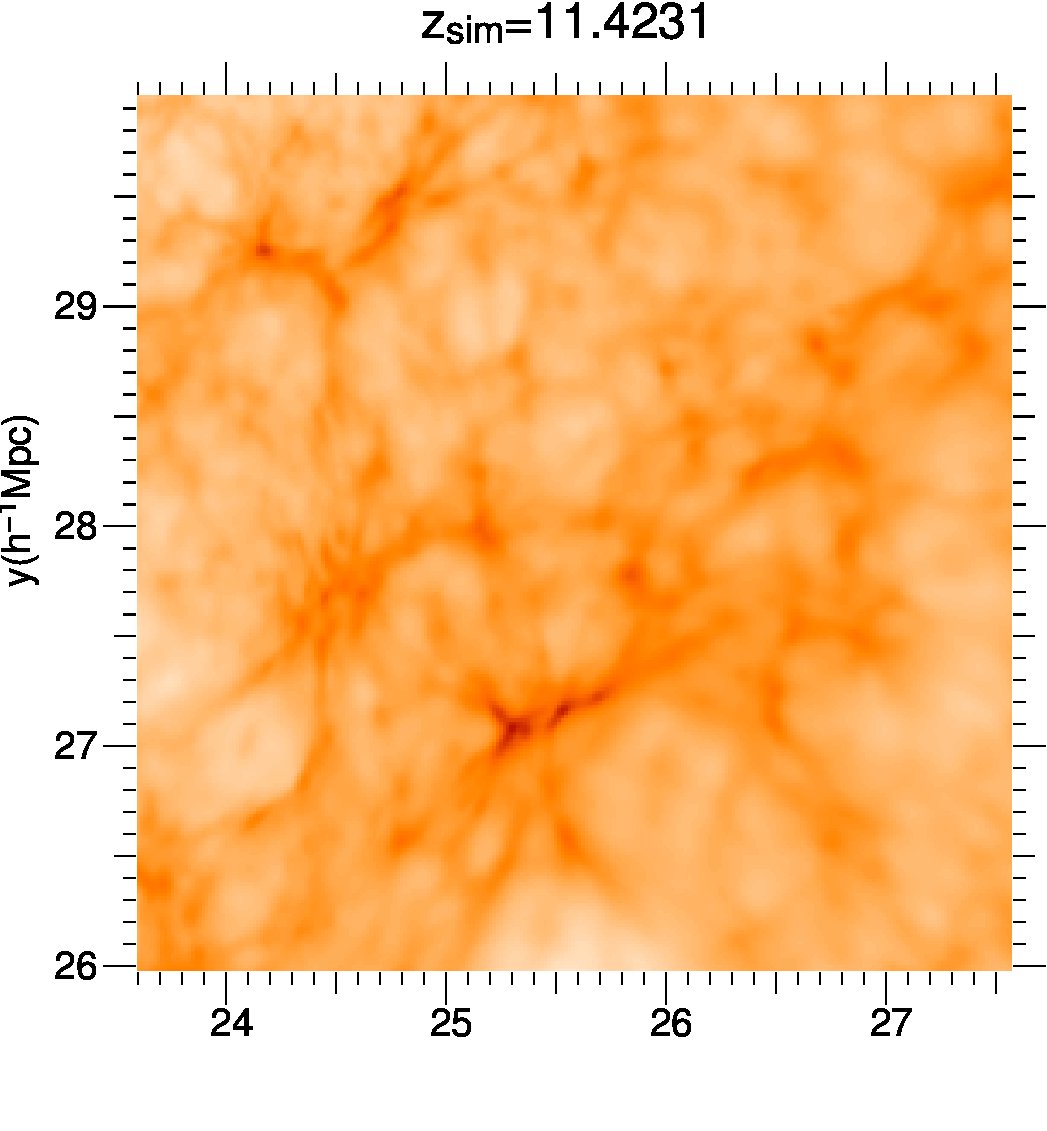}}&
{\includegraphics[height=5.7cm,clip=true]{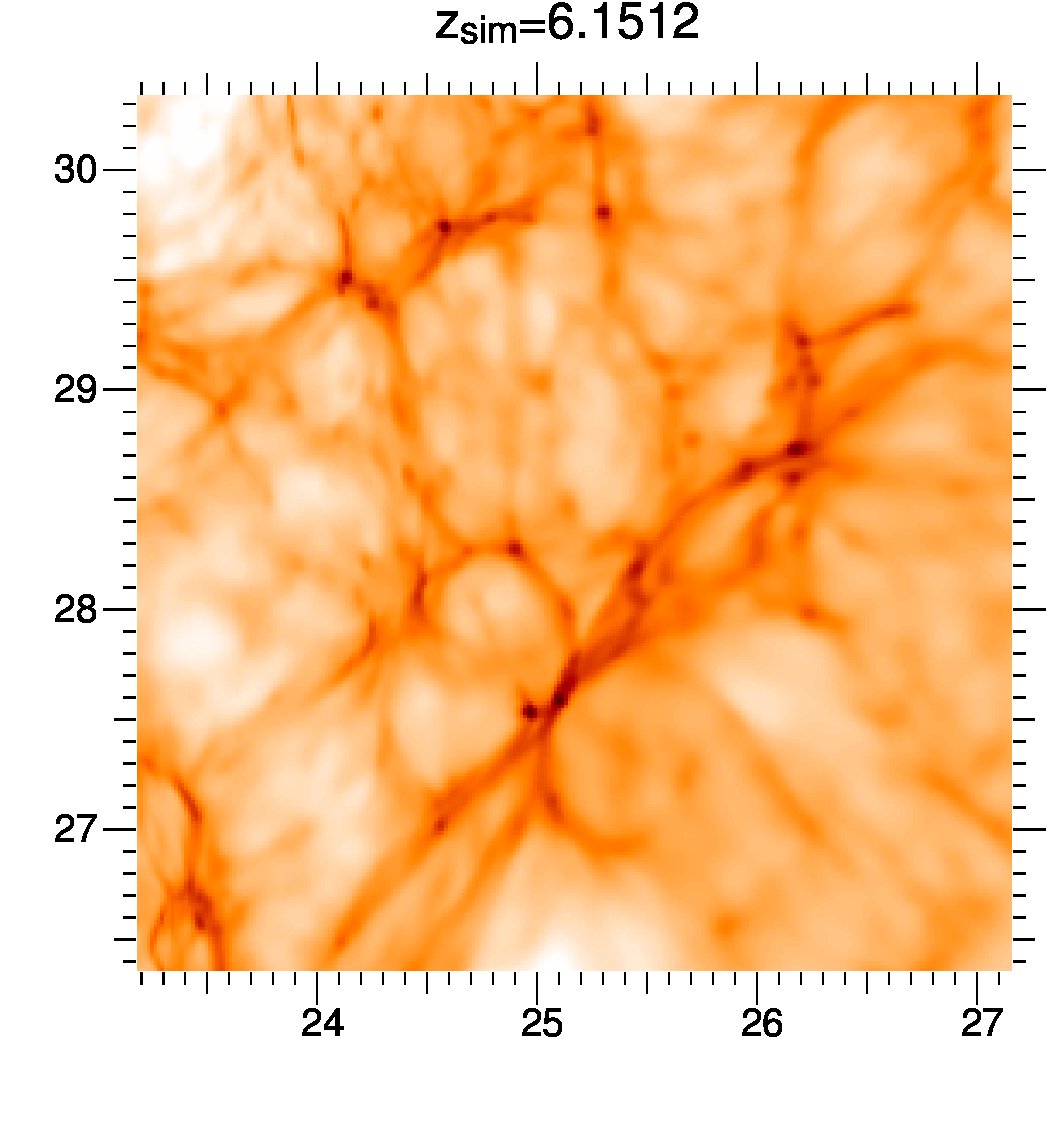}}&
{\includegraphics[height=5.7cm,clip=true]{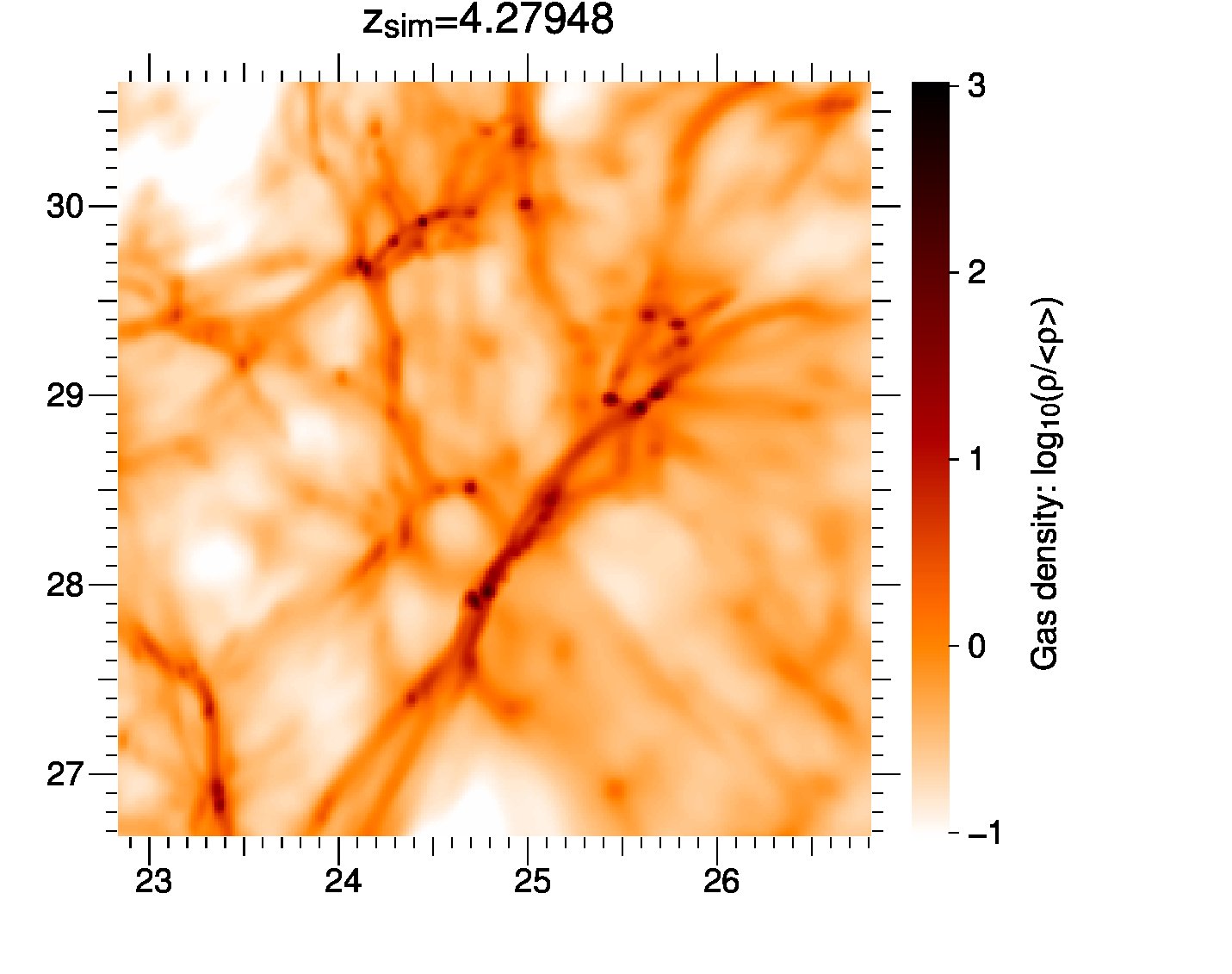}}
\\
{\includegraphics[height=5.5cm,clip=true]{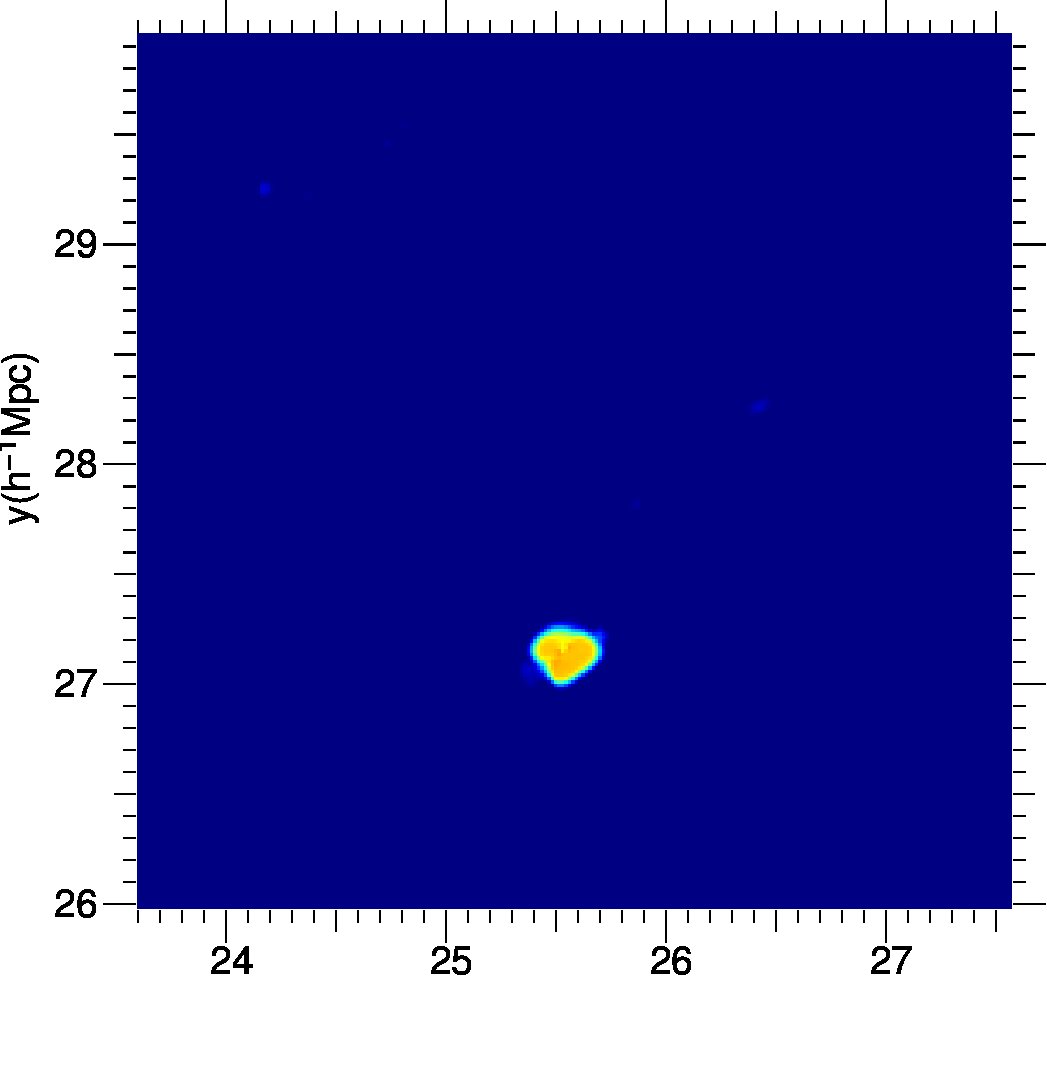}}&
{\includegraphics[height=5.5cm,clip=true]{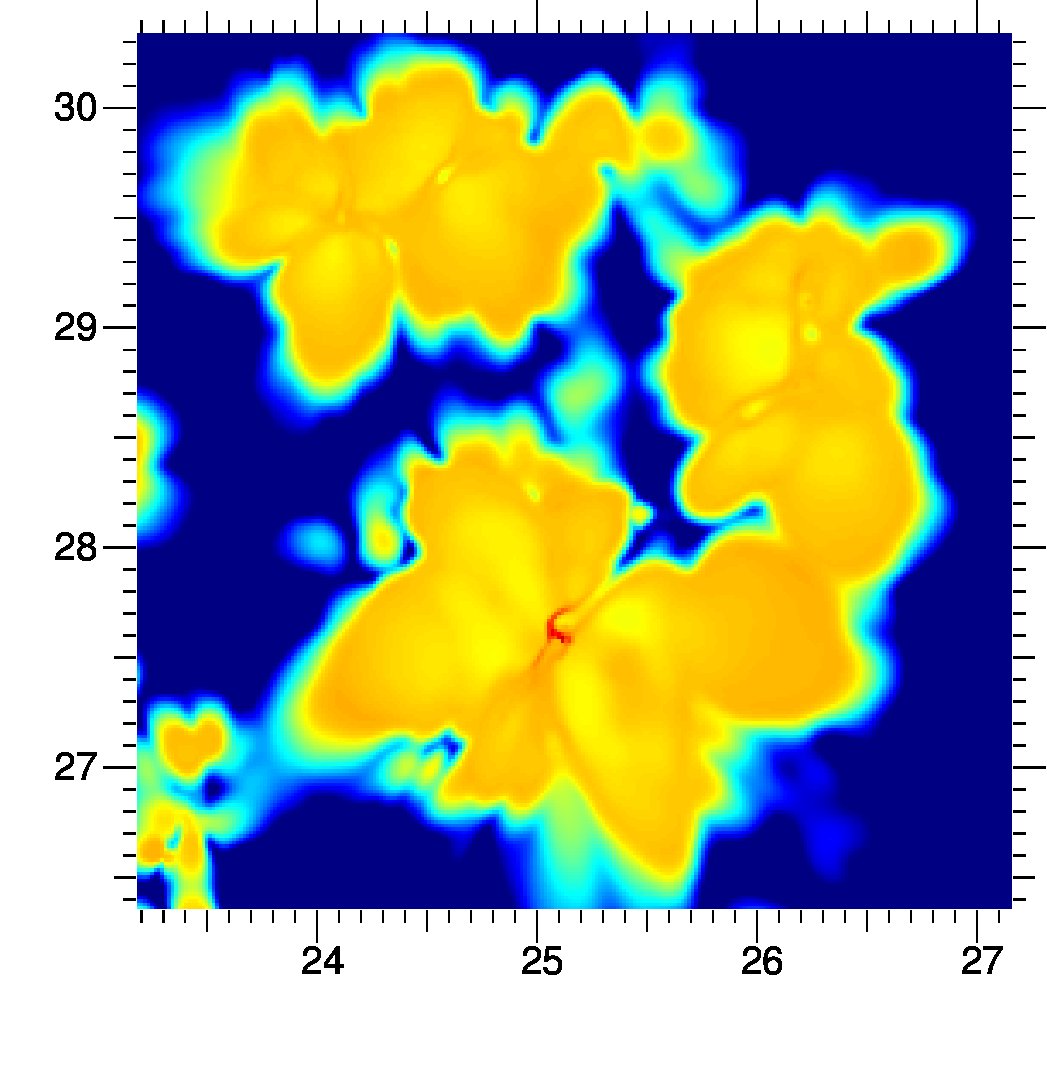}}&
{\includegraphics[height=5.5cm,clip=true]{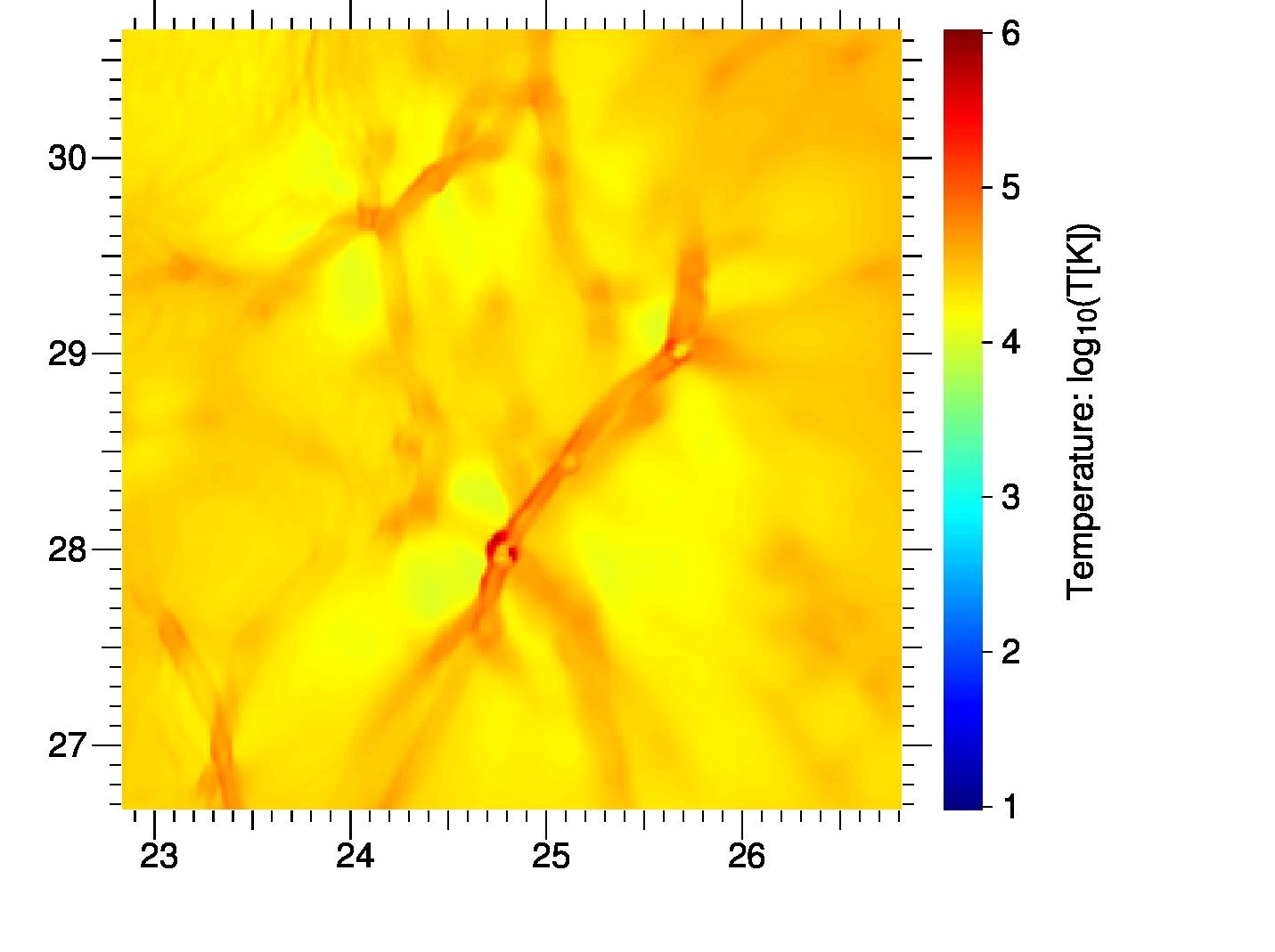}}
\\
{\includegraphics[height=5.5cm,clip=true]{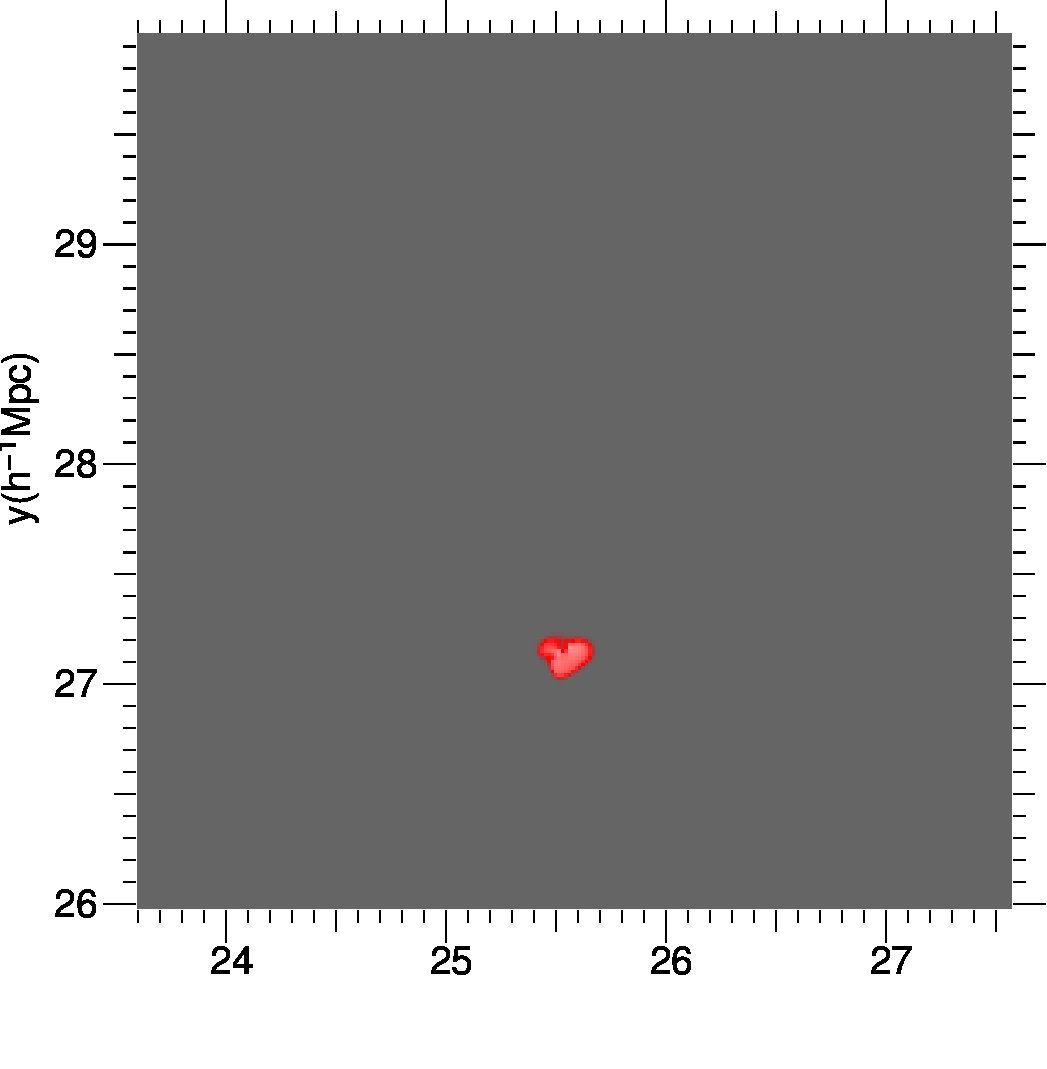}}&
{\includegraphics[height=5.5cm,clip=true]{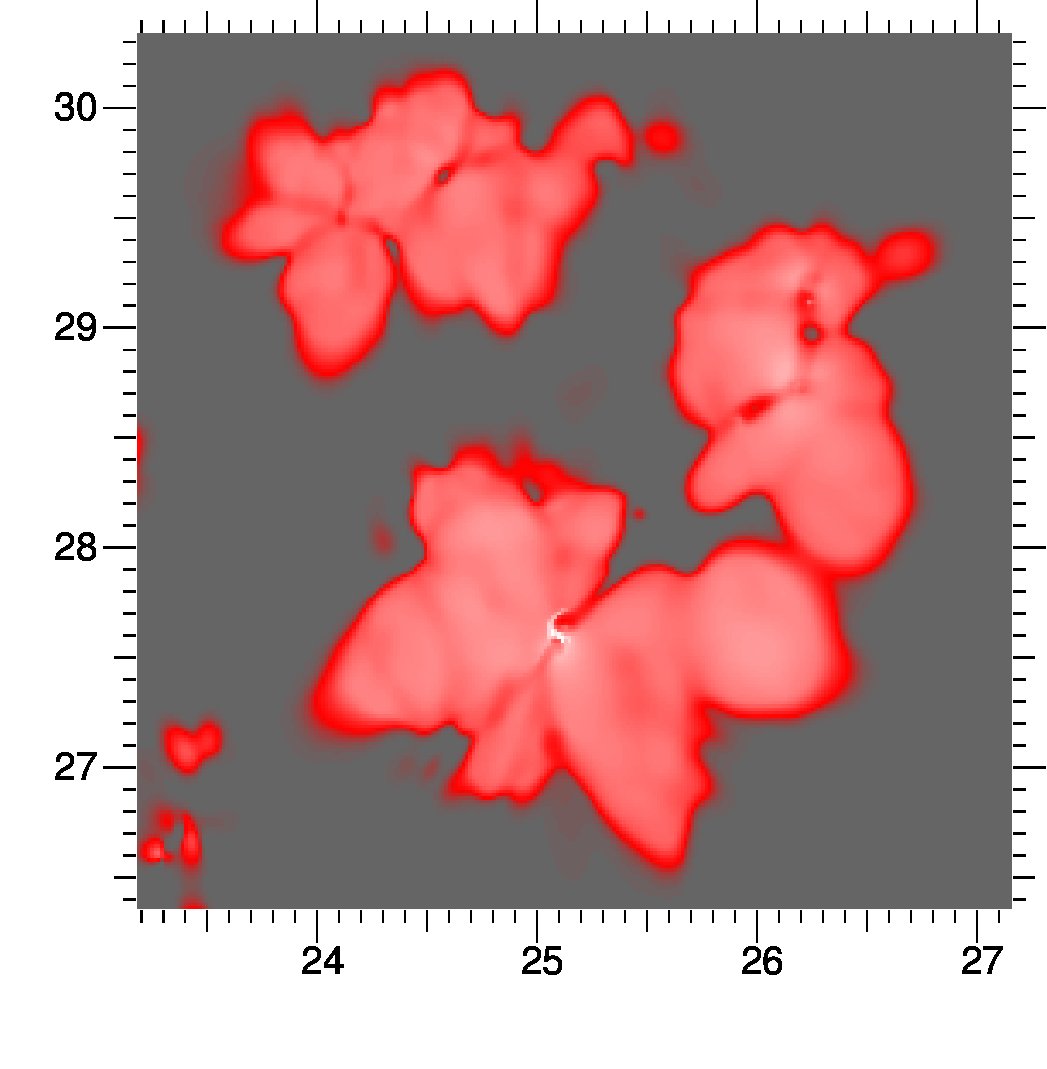}}&
{\includegraphics[height=5.5cm,clip=true]{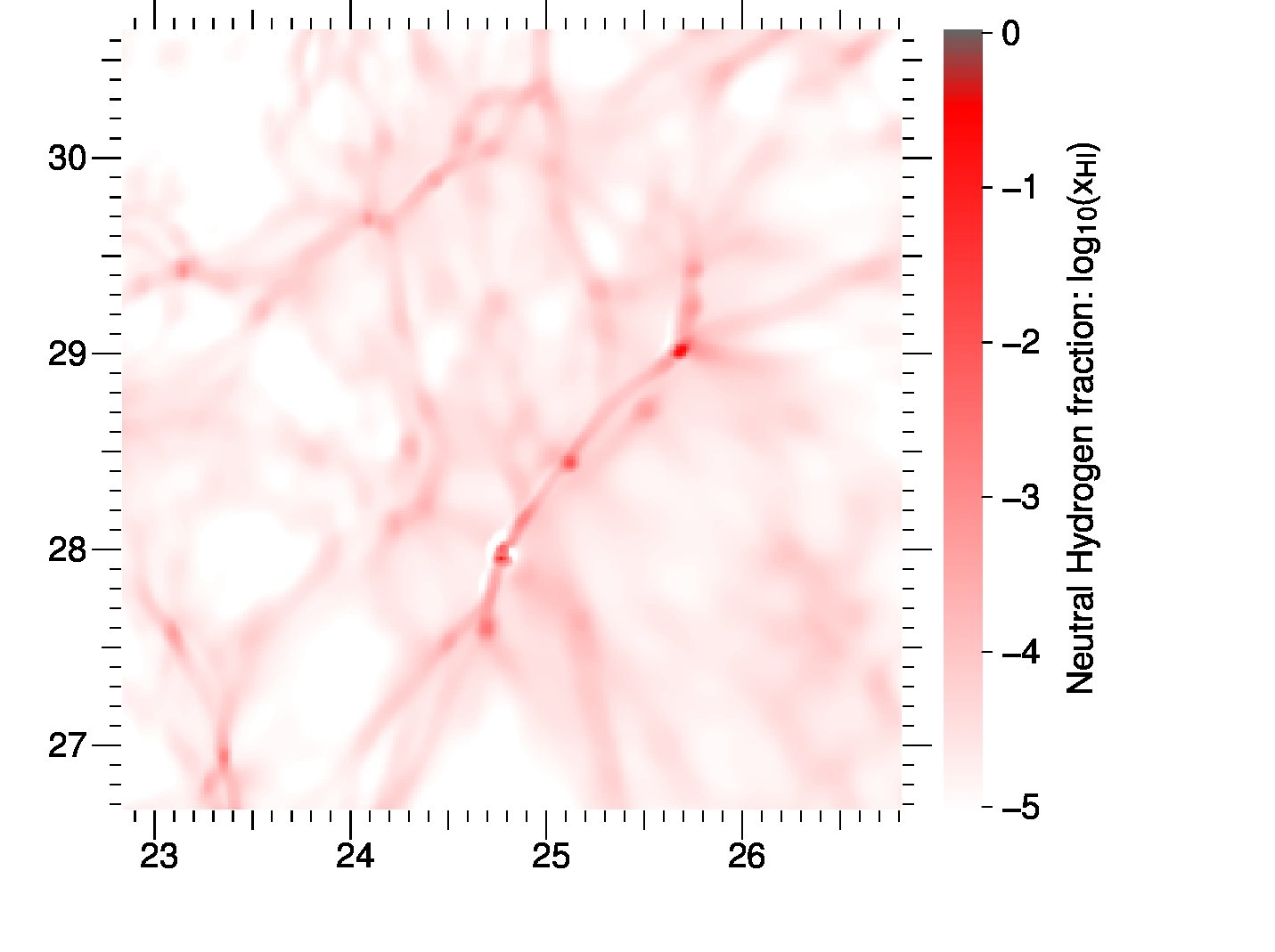}}
\\
{\includegraphics[height=5.5cm,clip=true]{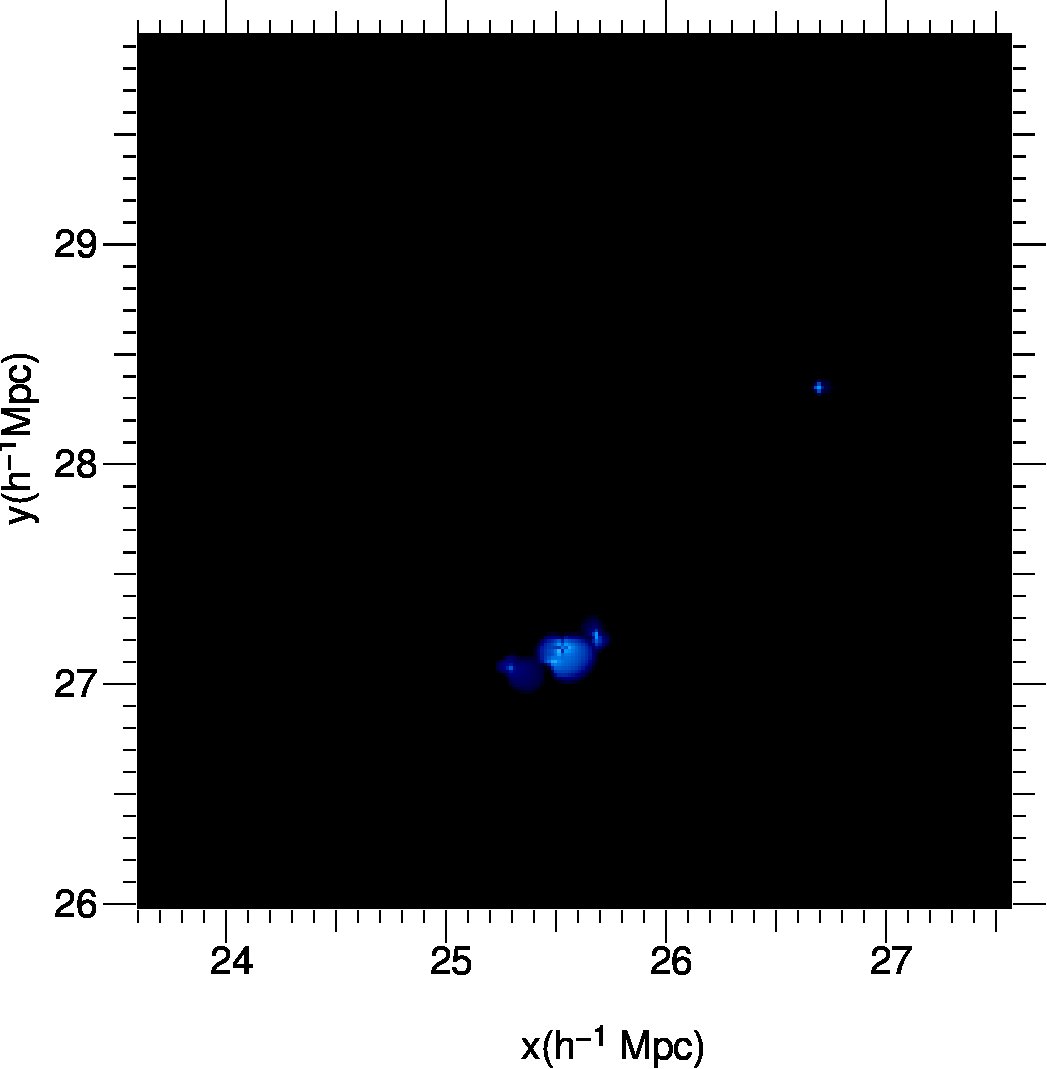}}&
{\includegraphics[height=5.5cm,clip=true]{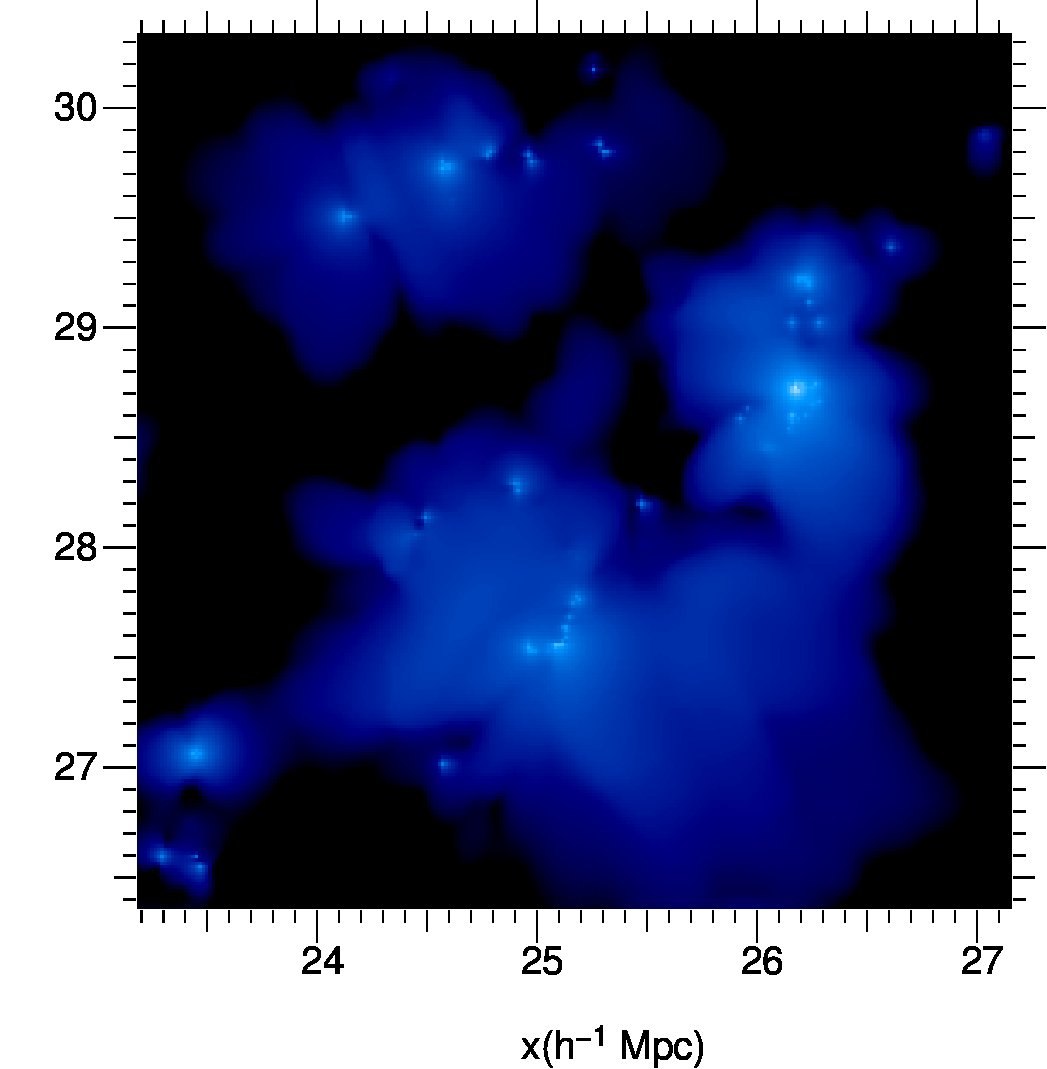}}&
{\includegraphics[height=5.5cm,clip=true]{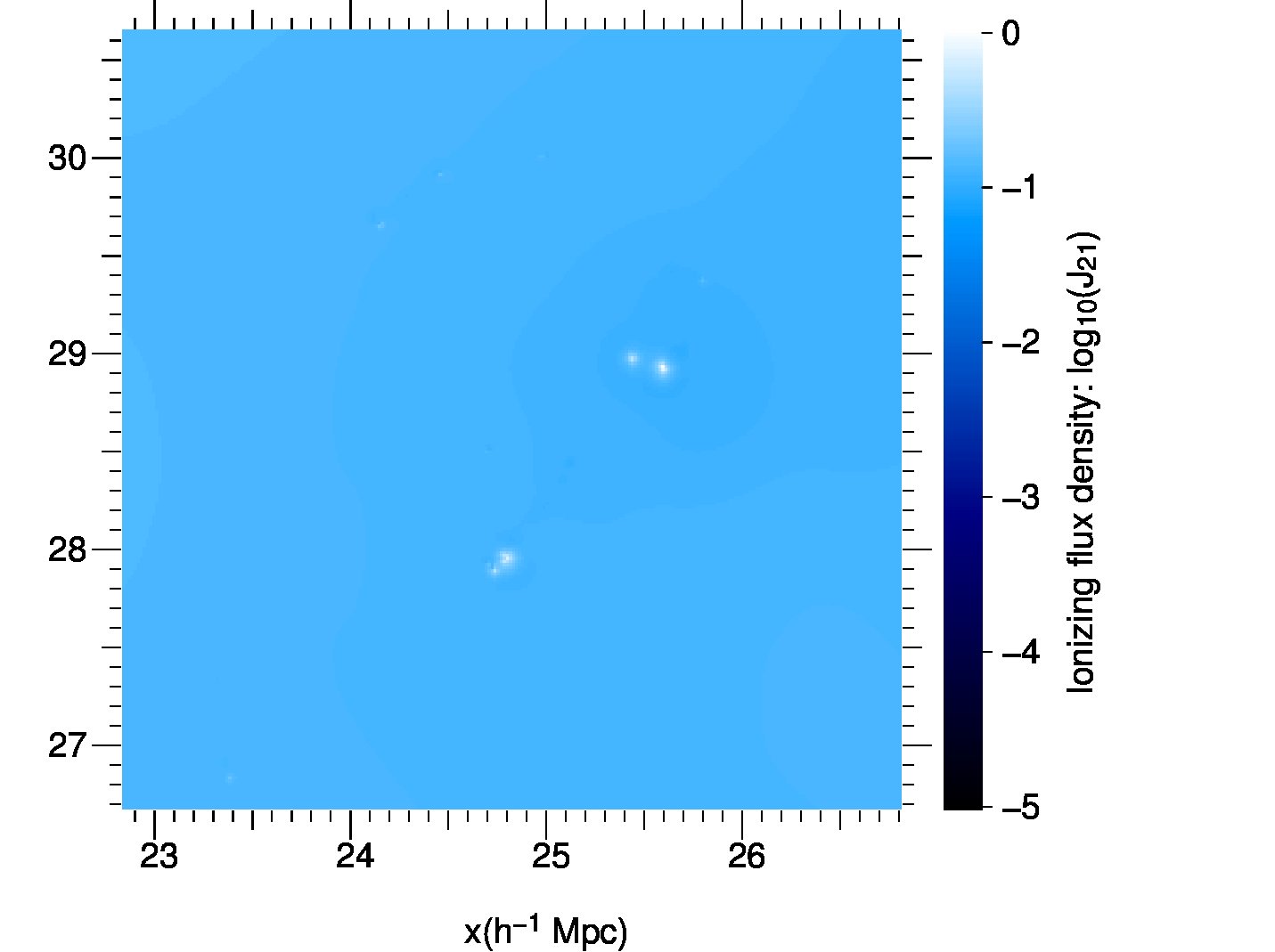}}
\\

\end{tabular}
\caption{Maps of the physical properties properties of the gas in a slice through the comoving 4 h$^(-1)$ Mpc$^3$ subvolume that contains the progenitor of M31. From left to right: \zcoda$=11$ to \zcoda$=4.28$ (end of the simulation).{\em Top:} co-moving gas density,  {\em 2nd row:} Temperature, {\em 3rd row:} Neutral gas fraction, {\em bottom:} Ionizing flux density. Gas density and ionizing flux density are a projection based upon averaging over a slice 40 cells (or 625 h$^{-1}$ kpc) thick, while the temperature and neutral fraction shown are taken from a slice just 1 cell thick (i.e. 15.625 h$^{-1}$ kpc).}
\label{f:maps}
\end{figure*}

\begin{figure*}
{\includegraphics[height=6.4cm,clip=true]{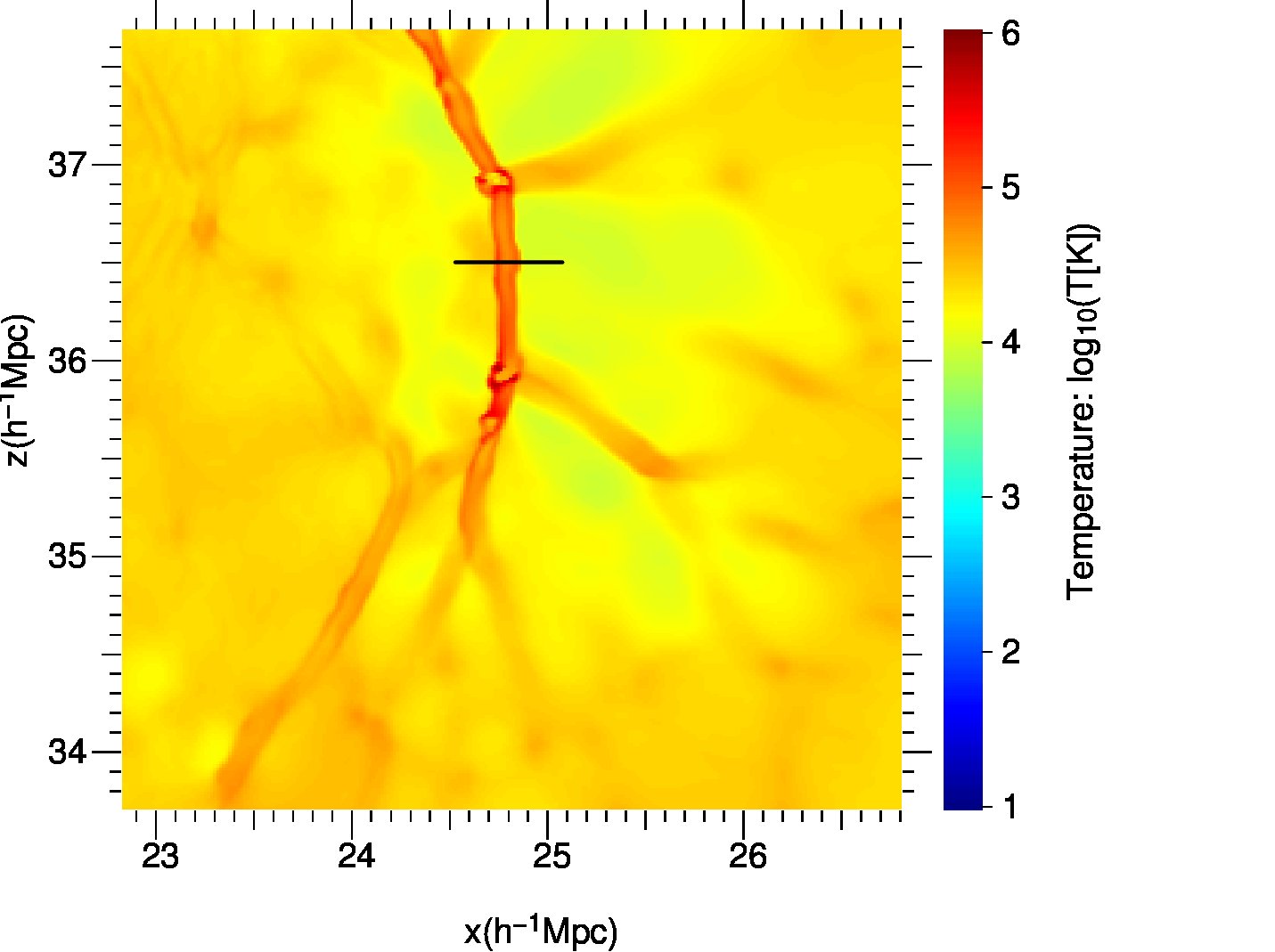}}
{\includegraphics[height=6.4cm,clip=true]{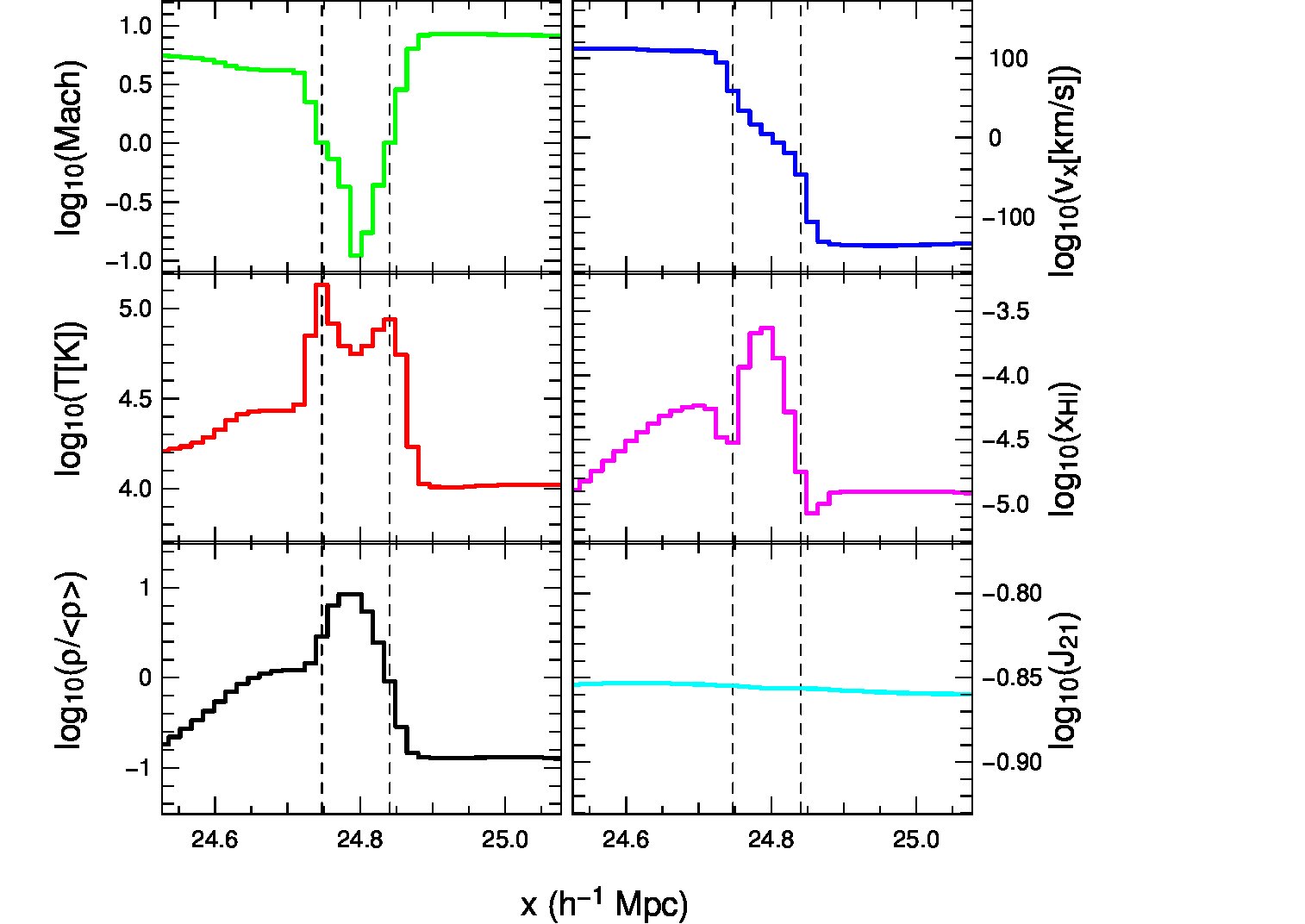}}
\caption{Illustration of the sheathed temperature structure of gas filaments. {\em Left:} a xz slab (1 cell thick) of the cutout of Fig. \ref{f:maps}, chosen to highlight the filamentary structure. {\em Right:} gas overdensity, temperature, Mach number, x axis velocity, neutral Hydrogen fraction and ionizing flux density $J_{21}$ along the black segment of the left panel. The vertical dashed lines mark the position of the two temperature peaks, for reference.}
\label{f:sheath}
\end{figure*}

\begin{figure*}
{\includegraphics[height=6.5cm,clip=true]{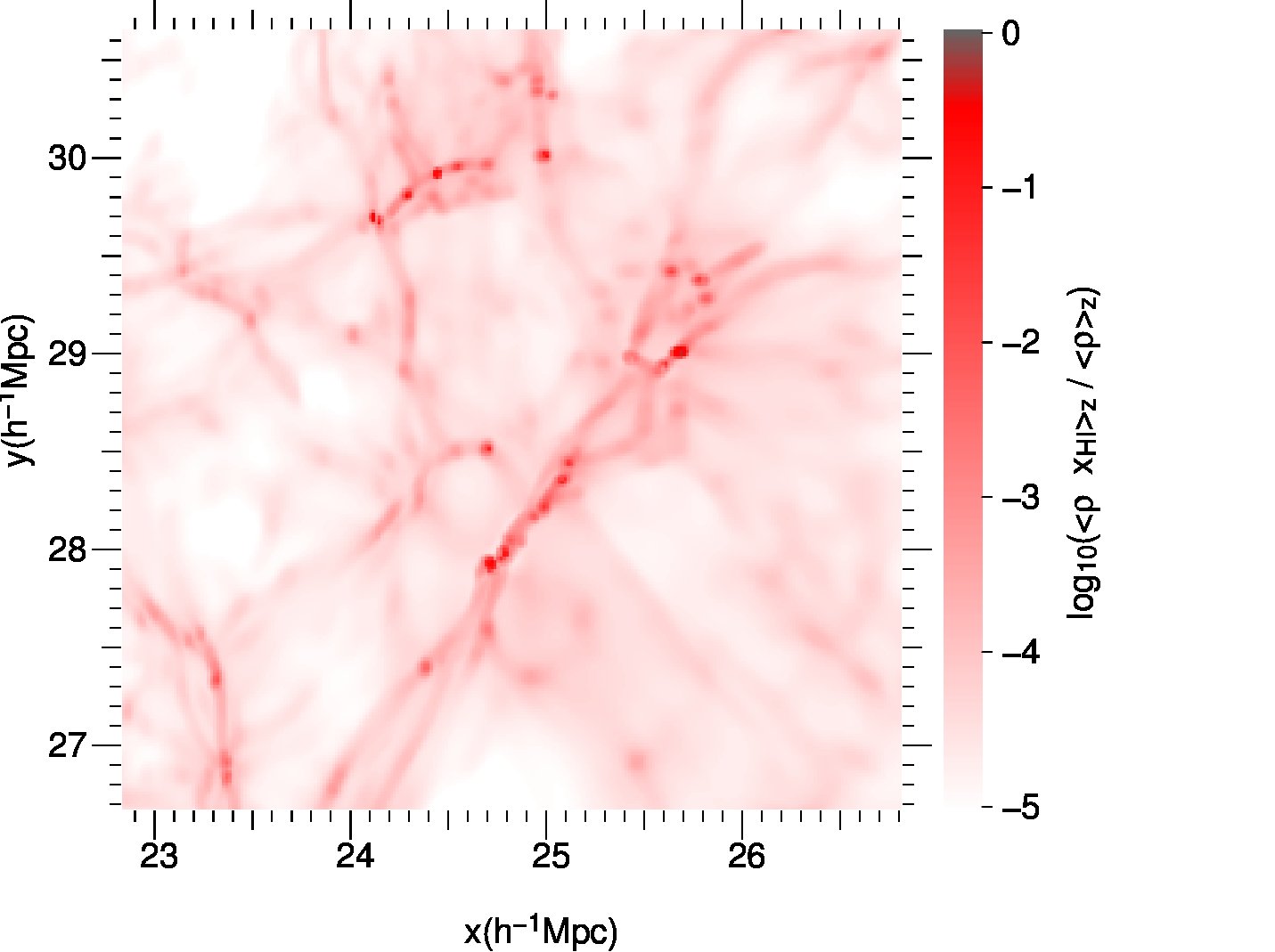}}
{\includegraphics[height=6.5cm,clip=true]{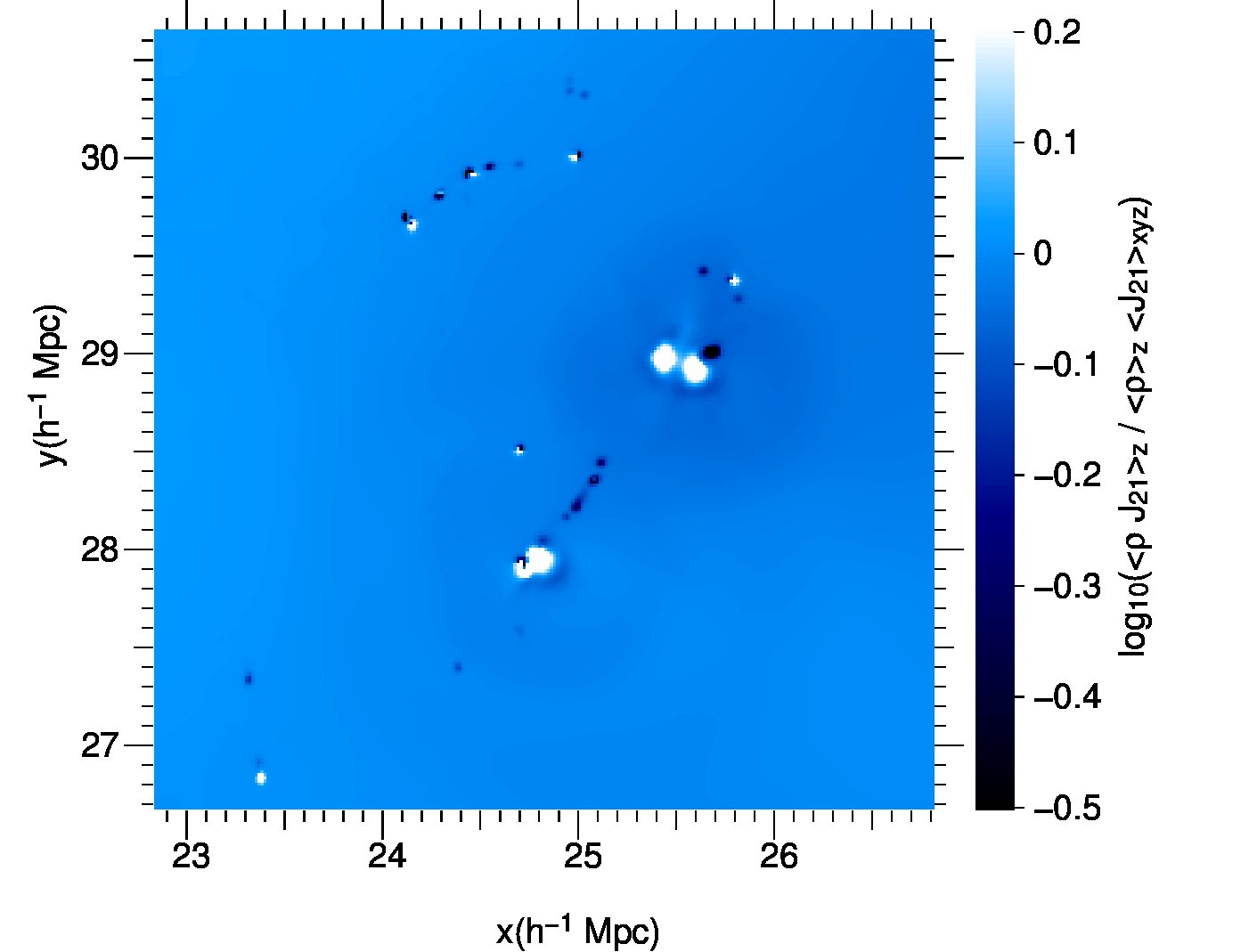}}
\caption{{Mass-weighted maps (xy projection) of the neutral fraction (left) mass-weighted fractional excess of the ionizing flux density (right) for the M31 cutout at \zcoda$=4.28$. The fields are averaged over a 40 cells (625 \hmkpc) thick slice.}}
\label{f:maps2}
\end{figure*}

The physical properties of the gas for the M31 progenitor cutout at 3 different epochs are shown in Fig. \ref{f:maps} . The gas density and ionizing flux density are averaged over a 625 \hmkpc thick slice, while the temperature and neutral fraction shown are taken from a 1 cell thick slice (i.e. 15.625 \hmkpc).
The gas is distributed along sheets, filaments and knots, which become more pronounced with time. The small scale structure visible in the high redshift maps progressively disappears, due to a combination of hierarchical merging and smoothing due to heating by UV radiation. Meanwhile, the voids get less dense, as their gas is pulled towards higher density regions. Let us recall that the filamentary structures seen in these maps are more often 2d planes extending perpendicular to the map than actual 1D filaments. Statistically, \cite{metuki2015} showed that the volume in gas sheets is 7-8 times larger than the total volume of filaments, although the latter end up hosting a similar fraction of the total dark matter mass. Nevertheless we will call these filaments because that is what they look like in a 2D map, but it should be kept in mind that these are often sheets rather than actual filamentary structures of gas.

In the high redshift cutout (\zcoda  $\sim 11$), the first star forms in the highest density clump, located at $(x,y)=(25.5, 27)$\hmpc, and immediately starts to photo-heat its surroundings, as can be seen in the leftmost temperature map. Since the UV radiation is responsible both for ionizing the hydrogen gas and heating it to a few $10^4$ K, it is not surprising that temperature, neutral fraction and ionizing flux density maps look very similar in structure. The ionization front propagates inside-out, faster in low density regions and slower in high density regions because the I-front speed is proportional to the number density ratio of ionizing photons to H atoms at the location of the front. This gives rise to the typical butterfly-shaped ionized regions seen in the middle temperature and neutral fraction maps. Radiation leaves the source's host halo, first into low-density regions, and later on manages to penetrate denser nearby gas sheets and filaments. 

The \zcoda$=6.15$ temperature map also demonstrates the very different scales of the feedback processes implemented: the yellow-orange regions are photo-heated, while the much hotter, reddest region near the highest density peak is also subject to supernova feedback. Therefore stellar formation in this simulation impacts the gas in two very different ways: UV radiation produces long range photo-heating, while supernovae form very local, but very hot bubbles. These supernovae-heated regions are significantly more highly ionized than regions which are only photo-heated.

The bottom middle panel shows that the photon propagation is restricted to photo-heated, ionized regions, as expected since the mean-free-path of ionizing UV radiation is very small in the neutral gas ahead of the I-fronts, so the boundaries of the H II regions are sharply defined. Additionally, this panel shows the location of young star particles, appearing as peaks in the ionizing flux density distribution. Their distribution follows closely the gas density peaks, in which they form. 
The last redshift (post-reionization) panel displays several interesting features, as follows. 

\subsubsection{Sheathed gas filaments}

First of all, the temperature map shows that gas filaments after reionization (or after reionization is completed locally) are rather thick, as also seen for instance in \cite{pawlik2009}, although using a uniform background, and in \cite{wise2014}.
Moreover, the filaments display a sheathed structure in temperature, with a hot, tube-like envelope surrounding a cooler core. This is also seen, for instance, in \cite{ocvirk2008}, although the aspect of the sheath looks thicker in the latter, perhaps because of the larger cell size in low density regions (i.e. the outskirts of the filament). More recently, \cite{kaurov2014} obtained sheathed filamentary structures very similar to CoDa's filaments in physical properties and thickness (about $\sim 200$ \hmkpc in both cases). 

We investigated the origin of this structure. It could simply be due to the difference in cooling rates between the denser, more neutral core and the more diffuse filament envelope. There could also be a more dynamical process causing this temperature gradient: the filament could expand radially due to the outward pressure force caused by its photo-heating. During this expansion, its outer layers would get shocked, forming a high temperature layer similar to the contact discontinuity seen in the adiabatic shock tube experiment \citep{toro1994}. To gain better insight into the process, we detail the properties of a gas filament in Fig. \ref{f:sheath}. The left panel shows an xz slab (one cell thick) of the cutout of Fig. \ref{f:maps}, chosen to highlight the filamentary gas structure. The right panel shows the gas properties of individual cells along the short black segment crossing the filament of the left panel. The density is highest at the filament's center, and decreases outwards. The temperature shows a double peaked profile, surrounding the cooler, high density center. The temperature peaks, combined with the decreasing density, result in 2 dips in the Hydrogen neutral fraction. Moreover, the gas velocity field reveals that all of the gas is in-flowing towards the filament, rather than out-flowing, and the temperature peaks are located at the position showing the strongest velocity gradients, i.e. they demark a strong deceleration region in the gas, and the transition from a supersonic to subsonic flow, as shows the Mach number plot. Therefore, this hot layer is not a contact discontinuity (which would be flowing outwards) but rather an accretion shock formed by gas accreting onto the filament or plane. This is similar to the accretion shocks found for 1D pancake collapse by \cite{shapiro1985}, in which dark-matter-dominated gravity drives a cold, supersonic baryonic infall toward the central density peak, and this causes strong shocks that decelerate the baryons and convert their kinetic energy of infall to thermal energy (i.e. high T).  It is also similar to the halo accretion shocks described by \cite{BD03}. However, instead of spherical accretion as considered by \cite{BD03}, we here have either planar, perpendicular accretion onto a sheet or radial, cylindrical accretion onto a filament.

\subsubsection{Self-shielding}

The filament's core is also significantly more neutral than its hot envelope. This is just the result of the higher density and therefore higher cooling and recombination rates: indeed, the bottom right panel of Fig. \ref{f:sheath} shows that the ionizing flux density is completely flat across the filament: the UV background does not ``see'' the filament. It is not able to self-shield.

Here we investigate further the occurrence of self-shielding in CoDa. The low redshift ionizing flux density map of Fig. \ref{f:maps} shows that except at the location of the 3 most massive density peaks still forming stars, the pockets of radiation seem to have given way to an almost perfectly uniform UV background. { To get a sharper view of this, we computed the ratio of the gas-mass-weighted ionizing flux density to the average $J_{21}$:}
\begin{equation}
  \langle \rho J_{21} \rangle_z / \langle \rho \rangle_z \langle J_{21}\rangle_{xyz} \, ,
\end{equation}

{where the subscript $_z$ denotes an average in the $z$ direction while $\langle J_{21} \rangle _{xyz}$ denotes the average value over the whole cutout volume. The resulting map is shown in the right panel of Fig. \ref{f:maps2}. Thanks to the mass-weighting, both {\em sources and sinks} appear. While the largest gas over-densities seem to host sources, the smaller ones tend to be dark and act as sinks. Interestingly, most haloes hosting sources seem to be also partly sinks, and the mass-weighted neutral fraction map shows that all photon sources are located next to rather neutral cells. This illustrates how galaxies are all at least marginally resolved in CoDa, i.e. even when they are just a few cells across, the code can still capture, albeit coarsely, their strongly heterogeneous internal structure.}

\begin{figure}
  {\includegraphics[width=1.15\linewidth,clip]{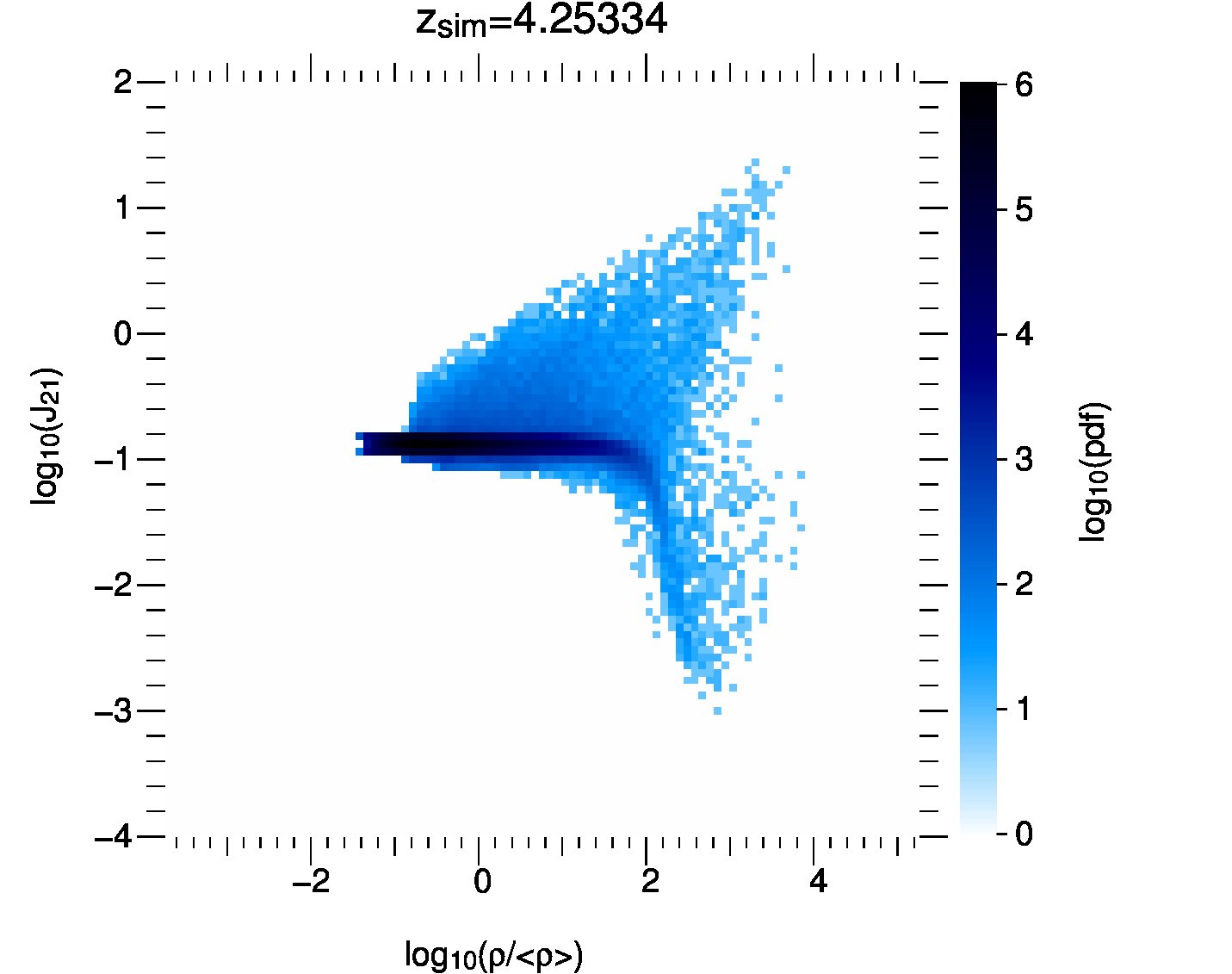}}
\caption{Distribution of the ionizing flux density versus gas overdensity for the M31 progenitor cutout, last snapshot.}
\label{f:cobra}
\end{figure}

A filament of gas joins the 2 main sources of the map. The lower half of it consists of a chain of small gas overdensities, and seems able to self-shield, as shows the photon-density map. On the other hand, the upper half is less dense, and does not appear self-shielded: its ionizing flux density is identical to the background. This link between density and self-shielding is made more obvious by the phase diagram ($J_{21}$ versus gas density) shown in Fig. \ref{f:cobra}: the majority of the cells see a UV background at $J_{21}\sim -1$. The high density part of the diagram has 2 branches. In the upper branch, the cells are under the influence of a nearby source or contain one.
The lower branch arises from self-shielding, and starts at an overdensity of $\sim 100$, similar to what was found by \cite{aubert2010}. This value is at the very high end of the density range of filaments, and most filaments have lower densities, including that of fig. \ref{f:sheath}. Therefore, while haloes are at least partially self-shielded, filaments are not.
Filaments can indeed resist the I-front propagation {\em at the onset of their own local reionization}, but not for long. The \zcoda$=6.15$ neutral fraction map indeed shows a few neutral clumps and filaments still partially neutral, surrounded by ionized gas and pockets of radiation. However this does not last and in the last snapshot they end up with neutral fractions between $x_{\rm HI}=10^{-2}-10^{-4}$.


\subsection{Global properties}
\label{s:glob}
\begin{figure*}
\begin{center}
\begin{tabular}{cc}
  {\includegraphics[height=7.8cm,clip]{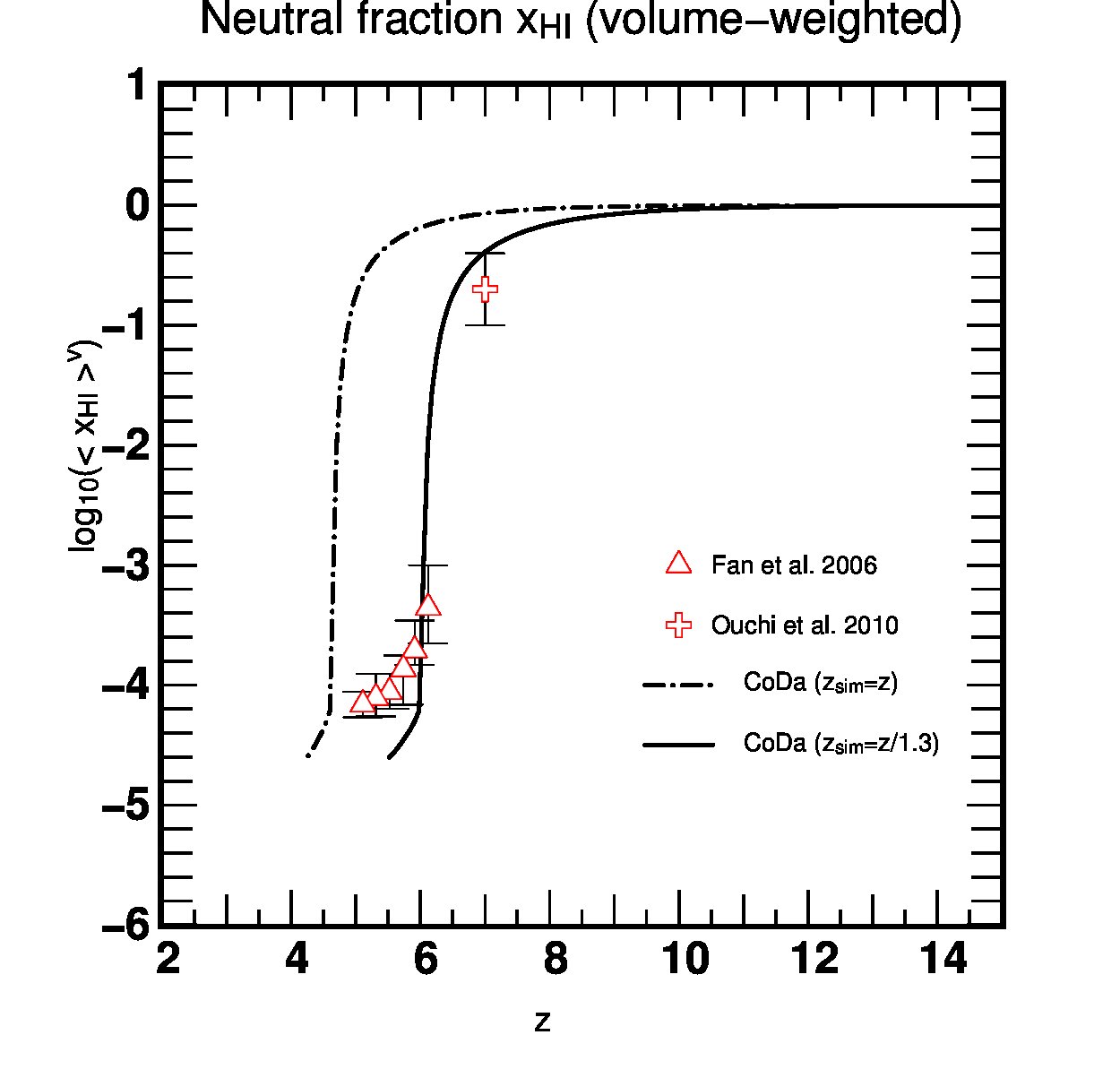}}&
  {\includegraphics[height=7.8cm,clip]{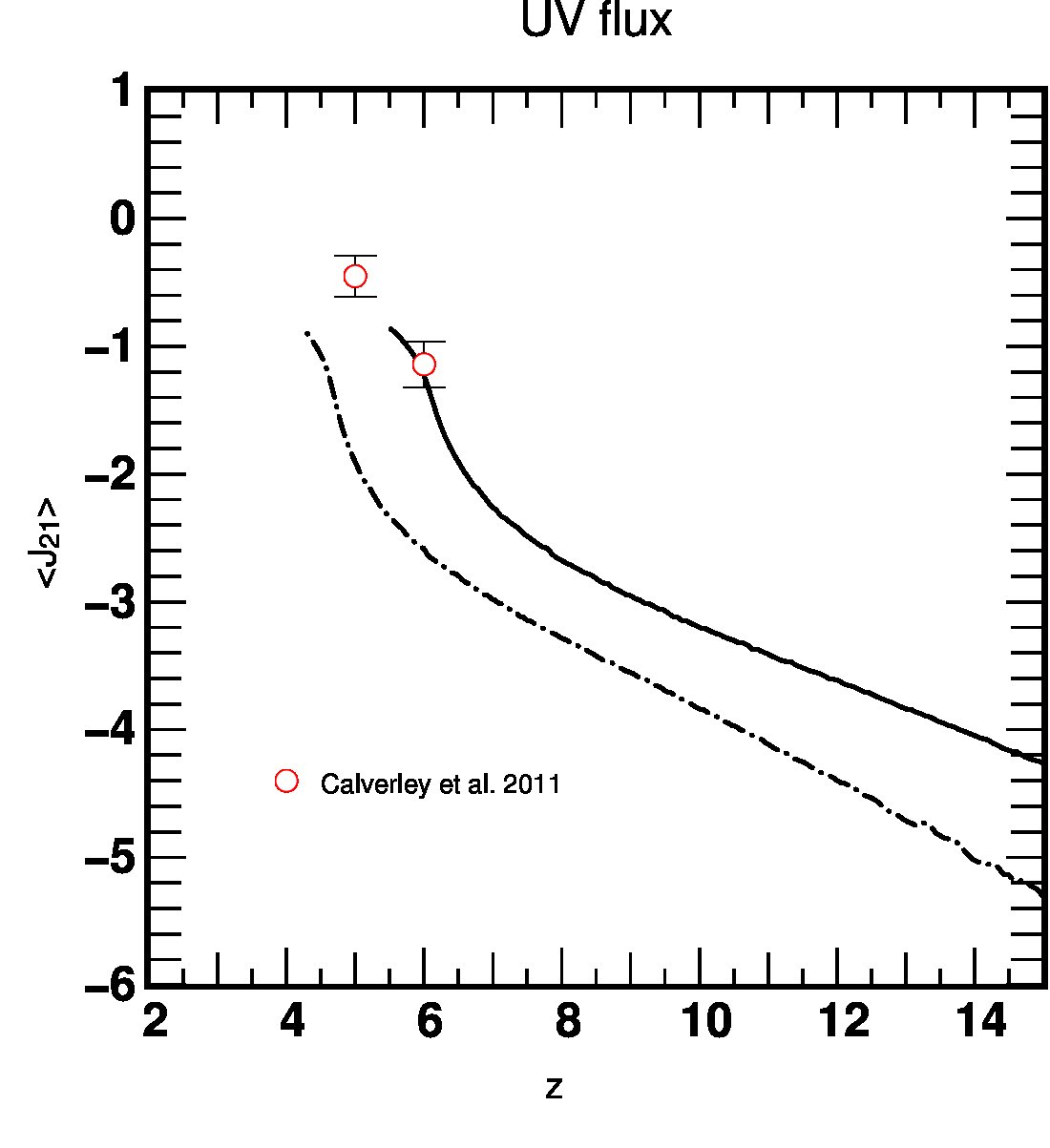}}\\
  {\includegraphics[height=7.8cm,clip]{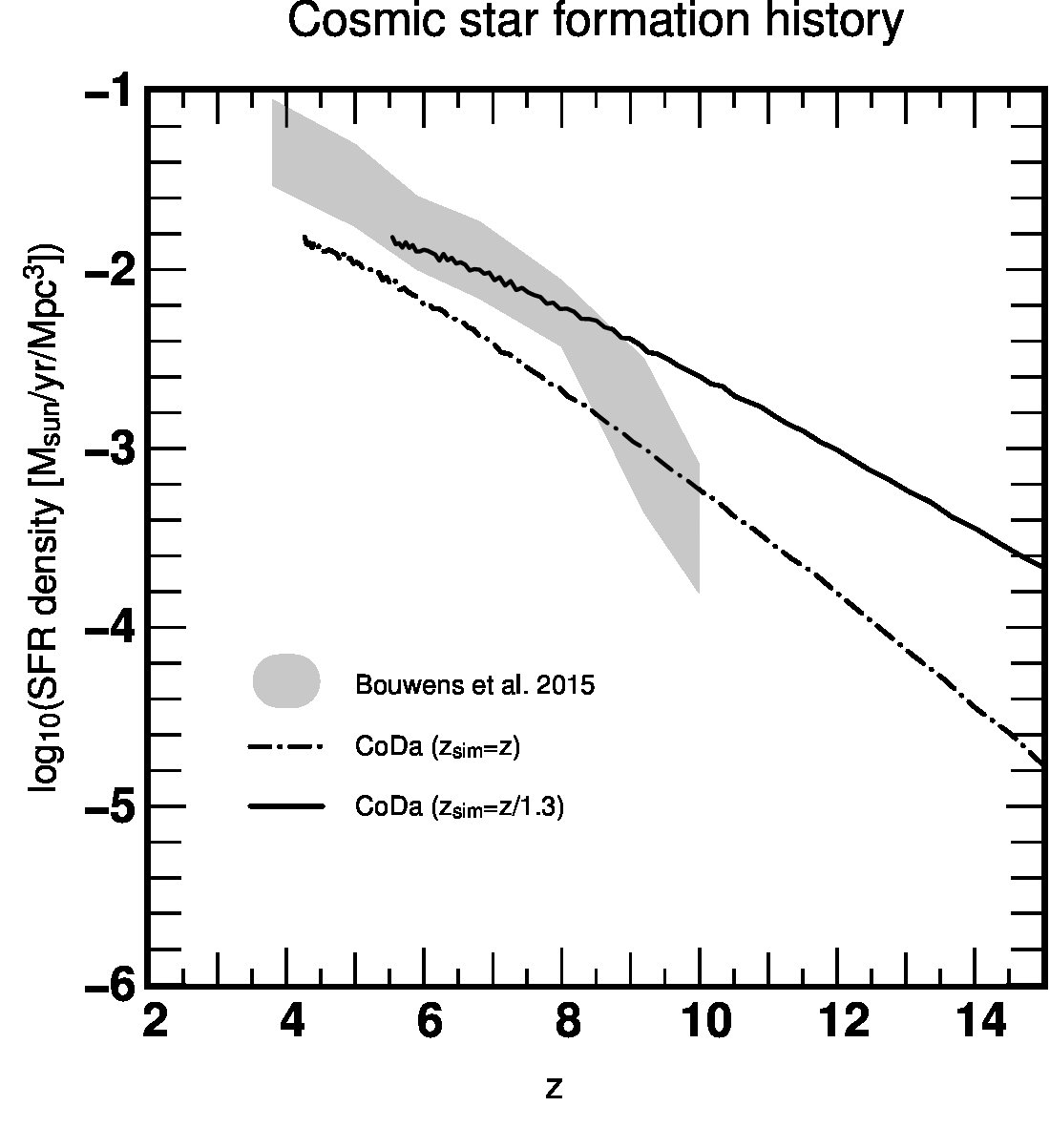}}&
  {\includegraphics[height=7.8cm,clip]{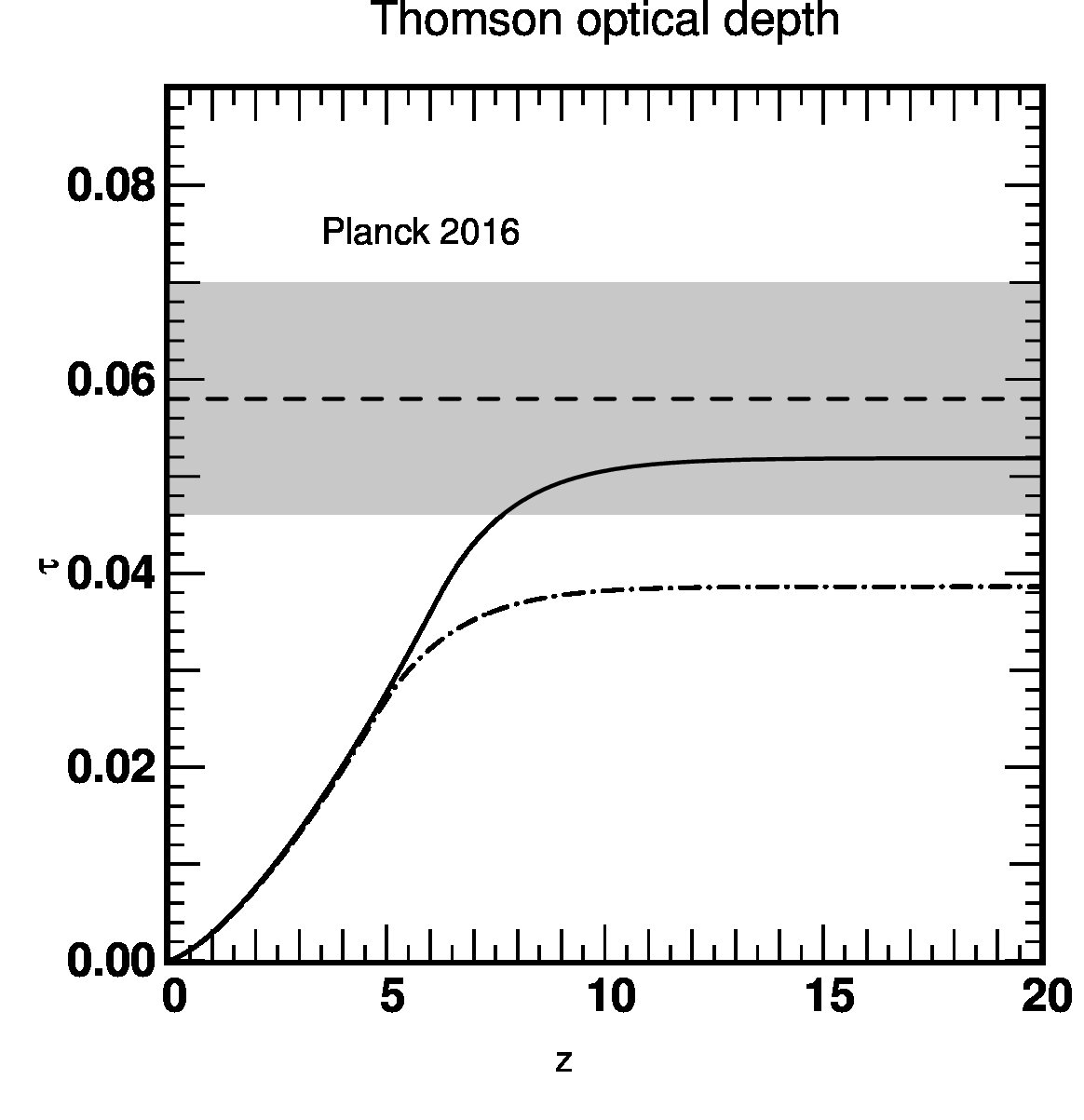}}
\end{tabular}
\end{center}
\caption{Global properties of the simulation, and comparison to observations. The dot-dashed black line shows the simulation results, while the solid black line shows a temporal rescaling of the CoDa simulation, i.e. for instance $x_{\rm HI}($\zcoda$=z/1.3)$, so as to shift the end of the EoR to z=6 instead of z=4.6. This transformation brings all the fields in agreement with the available high-z observations.}
\label{f:glob}
\end{figure*}

In this section we show that the global evolution of the simulation is physically consistent and in line with our theoretical understanding of the EoR, and to some extent with the observational constraints available.

The most basic quantities to consider when gauging the success of a global EoR simulation are the evolution of the cosmic means of the neutral fraction, ionizing flux density, and cosmic star formation history. These are shown in Fig. \ref{f:glob}, along with several observational constraints from \cite{fan2006}, \cite{ouchi2010}, \cite{calverley2011}, and the Planck CMB thomson optical depth $\tau$ \citep{planck2016}. The observed cosmic star formation rate (hereafter SFR) constraints are taken from  \cite{bouwens2014a}: the grey area shows the envelope including the dust-corrected and dust-uncorrected SFRs. As a first remark, we note that the cosmic SFR in CoDa increases at all times, unlike the simulation of \cite{so2014}, which shows a decline at late times. In the latter, the authors suggested this could be due to the small box size they used. This may indeed be the case, since CoDa is more than 95 times larger in volume.

The neutral fraction plot shows a characteristic very steep decrease, down to $x_{\rm HI}\sim 10^{-4.2}$, where the slope becomes more gentle. This transition marks the end of the EoR, and correlates with the end of the surge in ionizing flux density seen in the middle panel. Following this definition, reionization is complete in the simulation at redshift \zcoda$=4.6$, which is about 300 Myr later than indicated by observational constraints. This discrepancy is consistent with the lower level of star formation in the simulation as compared to the observations of \cite{bouwens2011}, shown in the right panel. 


{These observational constraints are global ones, of course, and do not directly constrain the history of the local universe simulated by CoDa, to the extent that the local universe might be statistically differentiated from the global average.}

In CoDa, the EoR lasts about 30\% longer than global EoR observational constraints suggest. This could be a problem when using CoDa for producing mock observations of the Lyman $\alpha$ forest or the  Gunn-Peterson trough for instance, for which the large size of the simulation is a desirable property. In an attempt to correct for this, we also show the effect of shrinking the redshift axis of the simulation: the solid black line in Fig. \ref{f:glob} shows the global properties of the CoDa simulation at the rescaled redshift \zcoda$=z/1.3$. { This simple rescaling mimics the effect of a modest increase of the star formation efficiency parameter. It brings the neutral fraction evolution in reasonable agreement with the observational constraints, as well as the ionizing flux density, the cosmic star formation history and the CMB Thomson scattering optical depth $\tau$. Therefore the simulation should remain viable for producing mock observations, although other quantities, such as the collapsed mass fraction in halos, could be offset because of this temporal shift. We plan to explore this in future papers.} In the rest of the paper, we carefully, explicitly state when such rescaled redshifts are used.

{The post-reionization neutral hydrogen fraction in CoDa is about 0.5 dex lower than observed. Such an offset is not uncommon and the literature shows that simulations of the EoR can exhibit a variety of such small departures from the observed evolution of the ionized fraction measured from quasar lines of sight (see for instance \cite{aubert2010,zawada2014,gnedin2014,aubert2015}), in particular in the fully coupled radiation-hydrodynamics regime, due to the cost and difficulty of tuning such simulations. In the case of CoDa, this offset may be reduced by adopting a lower black-body temperature for our stellar sources, i.e. less massive stars, leading to a smaller effective ionizing photon energy, and therefore a smaller injection of energy per photo-ionization event, resulting in a lower average temperature of the large scale IGM, which would make it slightly more neutral. Indeed, Fig. 9 of \cite{pawlik2015} shows that changing the heat injection per photo-ionization event has a strong impact on the post-reionization neutral fraction. However, decreasing the black body temperature of our sources to e.g. 50 000 K would also increase the effective H cross-section $\sigma_E$, leading to more ionizations and therefore may lead to an even larger ionized fraction at equilibrium. More simulations will be required to test this in future work. Finally, \cite{aubert2010} showed that modelling the gas clumping at sub-grid resolution could help in reproducing the observed trend. Just how much resolution is required to capture this additional gas clumping fully, beyond that of the CoDa simulation, was demonstrated, for example, by \cite{emberson2013}, who calculated the transfer of ionizing UV radiation through a static snapshot of the density field from a sub-Mpc-volume cosmological simulation. More recently, \cite{park2016} used a fully-coupled radiation-hydro 
simulation of such a small volume to show that hydrodynamical back-reaction must be included, as well, since this causes the clumping to be time-dependent; eventually, even the denser, self-shielded gas ionizes and photoevaporates.}

Large simulations such as CoDa, which can be run only once because of their cost, are very difficult to calibrate. As described in Section 2.1, for instance, tuning the subgrid star formation efficiency parameters by performing a large suite of small-box simulations (e.g. 4 \hmpc boxes) to find parameters that satisfy observational constraints does not guarantee that a large 91 Mpc box will follow the same evolution, because of cosmic variance, and because the large box will contain voids which can be orders of magnitude larger in volume than the small 4 \hmpc calibration run. Moreover, it is unclear how periodicity (photons exit through one face and come back through the opposite face) affects the calibration of such small boxes. Taking a step back, it now appears that the strategy for calibrating such a large run should be more hierarchical. In the future, we can use the CoDa simulation, itself, to help re-calibrate, as well.




\subsection{Impact of radiative feedback on galaxy formation}
\label{s:SFR}
\begin{figure*}
\begin{center}
  {\includegraphics[width=0.73\linewidth,clip]{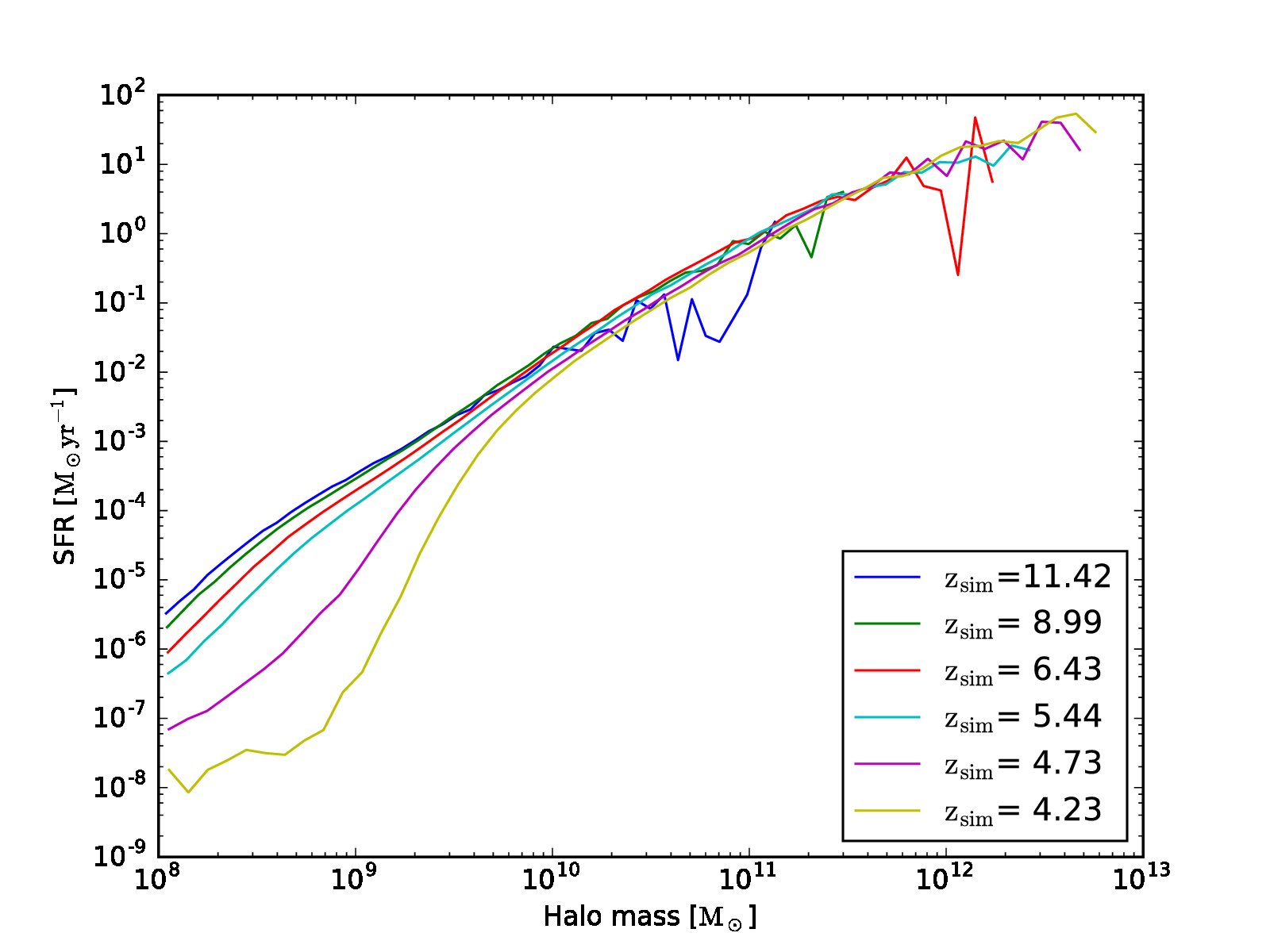}}
\end{center}
\caption{{Instantaneous star formation rate per halo as a function of instantaneous halo mass, for various redshifts. The instantaneous SFR is computed as the stellar mass formed within an $R_{200}$ radius sphere centered on the dark matter halo center of mass, during the last 10 Myr, divided by a duration of 10 Myr. Notice the sharp suppression at low mass.}}
\label{f:SFR}
\end{figure*}

In this subsection we investigate the impact of the radiation-hydrodynamics coupling on galaxy formation. Since early works of \cite{gnedin2000}, several authors have addressed the impact of radiative feedback on galaxy formation, at low and high mass and in a number of contexts, with a fixed uniform UV background from \cite{haardtmadau96,haardt2012}, and more recently with fully coupled RHD simulations \citep{pawlik2013,wise2014,jeon2014,rosdahl2015,aubert2015,pawlik2015}, as a reduction and possibly a suppression of star formation at low masses.
{We computed the instantaneous SFR of CoDa haloes as the stellar mass formed within a sphere of radius,
  $R_{200}$ (i.e. that within which the average density of the dark matter is 200 times the cosmic mean density) centered on the dark matter halo center of mass, during the last 10 Myr, divided by a duration of 10 Myr. Fig. \ref{f:SFR} shows the instantaneous SFR that results, as a function of the instantaneous mass of the dark matter halo, for several redshifts.}

There is a general trend for haloes at all masses to form fewer stars as time goes by, which is linked to a wide-spread decrease in accretion rate, as seen in \cite{ocvirk2008}. Here we find a trend: SFR $\propto M^{\alpha}$ for $M>10^{10}$ \Msun   with a slope $\alpha \sim 5/3$. This slope is compatible with the values found in the literature. i.e $1<\alpha<2.5$, in numerical and semi-analytical studies, such as \cite{tescari2009,hasegawa2013,yang2013,gong2014,aubert2015}. 
However, the most striking feature of Fig. \ref{f:SFR} is the very sharp decrease in SFR for the low mass haloes, around $\sim 2 \times 10^{9}$ \Msun. Before z$\sim6$, low mass haloes sit on the same global trend as the high mass haloes. However, during the EoR, they transition from this ``normal'' state to a strongly suppressed state: at z=4.2, the $10^9$ \Msun haloes form  almost 1000 times less stars than they did at z$=6$. {This suppression reflects the great reduction of the gas fraction inside the galaxies that is below 20,000 K (as required for star formation by our star formation criterion described in  Sec. \ref{s:ramses}), once the galaxy and its environment are exposed to photoionization during reionization. In contrast, the gas core of high mass haloes is dense enough to remain cool and/or cool down fast enough to keep forming stars, even if in bursts.}

\begin{figure*}
  {\includegraphics[width=0.48\linewidth,clip]{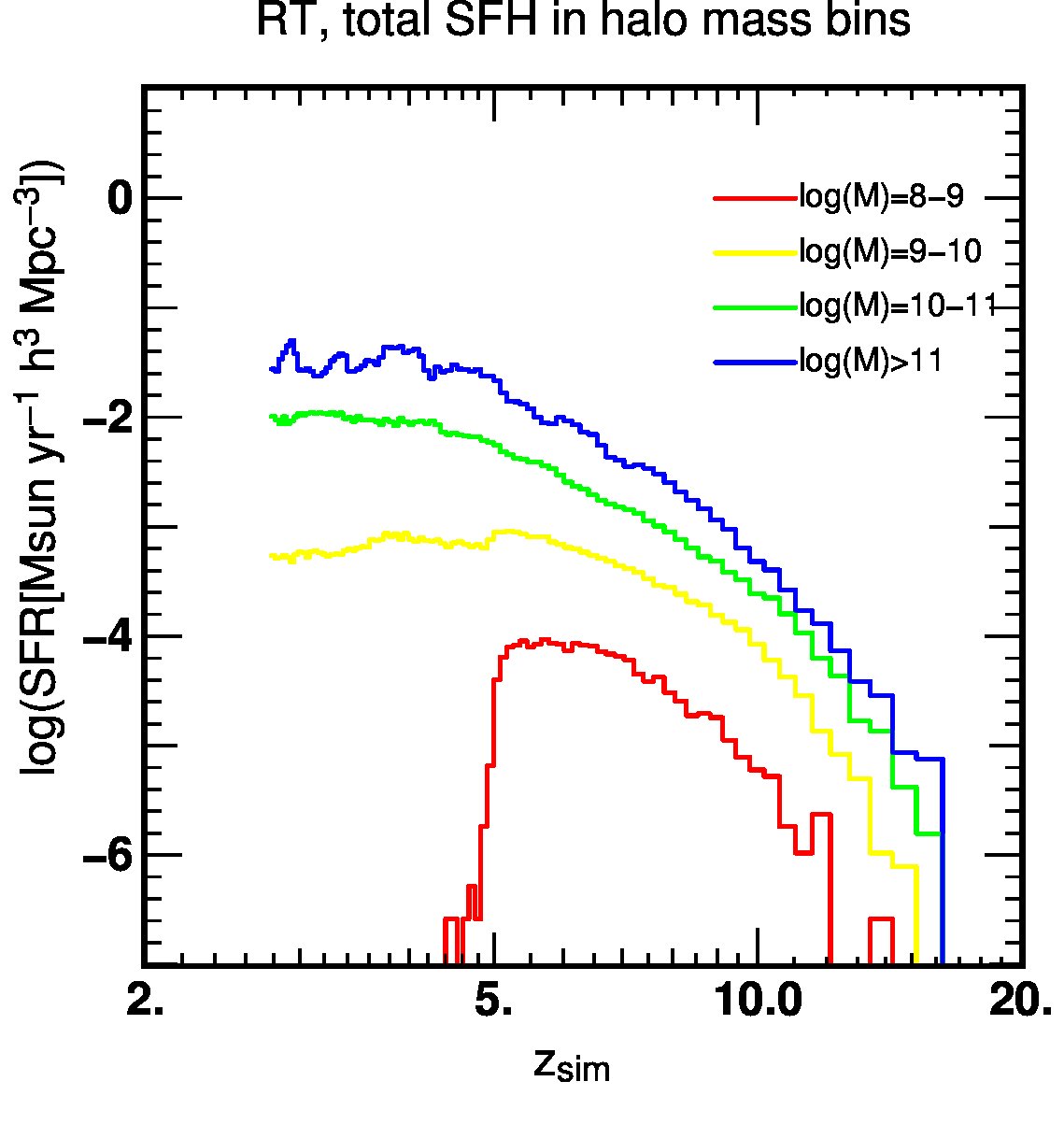}}
  {\includegraphics[width=0.48\linewidth,clip]{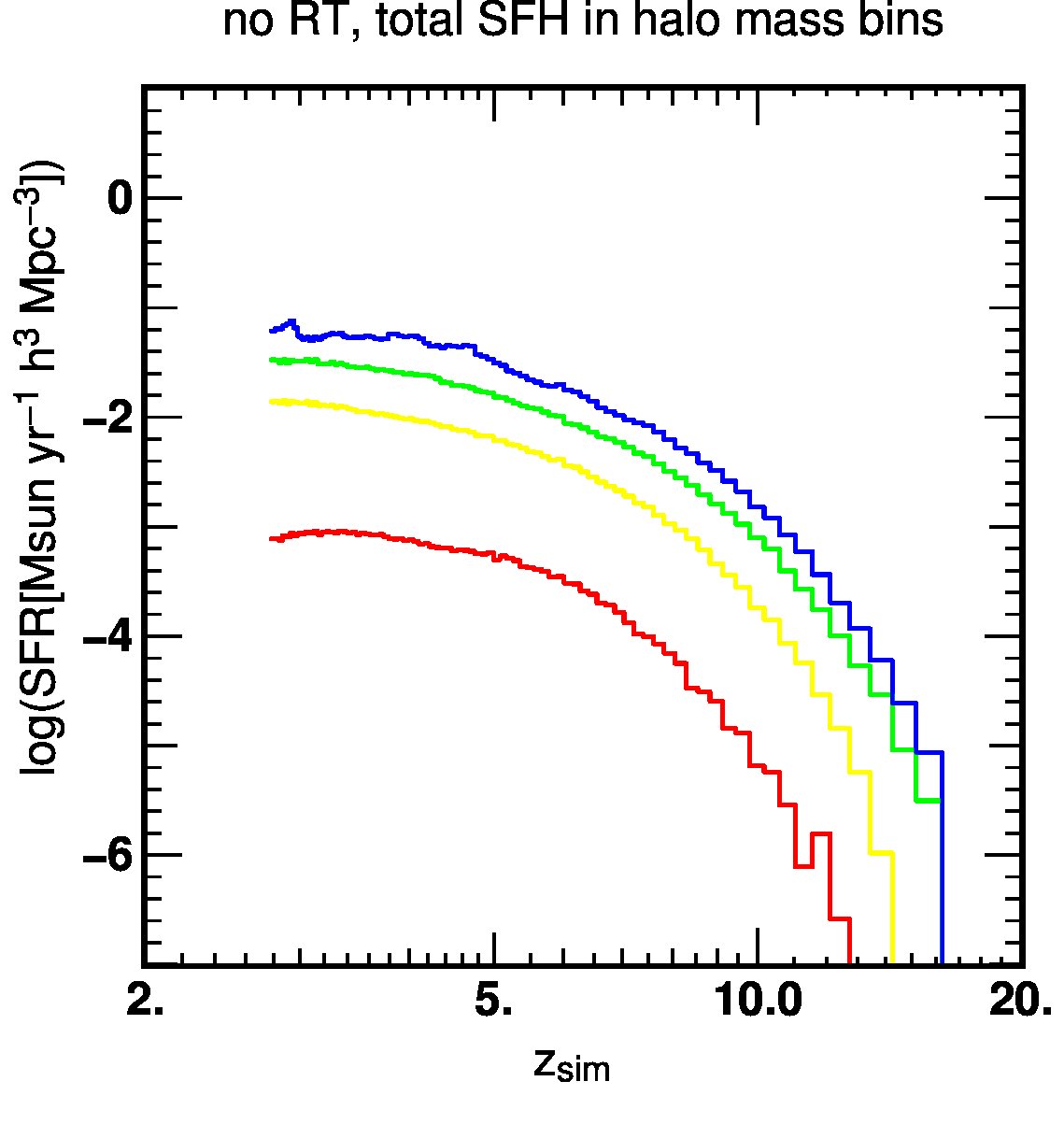}}
\caption{Total star formation histories of 4 halo mass bins for two simulations in a test box 8 $\hmo$ Mpc on a side. {\em Left:} with full radiation hydrodynamics. {\em Right:} SN feedback only, no radiation. The mass bins correspond to the haloes' final mass (i.e. halo mass measured in the final simulation timestep, at \zcoda$\sim 3$).}
\label{f:SFH}
\end{figure*}

To check this, we ran two 8 \hmpc RAMSES-CUDATON simulations: one with the exact same parameters and resolution ($512^3$ grid) as the CoDa simulation and a second without radiative transfer (therefore vanilla RAMSES, with SN feedback only), and no UV background. The star formation histories of haloes of various masses\footnote{the halo masses are those measured at the last snapshot, contrary to those of Fig. \ref{f:SFR}, which are instantaneous} are shown in Fig. \ref{f:SFH} for these 2 simulations. For the RHD simulation (left panel), there is again a very strong suppression of star formation at the end of the EoR for the low mass haloes. We note that more massive haloes are also affected, although in a less spectacular, if still significant way: even the $10^{10-11}$ \Msun   mass bin sees its star formation decrease by a factor of $\sim2$ when compared to the pure hydro run (right panel).
This experiment confirms that the addition of radiative feedback via photoheating associated with hydrogen photoionization is the cause of the sharp suppression of star formation seen in low-mass haloes.

While this trend is clear and extends well up into the intermediate halo mass range, far above the minimum mass resolved in our halo mass function, the quantitative details may be affected by the limits of our spatial resolution. {For instance, higher resolution simulations could form denser clumps inside a given halo, which would better resist ionization and photo-heating. Comparing with literature, we find for instance that \cite{oshea2015}, using the renaissance simulations, offering up to 10 times higher mass resolution and better spatial resolution, but in a volume thousands of times smaller than CoDa, reports SFR suppression at halo masses typically 10 times smaller, i.e. $\sim 2 \times 10^8$ \Msun. Meanwhile, \cite{gnedin2014}, using the  CROC simulations, find a very moderate impact of radiative feedback on the properties of low mass galaxies during reionization. However, they check for this effect by examining variations in the faint end of the UV luminosity function (hereafter LF). While this is valuable from an observational perspective, the impact of radiative feedback on star formation or the lack thereof is more easily assessed by looking at the SFR - halo mass relation as we do here, than at the UV LF. This difference in methodology makes a direct comparison with \cite{gnedin2014} difficult. Moreover, their treatment of star formation relies on modelling molecular hydrogen formation. This raises the density threshold for star formation, and therefore pushes star formation to higher mass haloes. The smallest galaxies, which are the ones most-strongly suppressed by radiative feedback in CoDa, would probably not have formed any stars at all within the CROC formalism. Therefore the question of the impact of radiative feedback on low mass haloes is tied, not only to spatial and mass resolution, but also to the choices made for the star formation recipe.

  While performing a higher resolution CoDa would be prohibitively expensive with the current code and facilities, we performed a simple resolution study spanning 2 times higher and up to 4 times coarser spatial resolution with respect to CoDa, using $512^3$ boxes of 4 to 32 \hmpc on a side. Details can be found in Sec. \ref{s:resstudy}. Preliminary results indicate that degrading resolution beyond CoDa's produces a radiative suppression of star formation at increasingly higher masses. However, the test boxes at CoDa resolution and 2 times higher resolution yield the same suppression mass scale of about $\sim 2 \times 10^9$ \Msun. We note that this is compatible with the results of \cite{pawlik2015}, obtained with a mass resolution comparable to CoDa but better spatial resolution due to the use of an SPH-based Lagrangean hydro code instead of CoDa's unigrid scheme. Their study reports a SFR suppression mass similar to ours, of $\sim 10^9 $ \Msun.}

\subsection{Contribution to the cosmic star formation density}

\begin{figure*}
\begin{center}
 {\includegraphics[width=0.79\linewidth,clip]{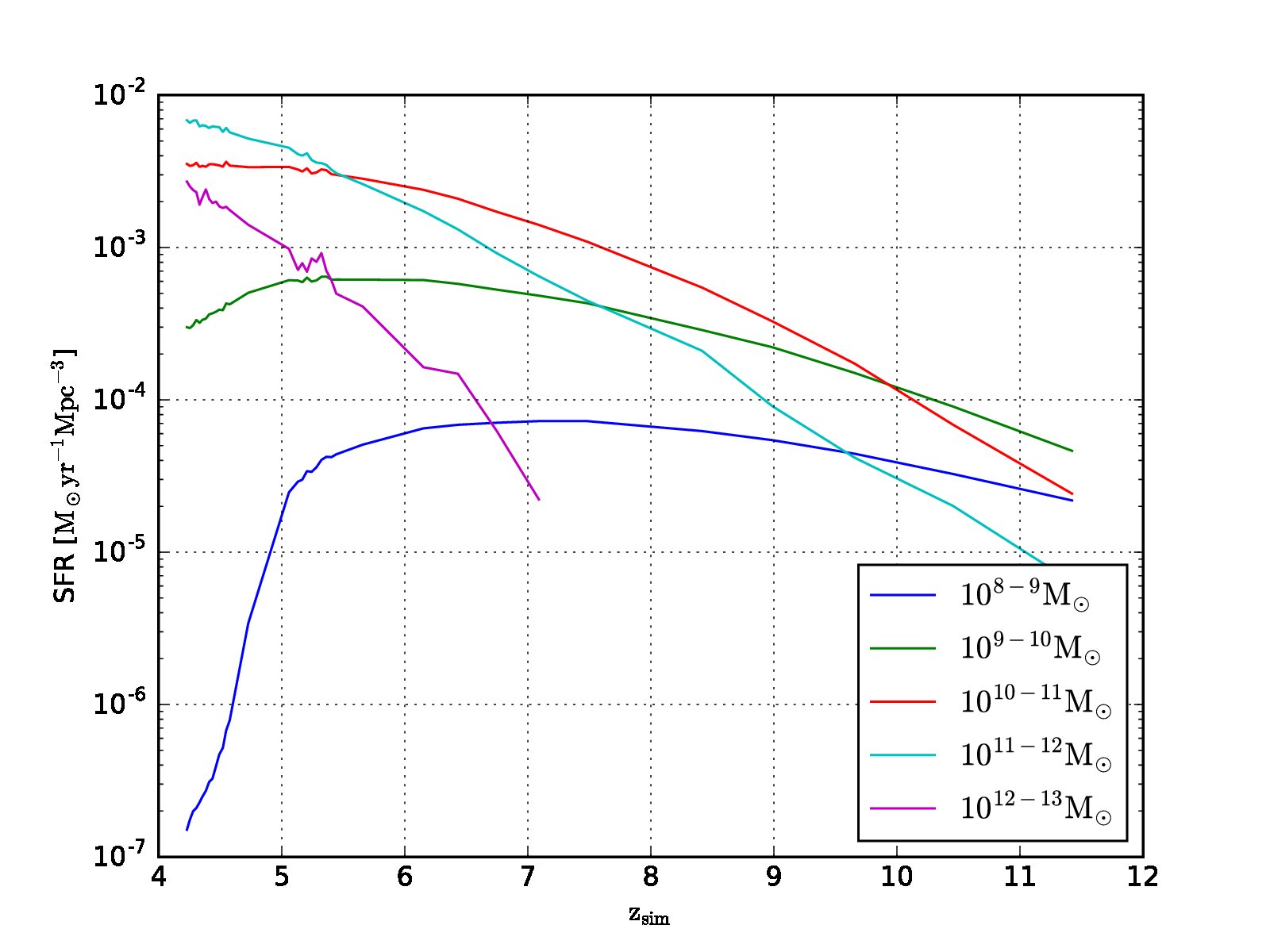}}
\end{center}
\caption{{Cosmic SFR density as a function of redshift for various bins of instantaneous halo mass: each line shows the contribution of all galaxies within a given mass bin to the total cosmic SFR density.}  
}
\label{f:SFRz}
\end{figure*}

We now investigate the contribution of haloes of various masses to the total star formation density of the box. While Fig. \ref{f:SFR} already made clear that individual halo SFR increases with mass, one needs to multiply their SFR by the halo number density to obtain their contribution to the star formation density. This is shown in Fig. \ref{f:SFRz}: each line shows the star formation density due to each mass bin (each mass bin is a decade wide). As expected, the lowest mass haloes contribute almost equally to the other mass bins at high z (\zcoda$>$11) but their contribution decreases sharply during the late stages of the EoR, as they become increasingly suppressed. This decrease, though more gentle, is also seen for the $10^{9-10}$ \Msun bin. However, for the 3 most massive bins, we find a constant rise of the star formation density. This rise is faster for increasing halo mass. This $10^{12-13}$ \Msun mass bin can only be traced at \zcoda$<7$, as before this time the simulation does not contain any halo that massive. The $10^{10-11}$ \Msun haloes dominate the cosmic star formation density for most of the simulation time, before being overtaken by the $10^{11}$ \Msun mass bin. The hierarchy between mass bins and the overall evolution is very similar to that shown in Fig. 2 of \cite{genel2014}, although the details of the methodology and physics implemented differ vastly. The total SFR is however lower than the observational estimate, as was already pointed out in Sec. \ref{s:glob}. This reflects the underestimated value adopted for the subgrid parameters which control the efficiency of star formation over-all, but not the relative contributions for different halo masses.

We will refrain, at this stage, from drawing conclusions as to which mass bin contributes most to large scale reionization. Indeed, while all stellar particles have the same intrinsic specific emissivity, we do not know how much of the ionizing photons produced made it into the IGM. We will come back to this question by measuring the circum-galactic escape fraction in a future study.

\subsection{UV luminosity function}

\begin{figure}
  {\includegraphics[width=1.\linewidth,clip]{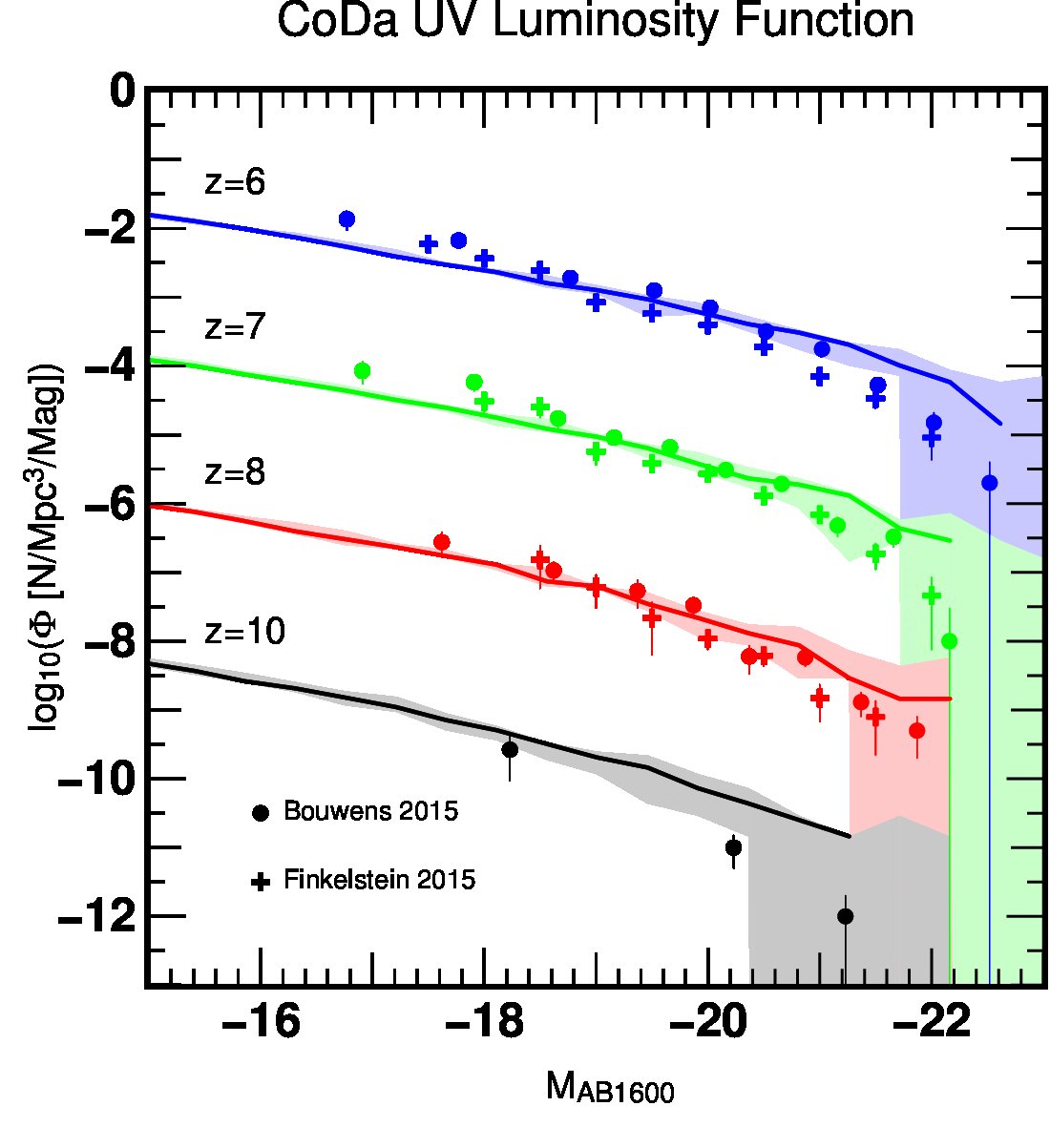}}
  \caption{UV luminosity functions and comparison with observations. The full circles and crosses with error bars are the observations from Bouwens et al. 2015 and Finkelstein et al. 2015, respectively, at z=$[6,7,8,10]$, while the shaded area and the thick line show the envelope and the median of the LFs of 5 equal, independent, rectangular sub-volumes taken in the CoDa simulation, at zcoda $=[6,7,8,10]/1.3$. For clarity, the LFs have been shifted downwards by 0, 2, 4 and 6 dex.}
  \label{f:uvlf}
\end{figure}


In order to gain more insight into the balance of bright vs faint galaxies in our simulation, we computed the luminosity function (hereafter LF) of CoDa haloes. We computed the $M_{AB1600}$ magnitudes using the lowest metallicity stellar population models of \cite{conroy2010}. 
We ensured consistency with RAMSES-CUDATON's source model by rescaling Conroy's single stellar population models in flux so as to obtain the same ionizing photon output over 10 Myr. The results are shown in Fig. \ref{f:uvlf}, along with observational constraints. The latter are taken from \cite{bouwens2015} and \cite{finkelstein2015}, which have been shown to be in broad agreement with a number of other studies including \cite{oesch2013} and \cite{bowler2015}.
The LFs have been shifted vertically for clarity. The shaded area shows the envelope of the LFs obtained for 5 rectangular independent sub-volumes of the CoDa simulation. Each of these sub-volumes spans $\sim 150,000$ Mpc$^3$ ($\Delta x=\Lbox/5$,  $\, \Delta y=\Delta z=\Lbox$, i.e. 1/5 of the full box volume), which is similar to the volume probed by CANDELS-DEEP at $z=6$. The resulting envelope therefore illustrates the expected effect of cosmic variance at $M>-20$. The thick solid line shows the median of these 5 LFs. The observed $z=6-8$ LFs are in rather good agreement with our simulation for $M>-21$. 

However, the simulation seems to predict an overabundance of bright galaxies, at all redshifts. Due to the small simulated volume compared to the observations (the survey volume at the bright end is 4 times larger than CoDa), it is not clear if this is a robust prediction of the simulation or just a statistical accident. There could however be several reasons for such an overabundance:
\begin{itemize}
\item{Missing physics:}
\begin{itemize}
\item{dust: the only impact of dust extinction in CoDa is through a constant stellar escape fraction $\fesc^{dust}=0.5$. However, dust extinction becomes increasingly important at the bright end \citep{kimm2013,gnedin2014a}. Moreover, observed high redshift galaxies, even during the EoR, could be fairly dusty \citep{watson2015}.}
\item{AGN feedback: the radiative and/or mechanical feedback from AGNs is believed to be responsible for the drop-off of the bright end of galaxy LF at low redshift. Could it be that early super-massive black holes in massive high redshift galaxies regulate the bright end of the galaxy LF as well? CoDa does not include AGN feedback, and it could help explain the overabundance in our LFs.}
\end{itemize}
\item{overlinking: the FoF algorithm we used for halo detection is notorious for producing too massive haloes at high redshift, when compared to other halo-finding algorithms, such as the spherical overdensity. This is known as the ``overlinking problem'' \citep{watson2013}. However, we checked that using a shorter FoF linking length b=0.15 instead of the usual b=0.2, which could be more adequate at high redshift, did not improve the LFs.}
\end{itemize}

\subsubsection{Halo Mass and UV luminosity:}
\label{s:hmuv}

\begin{figure*}
  {\includegraphics[width=0.49\linewidth,clip]{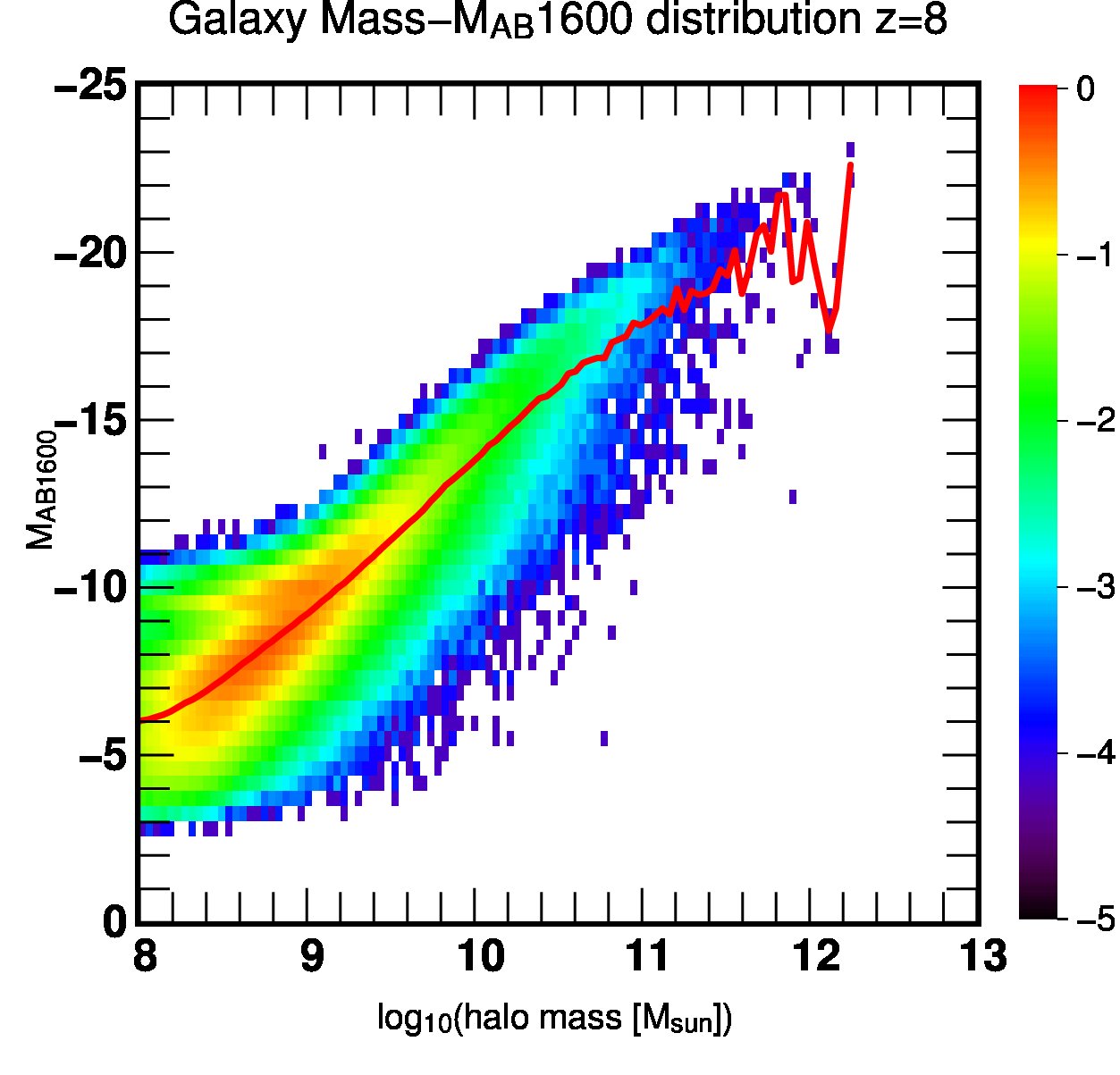}}
  {\includegraphics[width=0.49\linewidth,clip]{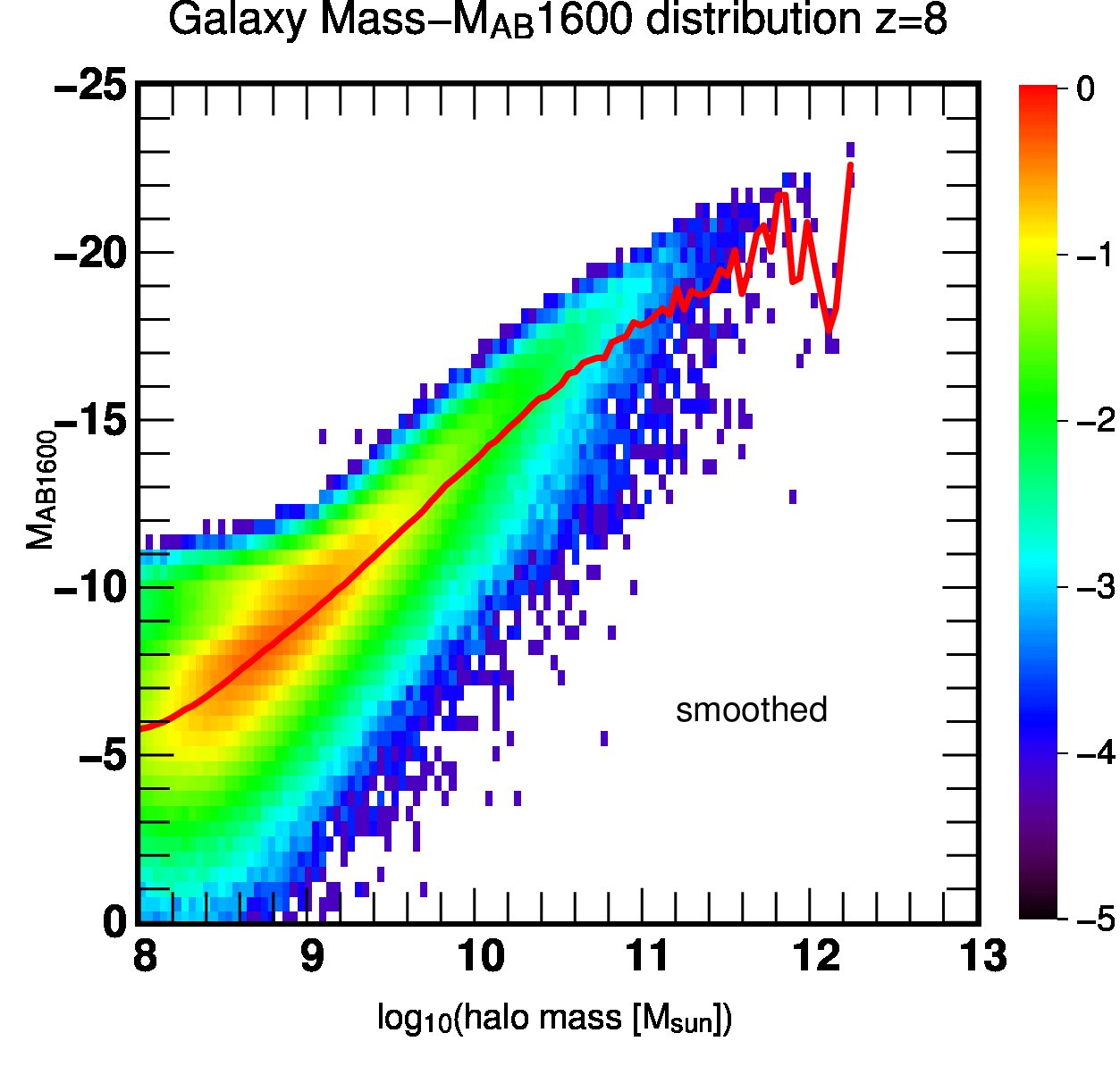}}
\caption{Galaxy mass - magnitude distribution at z=8 (i.e. \zcoda=6.15). The color indicates the galaxy number density in N/Mpc$^3$/Mag/log(M$_{\odot}$). The red line indicates the average Magnitude for each mass bin. Left: original stellar particle masses (i.e. quantized). Right: ``smoothed'' stellar particle masses.}
\label{f:massmag}
\end{figure*}

{
We also investigated the relation between UV galaxy luminosity and halo mass. There is a general trend of increasing UV continuum luminosity with halo mass. This is shown in the left panel of Fig. \ref{f:massmag}, which corresponds to the galaxy population of the $z=8$ UV LF of Fig. \ref{f:uvlf}. Moreover, for a given halo mass, the luminosity can fluctuate significantly, as shows the vertical spread of the distribution. The vertical dispersion increases with decreasing halo mass, and is largest for the mass range sensitive to radiative feedback. The horizontal overdensity at $(M_{AB1600} \sim -10,\, M<10^9$ \Msun$)$ is due to the quantization of stellar mass: the haloes located in this region have exactly one young UV-bright stellar particle, which has a fixed minimum mass. Below this limit, haloes are populated by star particles older than 10 Myr, therefore fainter in the UV.
Indeed, in RAMSES-CUDATON, the mass of a stellar particle is always a multiple of the elementary star particle mass $M_{\star}=3194$ \Msun. This quantization of the stellar mass leads to a quantization in the magnitudes and therefore to possible artifacts in the faint-end LF, whereas the bright end LF is not affected, due to the larger stellar masses involved.
We can mitigate the effect of this quantization by modifying the stellar particles' masses by adding to them a random number taken from a uniform distribution between $[-M_{\star}/2,M_{\star}/2]$, where $M_\star$ is the stellar particles elementary mass. A stellar particle of mass $M_\star$ is therefore assigned a mass between $0.5 M_{\star}$ and $1.5 M_\star$, while a stellar particle of mass $10 M_\star$ is assigned a mass between $9.5 M_\star$ and $10.5 M_\star$. Therefore this ``smoothing'' of the stellar particles' mass is stronger, relatively, for haloes with lower stellar mass, but on average the total added stellar mass is 0.
This simple procedure reduces quantization artifacts in the faint end of the LF, as shows the right panel of Fig. \ref{f:massmag}, and will help its interpretation.}

\section{Discussion}

\subsection{Observing the End of Reionization: Depressing the Faint-End of the Luminosity Function}

\begin{figure}
  {\includegraphics[width=1.\linewidth,clip]{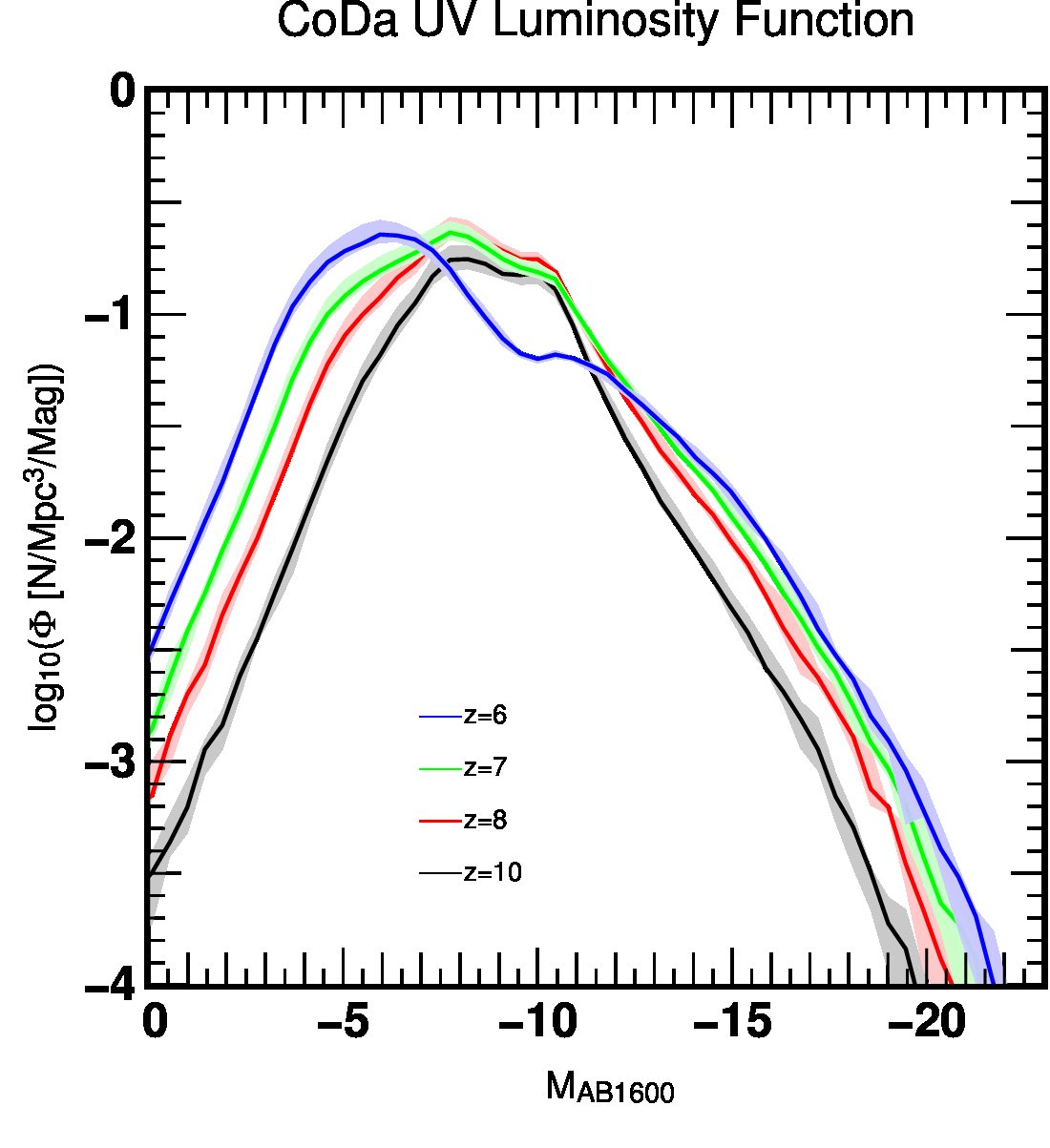}}
\caption{Same as Fig. 10, extended to include the faint end, using the ``smoothed'' stellar particle masses as explained in Sec. \ref{s:hmuv}}
\label{f:faint}
\end{figure}

{
It has long been thought that reionization must have exerted a negative
feedback on the ability of low-mass galactic haloes to form stars,
by suppressing the infall of intergalactic baryons from the photoheated IGM
following its local reionization (e.g. \citep{shapiro1994,thoul1996,quinn1996,kitayama2000,gnedin2000}.  As we mentioned in Section 1, this effect may help explain why
there are so many fewer dwarf galaxy satellites observed in the Local Group
than there are low-mass haloes predicted by N-body simulations of the CDM model.  
However, it has also been suggested that the suppression of star formation in
low-mass galaxies by reionization feedback might be observable in the evolution
of the global star formation rate and galactic luminosity function at high
redshift.  \cite{barkana2000}, for example, used a semi-analytical approach
to show that if the local IGM Jeans mass filter scale jumped up when a patch
became an H II region, a sharp drop in the global star formation rate would
appear as reionization overtook a significant fraction of the volume of the
universe.  Large-scale N-body + radiative transfer simulations of patchy
reionization like those by \cite{iliev2007,iliev2012}, \cite{ahn2012}, and
\cite{dixon2016}, for example, which also make this assumption about 
reionization suppression of low-mass galaxies inside H II regions find that 
reionization is, in fact, self-regulated, such that the rise of the suppressible 
galaxies limits their ability to finish reionization.  As a result, the dominant 
contribution, they find, comes from the larger-mass halos which are too big to be 
suppressed and whose number density is growing exponentially fast as reionization
approaches completion.  In that case, the global star formation rate does NOT 
decline sharply as reonization ends.  However, the signature of this low-mass
suppression may still be observable in ways that reflect the earlier onset and
longer ramping up of reionization when low-mass galaxies dominate the early 
phase, before they saturate and the more massive galaxies dominate (e.g. \cite{ahn2012,park2013}). Since the low-mass suppressible dwarf galaxies whose 
contribution eventually drops off are also the ones at the faint end of the 
luminosity function, these expectations are consistent with another, possibly
observable effect suggested by \cite{barkana2000}, of a drop in the galaxy 
LF at the faint end, marking the end of reionization.  Similar conclusions have
also been drawn more recently by, for example,  \cite{duffy2014}.
  
We can use our CoDa simulation results to address both of these questions:
Does the global star formation history show a strong drop as reionization approaches 
the end?  And is there a depression of the faint end of the LF at this time? 
  

According to our results for the global SFR density plotted
in Fig. \ref{f:glob}, there is no sharp downturn in the SFR as reionization
runs to completion, which is not surprising, given the
relative contribution to this SFR from haloes of different
mass plotted in Fig. 9.  The galaxies above 10$^{10}$ \Msun,
which Fig. 7 indicates are basically too massive to be
strongly suppressed by external radiative feedback alone
during reionization, are found to dominate that SFR.
Of course, we cannot say what the SFR history would have
been if we had neglected radiative feedback.  Our
comparison in Fig. 8 of the SFRs in the two test box simulations (SN-feedback only, no radiation, for one, and SN-feedback and radiative transfer in the other) as a function of the final masses of the
haloes within which the stars reside at the end of the simulations
at z = 3, shows that there can, in fact, be a depression of the
SFR even in the higher mass bins when radiation is included, but
much less dramatic than for the lower-mass haloes.  And whether
this is internal feedback or external reionization feedback is
not evident from that plot alone.  Nevertheless, it is a fact
in the CoDa simulations that the global SFR continues to rise
even up to the end of reionization, as seen in Fig. \ref{f:glob}, and contrary to the expectations
of Barkana and Loeb (2000).
However, since radiative star formation suppression is very efficient in CoDa, it results in a change of shape of the faint end of the galaxy UV LF, as shown in Fig. \ref{f:faint}, which uses the smoothed stellar particle masses, as explained in Sec. \ref{s:hmuv}. The $M_{AB1600} = -10 $ to $-12$ galaxies are hosted by haloes affected by radiative suppression by the UV background. As their SFR drops post-reionization, so does the abundance of galaxies in this magnitude range. Therefore, even though we find no sudden drop in the cosmic SFR density, we do see a sharp change in the shape of the LF at the very faint end, due to reionization, as proposed by e.g. \cite{barkana2000}. With respect to a similar exploration by \cite{gnedin2014}, we find, like them, that the total cosmic SFR evolution is smooth throughout reionization because it is dominated by the more massive, radiation-immune galaxies, but unlike them, we find that reionization affects the shape of the faint end LF. }

{

The bright end of the LF increases over time continuously, reflecting the convolution of the rising halo mass function with the halo-mass dependence of the star formation rate for haloes too massive to be suppressed by reionization.  At fainter magnitudes but still brighter than -10, the feedback from reionization on lower-mass haloes is reflected in the depression of the LF toward the end of reionization, in the magnitude range -10 to -12.
For magnitudes even fainter than -10 in Fig. \ref{f:faint}, the LF reflects the passively evolving stars inside the low-mass suppressible haloes which have stopped forming  new stars after they are overtaken by reionization.
The section fainter than -10 shifts the peak to fainter magnitudes over time, as the stars formed inside low-mass haloes before they were fully suppressed get older.
}

\subsection{The dark, not missing, satellites}
\label{s:missing}
\begin{figure}
  {\includegraphics[width=1.\linewidth,clip]{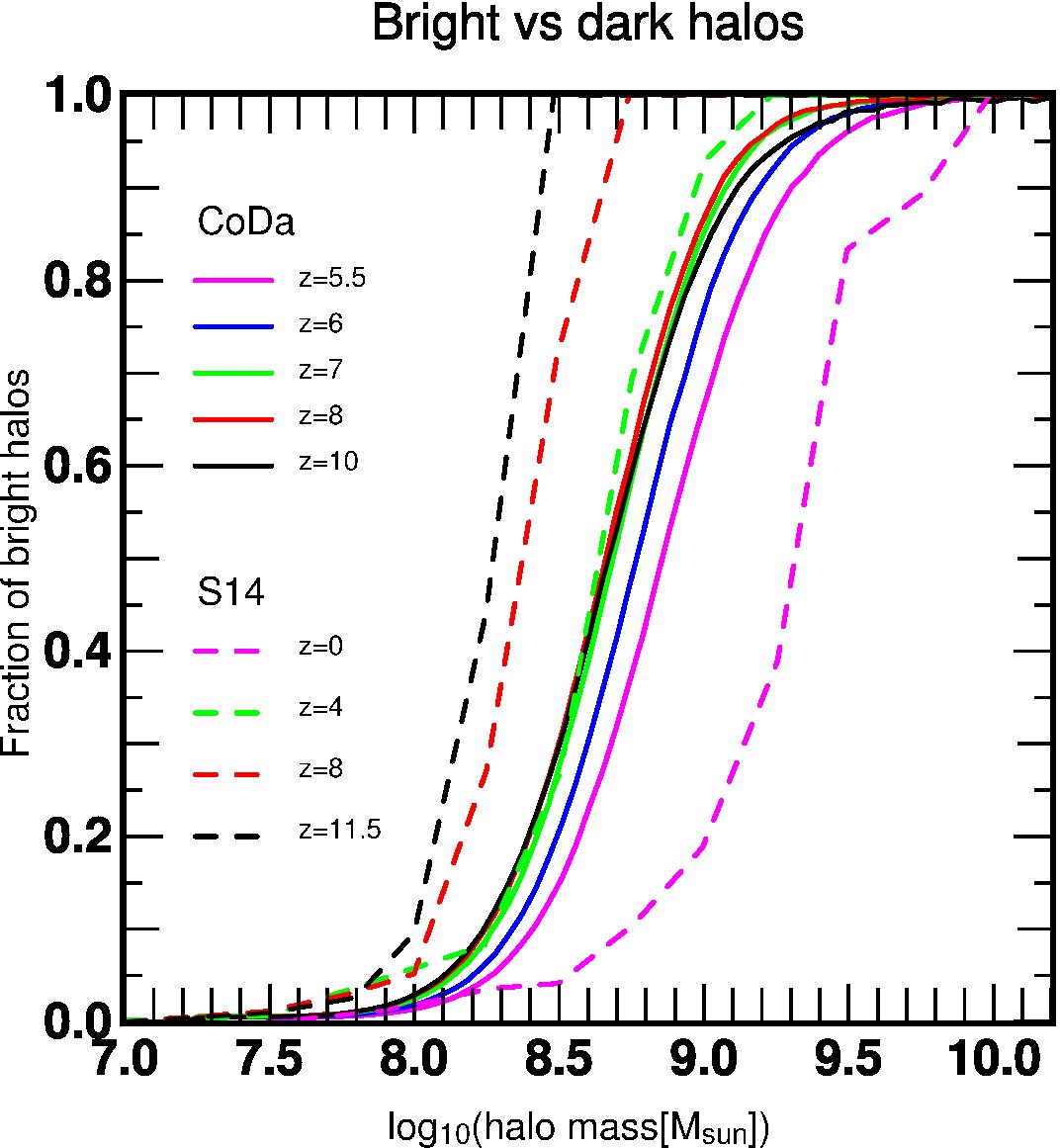}}
\caption{Fraction of bright halos as a function of mass and redshift. The original CoDa redshifts corresponding to the labels are \zcoda$=[5.5,6,7,8,10]$/1.3, i.e. as in Fig. \ref{f:uvlf}. The S14 curves show the results of Sawala et al. 2014.}
\label{f:fbright}
\end{figure}

The suppression of SFR we measured in low mass haloes is an important process in reducing the number of bright haloes, and therefore towards a remedy to the missing satellites problem. Indeed, Fig. \ref{f:fbright} shows that the fraction of bright haloes (i.e. haloes hosting at least one stellar particle) is a steep function of the dark matter halo mass. Below $10^8$ \Msun, more than 99\% of the haloes are dark. The transition between dark and luminous haloes takes place between $10^8$ and $10^9$ \Msun. The transition mass (i.e. the mass for which half of the haloes are dark) does not evolve much prior to reionization but shifts quickly to higher masses near the end of the EoR and just afterwards ($z=6$ and $z=5.5$). This evolution is similar to that seen in \cite{sawala2014}, although the latter used an instantaneous and uniform reionization model, implemented as a uniform heating of the gas (i.e. no coupled radiation-hydrodynamics was performed as opposed to CoDa). Thanks to this simplification in the treatment of the EoR, they were able to carry out their simulation down to z=0 and show that such a mass-dependent reduction in the fraction of luminous haloes is able to quantitatively match the observed abundance of satellites \citep{sawala2015}. Therefore, although CoDa did not run down to z=0, the trends we measured in the suppression of star formation in low mass haloes suggest that CoDa's z=0 low mass halo population will be predominantly dark, alleviating the missing satellites problem.

\section{Conclusions}
\label{s:conclusions}

CoDa (Cosmic Dawn) is a very large, fully coupled radiation-hydrodynamics simulation of galaxy formation in the local universe during the Epoch of Reionization. It was performed on Titan at Oak Ridge National Laboratory using RAMSES-CUDATON deployed on 8192 nodes, using 1 core and 1 GPU per node. This is the first time a GPU-accelerated, fully coupled RHD code has been used on such a scale.
The simulation accurately describes the properties of the gas and its interplay with ionizing radiation, in particular the growth of typical butterfly-shaped ionized regions around the first stars and first galaxies, accompanied by photo-heating and the subsequent progressive smoothing of small scale gas structures. Gas filaments tend not to be self-shielded once the reionization radiation sweeps across them; the flux density of the ionizing radiation internal to these filaments thereafter is identical to that of the background. However, they are indeed slightly more neutral than surrounding voids. Furthermore, they develop a sheathed temperature structure, showing up as a hot tubular envelope surrounding a cooler core, similar in nature to the more spherical accretion shocks seen around forming galaxies. On the other hand, haloes hosting gas denser than  $100 \, \langle \rho \rangle$, when they do not host an ionizing source, show up as photon sinks in ionizing flux density maps, and are therefore self-shielded. 

{The low star formation efficiency assumed in our simulation leads to late reionisation. This can be corrected by a simple contraction of the redshift axis, designed to make reionization complete at $z=6$. This remapping of the redshifts mimics the effect of a modest increase of the star formation efficiency parameter, and brings the simulation in agreement with several other observational constraints of the EoR, including the Thomson scattering optical depth measured by the Planck mission, the evolution of the ionizing flux density and the cosmic star formation rate.}

The star formation rate of individual galaxies of a given mass is on average higher at high redshift and decreases as the Universe expands.
Galaxies below $\sim 2.10^{9}$ \Msun are strongly affected by the spreading, rising UV background: their star formation rate drops by a factor as high as 1000.  {This suppression reflects the great reduction of the gas fraction inside the galaxies that is below 20,000 K, once the galaxy and its environment are exposed to photoionization during reionization.}

{
  This produces a large number of sterile, low-mass haloes by the end of reionization, and a corresponding depression of the faint end of the UV luminosity function in the magnitude range M$_{\rm AB1600} = [-12, -10]$. The latter is similar to an effect suggested by \cite{barkana2000}, although the accompanying sharp drop in the overall star formation rate they also suggested is not supported by the CoDa simulation results. This is explained by the fact that the total star formation rate is dominated at that time by more massive galaxies, too massive to be subject to this strong reionization feedback suppression.

  The halo mass scale below which the CoDa simulation finds galactic halo star formation to be reduced by radiative feedback during the EOR was checked for numerical resolution effects by a set of smaller-box simulations of different mass resolution, described in Appendix B.  The results showed that, while lower resolution than that of the CoDa simulation would increase the value of the threshold halo mass for suppression, a finer resolution by a factor of two in spatial resolution and eight in mass resolution (replacing one cell by 8 cells) finds the same threshold mass, suggesting that the value of the suppression mass is not an artifact of limited numerical resolution.}

Although CoDa did not run down to z=0, the trends we measured in the suppression of star formation suggest that CoDa's z=0 low mass halo population would be predominantly dark, alleviating the missing satellites problem.
In contrast, the gas core of high mass haloes is dense enough to remain cool and/or cool down fast enough to keep forming stars, even if in bursts.
Overall, star formation in the whole box is dominated by $10^{10}$ \Msun haloes for most of the EoR, except at the very end where $10^{11}$ \Msun haloes become frequent enough to take the lead.
The CoDa luminosity functions are in broad agreement with high redshift observations, except for the most luminous objects, where the number counts are subject to strong cosmic variance and may be affected by additional processes, such as evolving dust content and AGN feedback.



\section*{Acknowledgements}
This study was performed in the context of several French ANR (Agence Nationale de la Recherche) projects. PO acknowledges support from the French ANR funded project ORAGE (ANR-14-CE33-0016). NG and DA acknowledge funding from the French ANR for project ANR-12-JS05-0001 (EMMA). The CoDa simulation was performed at Oak Ridge National Laboratory / Oak Ridge Leadership Computing Facility on the Titan supercomputer (INCITE 2013 award AST031). Processing was performed on the Eos, Rhea and Lens clusters. Auxiliary simulations used the PRACE-3IP project (FP7 RI-312763) resource curie-hybrid based in France at Tr\`es Grand Centre de Calcul. ITI was supported by the Science and Technology Facilities Council [grant number ST/L000652/1]. SG and YH acknowledge support by  DFG grant GO 563/21-1. YH has been partially supported by the Israel Science Foundation (1013/12).
AK is supported by the {\it Ministerio de Econom\'ia y Competitividad} and the {\it Fondo Europeo de Desarrollo Regional} (MINECO/FEDER, UE) in Spain through grants AYA2012-31101 and AYA2015-63810-P as well as the Consolider-Ingenio 2010 Programme of the {\it Spanish Ministerio de Ciencia e Innovaci\'on} (MICINN) under grant MultiDark CSD2009-00064. He also acknowledges support from the {\it Australian Research Council} (ARC) grant DP140100198. GY also acknowledges support from MINECO-FEDER under research grants AYA2012-31101 and AYA2015-63810-P. PRS was supported in part by U.S. NSF grant AST-1009799, NASA grant NNX11AE09G, NASA/JPL grant RSA Nos. 1492788 and 1515294, and supercomputer resources from NSF XSEDE grant TG-AST090005 and the Texas Advanced Computing Center (TACC) at the University of Texas at Austin. PO thanks Y. Dubois, F. Roy and Y. Rasera for their precious help dealing with SN feedback in RAMSES and various hacks in pFoF. NG thanks J. Dorval for useful discussions regarding k-d trees which helped with the analysis of this simulation.

\bibliographystyle{mnras}
\bibliography{mybib}

\appendix
\section{CoDa halo mass functions}

\label{s:aMFs}
\begin{figure*}
\begin{center}
\begin{tabular}{cc}
  {\includegraphics[width=0.45\linewidth,clip]{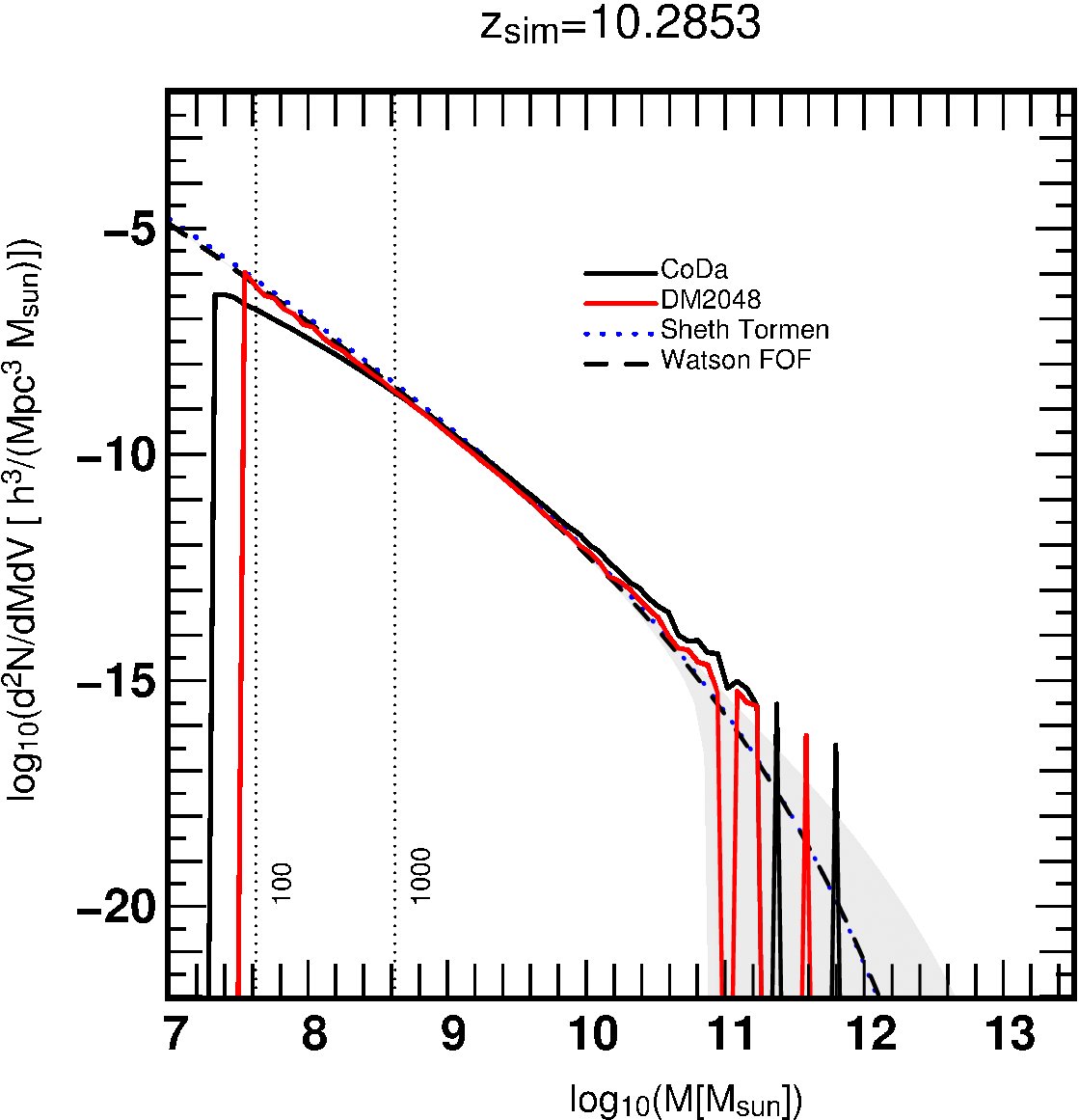}}&
  {\includegraphics[width=0.45\linewidth,clip]{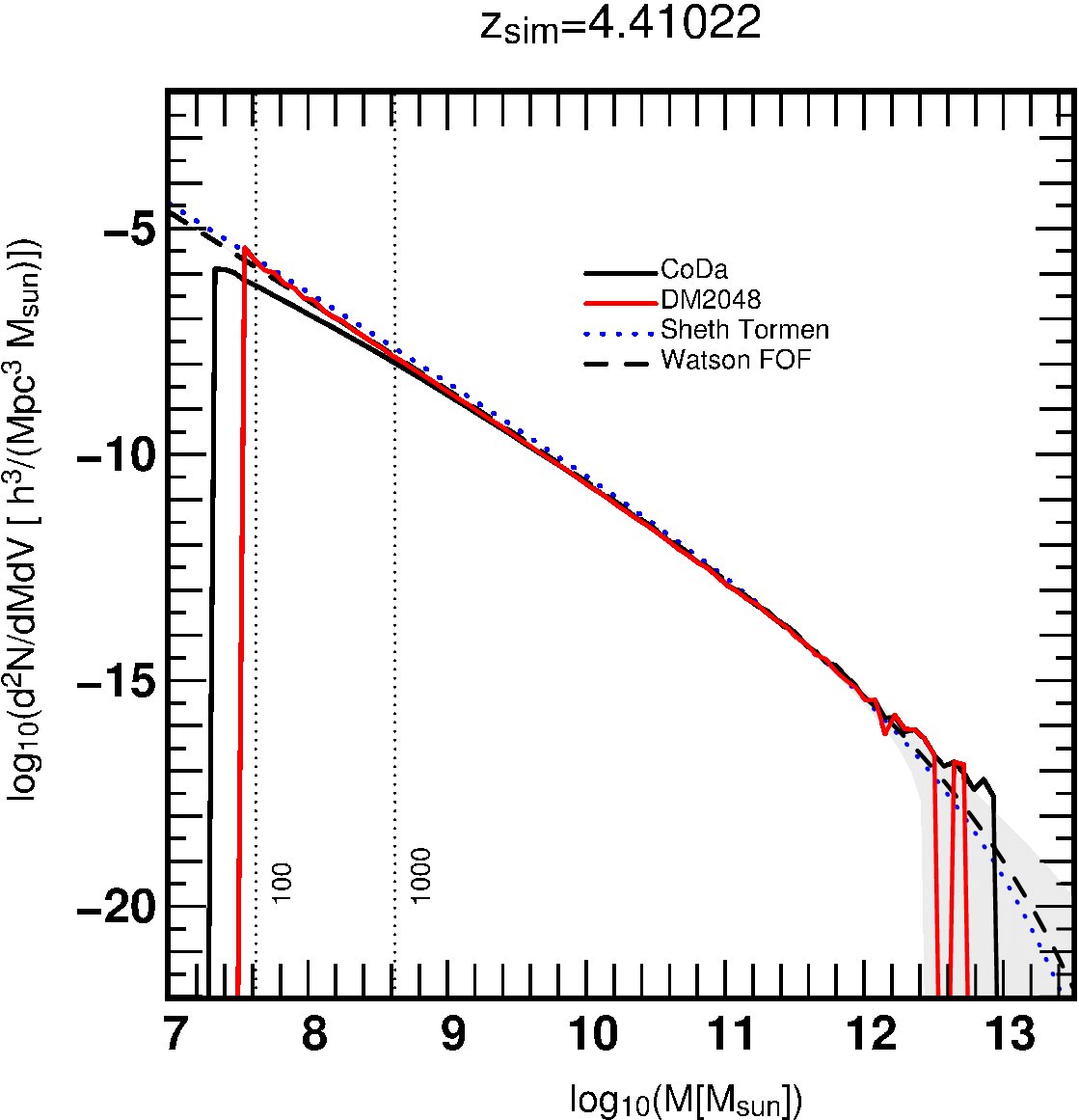}}\\
  {\includegraphics[width=0.45\linewidth,clip]{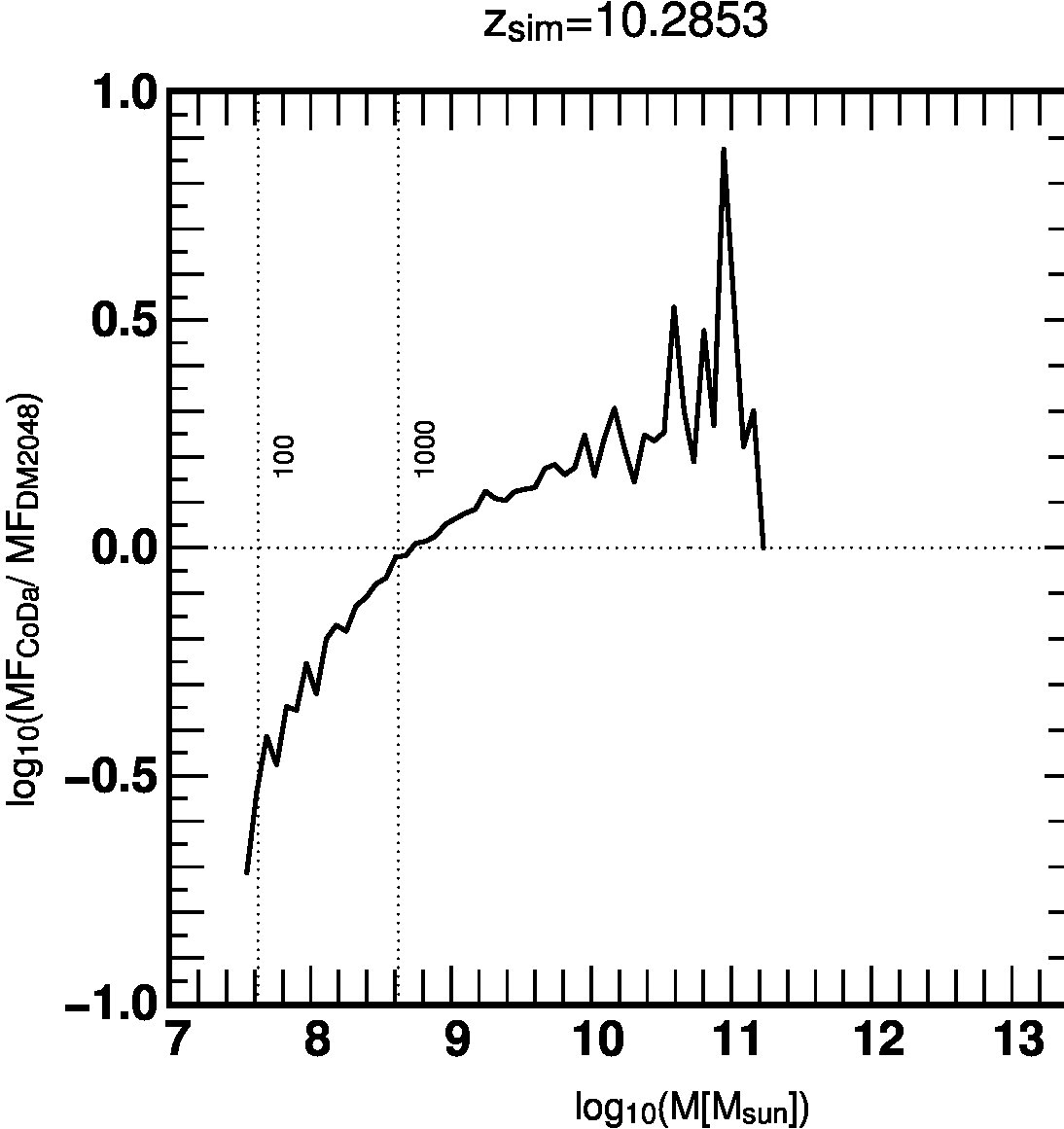}}&
  {\includegraphics[width=0.45\linewidth,clip]{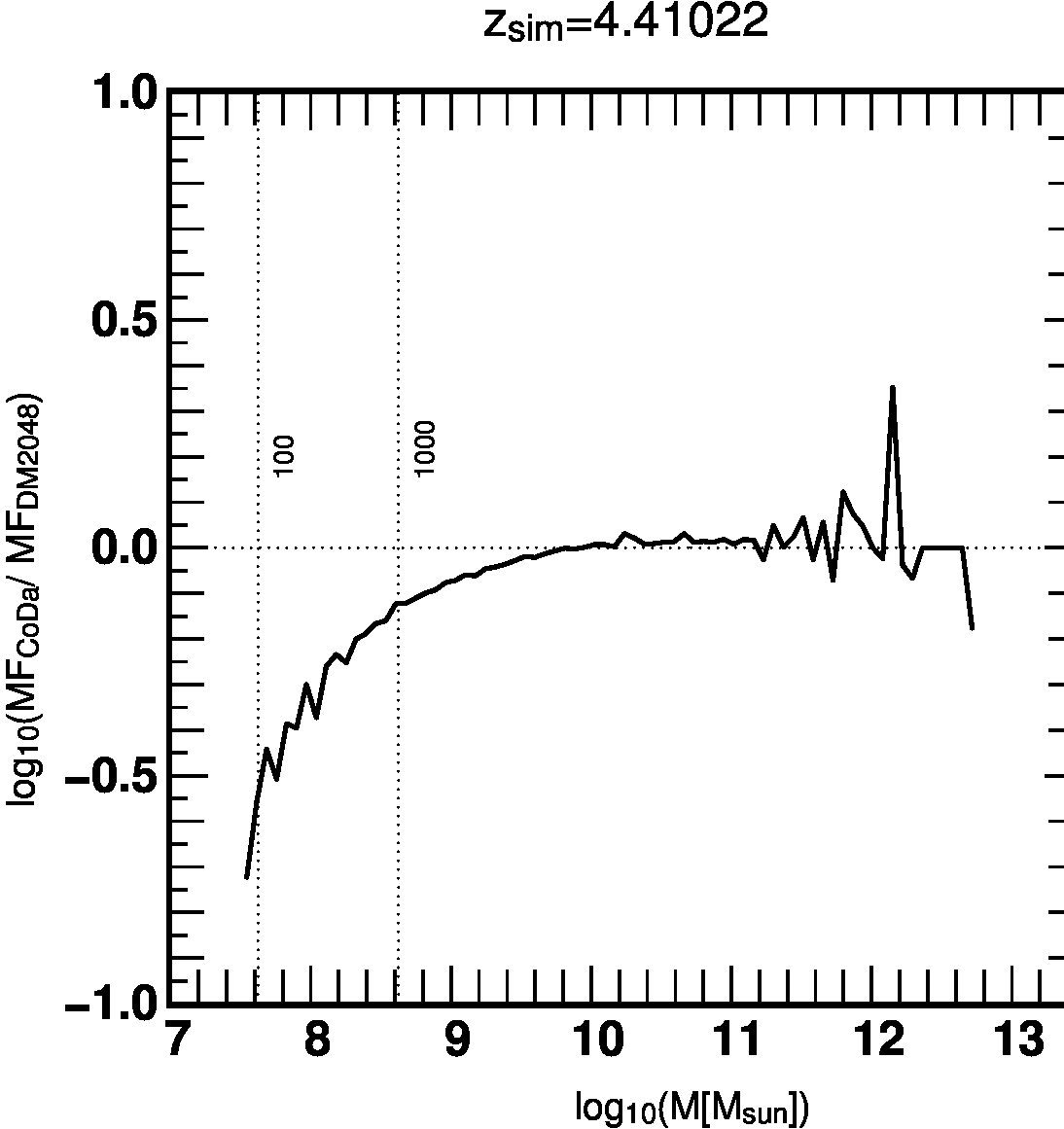}}
\end{tabular}
\end{center}
\caption{Mass functions of our simulations and comparison to literature, at \zcoda=10.28 and \zcoda=4.41. {\em Top:} the black and red solid lines show the mass functions obtained for CoDa and DM2048 respectively. Over-plotted are the theoretical mass function of \citet{sheth2002} and the \citet{watson2013} FoF universal fit (dotted and dashed lines, respectively). The gray area shows the Poisson error bars expected for the Watson FoF mass function. The vertical lines show the mass corresponding to 100 and 1000 particles haloes in CoDa.
{\em Bottom:} The ratio of the the CoDa and the DM2048 mass functions is shown (black solid line).}
\label{f:MFs}
\end{figure*}

The mass functions (hereafter MF) obtained with FoF \citep{roy2014} at \zcoda=10.28 and \zcoda=4.41 are shown in Fig. \ref{f:MFs}, along with a Sheth-Tormen MF \citep{sheth2002} and a FoF universal fit from \cite{watson2013}. At \zcoda=4.41, the CoDa MF is fairly well represented by both fits for haloes larger than 1000 particles, but sits slightly above the Poissonian error bars of the Watson FoF MF (gray area) at the high mass end. To check the origin of this excess, we also plot the MF of the DM2048 simulation, a dark matter only companion simulation run with the N-body code Gadget 2 using the same CLUES initial conditions as CoDa degraded to $2048^3$ resolution. This comparison is useful because DM2048 has about $\sim 20$ times better force resolution than CoDa: tree codes such as Gadget 2 \citep{springel2005} perform well with a force resolution set to $1/20$ to $1/40$ of their average interparticle distance, while CoDa's force resolution is equal to $\sim 1.5$ times the cell size (equal to the average interparticle distance), due to the unigrid scheme, as shown for instance in Fig. 1 of \cite{teyssier02}. The DM2048 can therefore be used as reference: it can inform us on possible artefacts due to CoDa's unigrid gravity solving. The DM2048 MF also displays some excess at the high mass end, comparable to CoDa's. This is confirmed by the bottom panel, showing the ratio of CoDa MF to DM2048 MF, which is close to 1 at the high mass end. 
In contrast, the MFs differ significantly at low masses, with CoDa showing an increasing deficit of low mass haloes compared to DM2048.
We attribute this deficit to CoDa's limited force resolution, which hinders the proper resolution and detection of the smallest haloes. This deficit is present in the \zcoda$=10.28$ MFs as well (left panels), with a similar amplitude. Above 1000 particles, on the other hand, CoDa displays an excess of haloes compared to DM2048. This is also likely caused by the limited force resolution of CoDa: the unresolved low mass haloes provide a large pool of ``free'', untagged particles which the FoF may spuriously link to massive haloes and therefore increase their particle numbers. All in all, in order to mitigate these effects in our analysis, we will refrain from analysing haloes less massive than $10^8$ \Msun. Above this mass, our MF is uncertain by no more than a factor of two on average, but much better in general, in particular above 1000 particles and at lower redshifts.

\section{Numerical Resolution and Suppression Mass}
\label{s:resstudy}

\begin{table}
  \begin{center}
\begin{tabular}{ccc}
  \hline
  Box size & Grid size & Suppression mass\\
  (\hmpc) & & ($10^9$ \Msun) \\
  \hline
  4&$512^3$&1.7\\
  \rowcolor{gray!25}
  8&$512^3$&1.7\\
  16&$512^3$&8  \\
  32&$512^3$&40 \\
  \hline
\end{tabular}
\caption{Parameters of the simulations of the resolution study. The gray row corresponds to the CoDa resolution. The suppression mass is defined as the intersection between the high mass SFR fit and the low mass SFR fit of each simulation.}
  \end{center}
  \label{t:reso}
\end{table}

As described above, CoDa finds that star formation is suppressed in 
low-mass haloes by the feedback associated with reionization, for haloes in
the range roughly below $\sim 2 \times 10^9$ \Msun.  To investigate the dependence
of this suppression mass on the size and mass resolution of the simulation, 
we performed a series of smaller-box simulations,
with input parameters identical to those of CoDa, but a range of resolutions,
both higher and lower than that of CoDa. The simulation parameters are summarized   
in Table \ref{t:reso}, along with the resulting suppression mass obtained in each case.
As seen in Table \ref{t:reso}, a fixed number of particles and cells are adopted for
a hierarchy of different box sizes, so their (space, mass)-resolutions
range from (2,8) to (1/4,1/64) times those of CoDa.  

\begin{figure*}
  {\includegraphics[width=0.9\linewidth,clip]{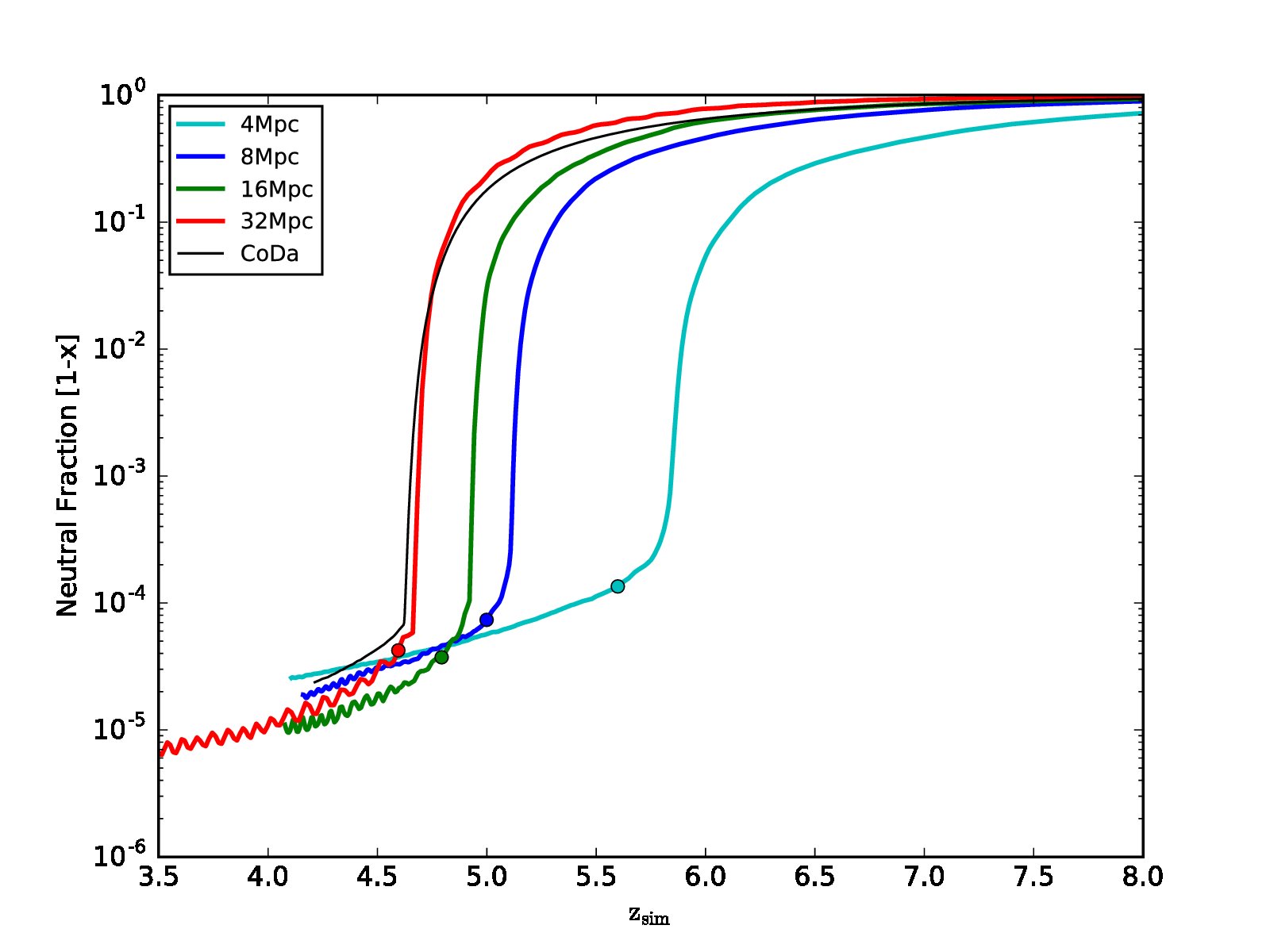}}
  {\includegraphics[width=0.9\linewidth,clip]{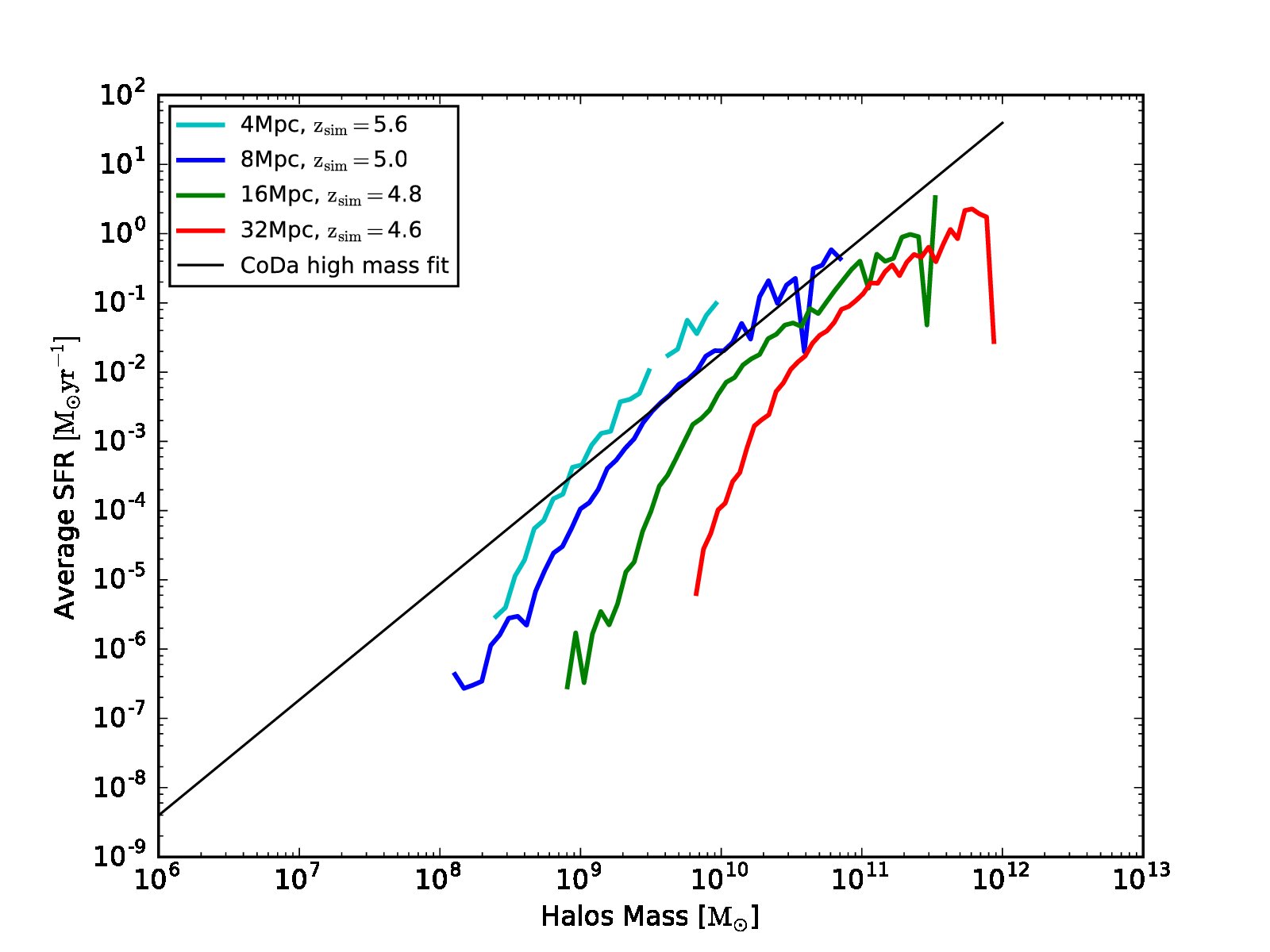}}
  \caption{Neutral fraction and average SFR per halo for the resolution study.}
  \label{f:res1}
\end{figure*}

Since we kept the input parameters for star formation efficiency the same
in all simulations, the higher resolution cases produced a higher star formation
rate (by resolving higher density gas and thereby triggering the subgrid star formation
criterion more often) and, hence, ended reionization earlier,
as shown in Fig. \ref{f:res1}.  In order to compare the halo mass scales of suppression at
different resolutions, therefore, it is necessary to make an adjustment for this
displacement of the reionization time-histories relative to each other
for the different cases.  Fortunately, this is a well-defined operation.
The global reionization histories in Fig. \ref{f:res1} all have in common a very sharp
drop in the neutral fraction at the end of reionization, followed by a much
flatter, slow decline thereafter in the post-reionization era, controlled then
by the average UV background and IGM density.  A well-defined epoch of comparison
is that which corresponds to a fixed interval of time just after this sharp end
of reionization.  Otherwise, if we compared them, instead, at the same cosmic
time, haloes in different cases would have spent a very different amount of time
experiencing the feedback effects of reionization, exposed to the UV background
radiation.  For instance, the end of reionization in the 32 \hmpc simulation happens
at \zcoda $\sim$ 4.6, so at \zcoda = 4.5, haloes have seen the post-reionization
UV background for about 35 Myr.  By contrast, for the highest resolution simulation,
in the 4 \hmpc box, reionization ends at \zcoda $\sim$ 5.8, so at \zcoda = 4.5, haloes would
have seen the post-reionization UV background for about 320 Myr, i.e. almost ten times
longer.  In order to make a meaningful comparison, then, we pick the time of comparison
to be at the same interval of time just after each case's reionization ends.  The
redshifts chosen to compare halo properties for different cases are shown by the
dot on each simulation's neutral fraction evolution (Fig. \ref{f:res1}, left panel) and
listed on the plot in the right panel, of the average SFRs as a function of halo
mass for each case at its corresponding redshift.  

\begin{figure*}
  {\includegraphics[width=1.05\linewidth,clip]{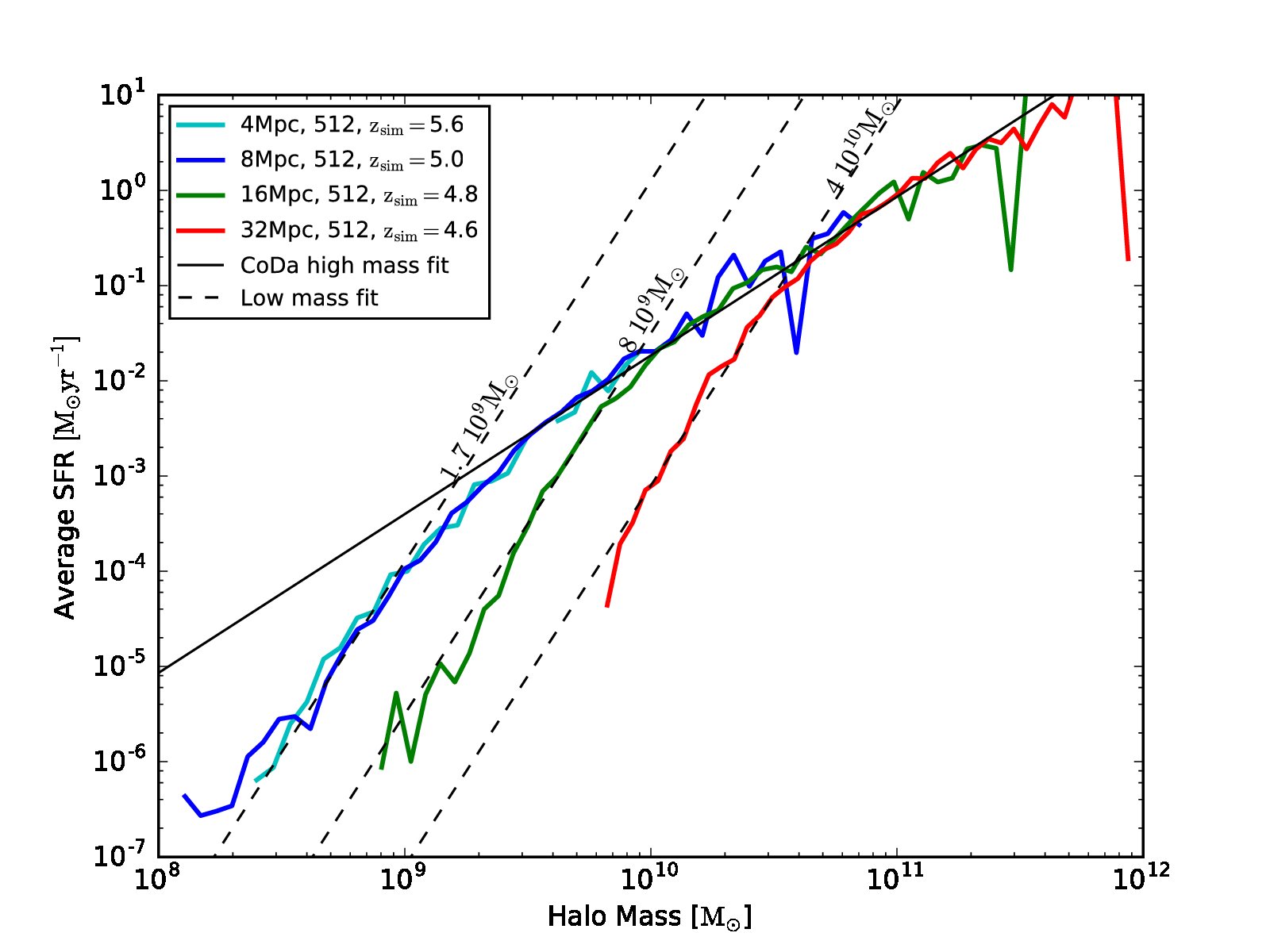}}
  \caption{Average SFR per halo renormalized to match the high-mass branch of the 8 \hmpc simulation (CoDa resolution, solid black line, given by $\log_{10}({\rm SFR})=5/3 \log_{10}({\rm M}) -18.6$). The dashed lines show a power law fit to the low mass branch of each simulation, with a slope ${\rm M}^4$. The numbers next to the dashed lines give the suppression mass as the mass where the low mass fit intersects the high mass fit.}
  \label{f:resresc}
\end{figure*}

The higher star formation rates for cases with higher resolution also mean that
the vertical scale of the SFRs in Fig. \ref{f:res1} should be adjusted to make a direct
comparison of the mass-dependence of the SFRs for different cases.  This, too, is
straightforward, since all of the cases have in common a universal shape for this
mass dependence, which shares the slope of the SFR at high mass, above the
mass scale where suppression occurs, and a turn-over at low-mass where suppression
is occurring.  This is obvious in Fig. \ref{f:resresc}, where we have renormalized the curves
in Fig. \ref{f:res1} (right panel) by adjusting their vertical heights so as to make their
high-mass SFRs lie on top of the power-law fit to this high-mass end of the CoDa SFR, given by $\log_{10}({\rm SFR})=5/3 \log_{10}({\rm M}) -18.6$, and shown as a straight line on the log-log plot in both figures.  It is clear from
Fig. \ref{f:resresc}, now, as we compare the cases from lowest resolution to highest resolution,
that the turn-over, reflecting the suppressed mass range (where Fig. \ref{f:resresc} labels the
value of the mass at which the low-mass turn-over segment joins the high-mass segment
in each case), moves to successively lower mass, until the CoDa resolution is reached in
the 8 \hmpc box.  When the resolution is increased yet again so as to exceed that
of the CoDa simulation, by a factor of two in length and a factor of eight in mass,
in the 4 \hmpc box, the SFR dependence on mass is identical with that
for the CoDa resolution case, as the curves for both cases completely overlap at
all halo masses. This demonstrates that the suppression mass found by the CoDa simulation
is well-enough resolved and not an artifact of numerical resolution.

\bsp	
\label{lastpage}
\end{document}

%% file: abstract_paul.mn.tex
Cosmic reionization by starlight from early galaxies affected
their evolution, thereby impacting reionization, itself.
Star formation suppression, for example, may explain
the observed underabundance of Local Group dwarfs
relative to N-body predictions for Cold Dark Matter.
Reionization modelling requires simulating volumes large enough
$[\sim (100\, {\rm Mpc})^3]$ to sample reionization "patchiness",
while resolving millions of galaxy sources above $\sim 10^8$ \Msun,
combining gravitational and gas dynamics with radiative transfer.
Modelling the Local Group requires initial cosmological
density fluctuations pre-selected to form the well-known
structures of the local universe today.
     Cosmic Dawn ("CoDa") is the first such fully-coupled,
radiation-hydrodynamics simulation of reionization of the local
universe. Our new hybrid CPU-GPU code, RAMSES-CUDATON,
performs hundreds of radiative transfer and ionization
rate-solver timesteps on the GPUs for
each hydro-gravity timestep on the CPUs. CoDa simulated
$(91 {\rm Mpc})^3$ with $4096^3$ particles and cells, to redshift 4.23,
on ORNL supercomputer Titan, utilizing 8192 cores and 8192 GPUs.
Global reionization ended slightly later than observed.
However, a simple temporal rescaling which brings the evolution of
ionized fraction into agreement with observations also reconciles
ionizing flux density, cosmic star formation history, CMB electron
scattering optical depth and galaxy UV luminosity function
with their observed values. Photoionization heating suppressed the star formation of haloes below $\sim 2 \times 10^9$ \Msun, {decreasing the abundance of faint galaxies around $M_{AB1600}=[-10,-12]$.}
For most of reionization, star formation was dominated by haloes between $10^{10} - 10^{11}$ \Msun, so low-mass halo suppression was not reflected by a distinct feature in the global star formation history. Intergalactic filaments display sheathed structures,
with hot envelopes surrounding cooler cores, but do not self-shield,
unlike regions denser than $100 \, \langle \rho \rangle$.